\documentclass[aps,prd,final,twocolumn,superscriptaddress,%
floatfix,preprintnumbers,nofootinbib]{revtex4}

\usepackage{dcolumn} 
\usepackage{amssymb}
\usepackage{amsbsy}
\usepackage[tbtags]{amsmath}
\usepackage{mathtools}

\usepackage{graphicx}

\usepackage{floatflt}
\usepackage[caption=false]{subfig}
\usepackage{overpic}
\usepackage{placeins} % defines the \FloatBarrier

\usepackage{supertabular}

\usepackage{xcolor}
\usepackage{ulem}
\usepackage{wrapfig}

\usepackage{xspace}
\usepackage{slashed}

\usepackage{hyperref}  
\usepackage{verbatim}

\usepackage{dsfont}

\newcommand{\Eins}{\mathds{1}}

\def\clap#1{\hbox to 0pt{\hss#1\hss}}

\def\mathrlap{\mathpalette\mathrlapinternal}
\def\mathclap{\mathpalette\mathclapinternal}

\def\mathrlapinternal#1#2{%
\rlap{$\mathsurround=0pt#1{#2}$}}
\def\mathclapinternal#1#2{%
\clap{$\mathsurround=0pt#1{#2}$}}

\newenvironment{itemize*}%
  {\begin{itemize}%
    \setlength{\itemsep}{0pt}%
    \setlength{\parskip}{0pt}}%
  {\end{itemize}}

\newcommand{\bc}{\begin{center}}
\newcommand{\ec}{\end{center}}
\def\t0{t\text{$=$}0}
\def\ix0{\xi\text{$=$}0}

\newcommand{\bea}{\vspace{-0mm}\begin{eqnarray}}
\newcommand{\eea}{\end{eqnarray}}

\newcommand{\mev}{\operatorname{MeV}}
\newcommand{\fm}{\operatorname{fm}}

\newcommand{\units}[1]{\ensuremath{\,\mathrm{#1}}}
\newcommand{\bra}[1]{\left\langle #1 \right|}
\newcommand{\ket}[1]{\left| #1 \right\rangle}

\newcommand{\Tr}{\mathrm{tr}}

\newcommand{\elll}{l}
\newcommand{\latcfn}{\ensuremath{C}}

\newcommand{\tpm}{\ensuremath{{\,\pm\,}}}

\newcommand{\prp}{\perp}
\newcommand{\transp}{\mathsf{T}}
\newcommand{\GammaOp}{\Gamma}

\newcommand{\GammaTwo}{\ensuremath{\Gamma^\text{2pt}}}
\newcommand{\GammaThr}{\ensuremath{\Gamma^\text{3pt}}}
\newcommand{\GammaDiq}{\ensuremath{\Gamma^\text{diq}}}
\newcommand{\lat}{{\text{lat}}}
\newcommand{\ren}{{\text{ren}}}
\newcommand{\unren}{{\text{unren}}}

\newcommand{\tcdot}{{\cdot}}

\newcommand{\myRe}{\ensuremath{\mathrm{Re}}}
\newcommand{\myIm}{\ensuremath{\mathrm{Im}}}

\newcommand{\dlangle}{{\rlap{\big\langle}\hskip 0.1em\big\langle}}
\newcommand{\drangle}{{\rlap{\big\rangle}\hskip 0.1em\big\rangle}}

\newcommand{\Wline}[1]{\ensuremath{{\mathcal{U}}[#1]}}
\newcommand{\WlineC}[1]{\ensuremath{{\mathcal{U}}{[#1]}}}
\newcommand{\WlineClat}[1]{\ensuremath{{\mathcal{U}}^\text{lat}{[#1]}}}

\newcommand{\Wlineren}[1]{\ensuremath{{\mathcal{U}}^\text{ren}[#1]}}

\newcommand{\vect}[1]{\ensuremath{\boldsymbol{#1}}}
\newcommand{\vprp}[1]{\vect{#1}_\prp}

\newcommand{\mmin}{{\text{min}}}

\newcommand{\quark}{q}
\newcommand{\nucl}[1]{{#1}}

\newcommand{\Afield}{A}

\newcommand{\tAmp}{\widetilde{A}}

\newcommand{\renZ}{Z}

\newcommand{\len}{\ell}

\newcommand{\nplus}{\ensuremath{\bar n}}
\newcommand{\nminus}{\ensuremath{n}}

\newcommand{\TMD}{TMD\xspace}
\newcommand{\TMDs}{TMDs\xspace}

\newcommand{\sW}{\ensuremath{\text{sW}}}

\newcommand{\toddmark}[1]{\ensuremath{\Bigg[#1\Bigg]_{\text{\tiny{odd}}}}}
\newcommand{\Eu}[1]{#1}

\newcommand{\fourint}{\ensuremath{\int\hspace{-1.1em}\mathcal{F}\hspace{0.0em}}}
\newcommand{\xfourint}{\ensuremath{\int\hspace{-1.2em}\mathcal{M}\hspace{0.2em}}}

\newcommand{\conherm}{\ensuremath{(\dagger)}}
\newcommand{\conpar}{\ensuremath{(\mathrm{P})}}
\newcommand{\contime}{\ensuremath{(\mathrm{T})}}
\newcommand{\myeps}{\ensuremath{\epsilon}}

\newcommand{\xmom}[1]{{\ensuremath{[#1]}}}
\newcommand{\ktmom}[1]{{\ensuremath{(#1)}}}
\newcommand{\xktmom}[2]{{\xmom{#1}\ktmom{#2}}}

\newcommand{\sgn}{\mathrm{sgn}}
\newcommand{\nbdash}{\protect\nobreakdash-\hspace{0pt}}

\newcounter{fnnumber}
\newcounter{fnnumberamp}
\newcommand{\MSbar}{\overline{\text{MS}}}

\begin{document}  

\title{Exploring quark transverse momentum distributions with lattice QCD}

\preprint{MIT-CTP~4178, JLAB-THY-10-1266}

\author{B.U.~Musch}
  \affiliation{Theory Center, Jefferson Lab, Newport News, VA 23606, USA}
  \email{bmusch@jlab.org}
\author{Ph.~H\"agler}
  \affiliation{Institut f\"ur Theoretische Physik T39,
   Physik-Department der TU M\"unchen, 85747 Garching, Germany}
  \affiliation{Institut f\"ur Theoretische Physik, Universit\"at
  Regensburg, 93040 Regensburg, Germany}
   \email{phaegler@ph.tum.de}
\author{J.W.~Negele}
  \affiliation{Center for Theoretical Physics, Massachusetts Institute of Technology, Cambridge, Massachusetts 02139, USA}
\author{A.~Sch\"afer}
  \affiliation{Institut f\"ur Theoretische Physik, Universit\"at
  Regensburg, 93040 Regensburg, Germany}

\date{May 25, 2011}

\begin{abstract} 
We discuss in detail a method to study transverse momentum dependent parton distribution functions (\TMDs) using lattice QCD.
To develop the formalism and to obtain first numerical results, we directly implement a bi-local quark-quark operator connected by a straight Wilson line, allowing us to study T-even, ``process-independent'' TMDs. Beyond results for $x$-integrated \TMDs and quark densities, we present a study of correlations in $x$ and $\vprp{k}$. Our calculations are based on domain wall valence quark propagators by the LHP collaboration calculated on top of gauge configurations provided by MILC with $N_f = 2{+}1$ asqtad-improved staggered sea quarks.
\end{abstract}

\maketitle

%\begin{comment}

\section{Introduction}
\label{sec-intro}

The modern approach to the intrinsic quark and gluon structure of hadrons, in particular the nucleon, rests on two pillars, the
generalized parton distributions (GPDs) \cite{Goeke:2001tz,Diehl:2003ny,Belitsky:2005qn,Boffi:2007yc}
and the transverse momentum dependent distribution functions (\TMDs)\footnote{also denoted as ``unintegrated'' PDFs. For an overview and more references, see also \cite{Collins:2003fm}.} \cite{Collins:1977iv,Ralston:1979ys,Collins:1984kg,Mulders:1995dh,Boer:1997nt}.
The theoretical status of GPDs is fairly clear: 
They can be analyzed within the framework of collinear factorization, and
have exact and unambiguous definitions based on off-forward hadron matrix elements
of gauge-invariant quark and gluon operators that are bi-local along the light cone.
Transformed to coordinate (impact parameter, \mbox{$\vprp{b}$-)} space, GPDs have standard interpretations as partonic probability densities 
in the longitudinal momentum fraction $x$ and $\vprp{b}$ \cite{Burkardt:2000za}.
Moreover, they fully incorporate the well-known hadronic form factors and the PDFs, which can be obtained from the 
GPDs from integrations over $x$ and in the forward limit (i.e., by integration over $\vprp{b}$), respectively.
Importantly, at leading-twist level, all-order QCD-factorization theorems have been established that directly relate the GPDs to
particular hard exclusive scattering processes like deeply virtual Compton scattering (DVCS) \cite{Collins:1998be}.
In this sense, the GPDs are process-independent, universal quantities.
Moments of GPDs have been studied in lattice QCD since 2002, and for a review we refer to \cite{Hagler:2009ni}.
A calculation of GPDs performed in the same lattice framework as employed in this work has been 
presented recently by the LHP collaboration in Ref. \cite{Bratt:2010jn}.

\begin{wrapfigure}{r}{0.2\textwidth}
  \begin{centering}
  \vspace{-4mm}
  \includegraphics[width=0.2\textwidth]{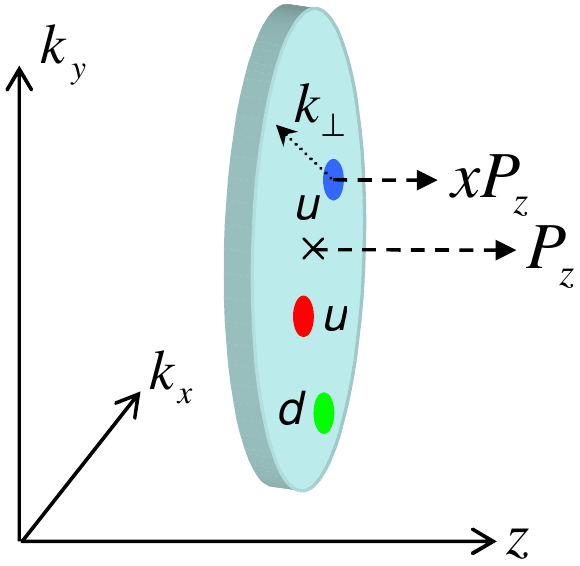}
  \caption{\label{fig-ill} Illustration of the transverse momentum distribution of quarks in the proton.}   
  \end{centering}
\end{wrapfigure}

Complementary information on the structure of hadrons is encoded in the TMDs.
Naively, they can be thought of as having a probabilistic interpretation and
describing the distribution of, e.g., quarks in a nucleon with respect to $x$ and 
the intrinsic transverse momentum $\vprp{k}$ carried by the quarks, as illustrated in Fig.~\ref{fig-ill}.
A great deal of the motivation to study TMDs hinges on their expected direct relation 
to the well-known "integrated" PDFs by an integration over $\vprp{k}$.
TMDs play a central role in the description and understanding of semi-inclusive deep inelastic scattering (SIDIS) processes
and re\-la\-ted single-spin asym\-metries.
Apart from this common folklore, however, one finds that the theoretical situation concerning TMDs is, in contrast to the GPDs, 
much more challenging. 
In contrast to the GPDs, the framework the TMDs are embedded in goes beyond collinear factorization, and the
theoretical concepts needed have not yet been fully developed\footnote{For a recent attempt in this direction see \cite{Collins:2007ph}}.
To explain some of the challenges in more detail, we begin with the definition of a basic, 
momentum dependent correlation function,
\begin{align}
  \Phi^{[\GammaOp]}_{\quark} (k,P,S;\mathcal{C}) & = \int \frac{d^4 \elll}{(2\pi)^4} \ 
  e^{-ik \cdot \elll} \nonumber \\ & \times 
  \underbrace{ \frac{1}{2} \bra{\nucl{P,S}}\ \bar \quark(\elll)\, \GammaOp\ \WlineC{\mathcal{C}_\elll}\ \quark(0)\ \ket{\nucl{P,S}} }_{\displaystyle \widetilde \Phi^{[\GammaOp]}_{\quark}(\elll,P,S;\mathcal{C}) }\ ,
  \label{eq-corr}
\end{align}
where $\ket{\nucl{P,S}}$ is a nucleon state of momentum $P$ and spin $S$ and $\GammaOp$ represents some Dirac matrix to be specified 
below\footnote{For better readability, we will frequently omit the arguments $\quark$, $P$, $S$ and $\mathcal{C}$ in the following.}.
The Wilson line $\WlineC{\mathcal{C}_\elll}$
is essential in order to ensure the gauge invariance of the expression.
As usual, it can be represented by a path ordered exponential, see Eq.~\eqref{eq-wlinecont}.
In a frame where the nucleon has a large momentum in $+$-direction (cf. appendix \ref{sec-conv}), $k^-$ is suppressed by a factor $\sim 1/P^+$,
and it is sufficient to consider the $k^-$-integrated correlator
\begin{equation}
  \Phi^{[\GammaOp]} (x,\vprp{k};P,S;\mathcal{C}) \equiv \int dk^-\,\Phi^{[\GammaOp]} (k,P,S;\mathcal{C})\vert_{k^+=xP^+}\,.
  \label{eq-corrkminusint}
\end{equation}
Based on its symmetry transformation properties (cf. appendix \ref{sec-symtraf}), this correlator can be parametrized in terms of real-valued \TMDs $f_{1,\quark}(x,\vprp{k}^2;\mathcal{C})$, $g_{1,\quark}(x,\vprp{k}^2;\mathcal{C})$, etc. \cite{Mulders:1995dh,Boer:1997nt,Goeke:2005hb}. 
Concrete examples will be given in section \ref{sec-parametrization}.
As we will see in the following, the correlator in Eq.~\eqref{eq-corr}, and in turn the \TMDs parametrizing it, will in general depend non-trivially
on the form of the path $\mathcal{C}$ along which the quark fields at the origin and at $\elll$ are connected.
The question that comes to mind is if the form of the path is in fact uniquely determined in some way, or 
in the other extreme completely arbitrary.
From a theoretical perspective, as long as the operator can be regularized and renormalized 
(including possible necessary modifications of the basic definition Eq.~\eqref{eq-corr}), 
we are in principle allowed to consider any path we like to probe the internal structure of the nucleon in such a framework.
Of strong immediate interest are of course the types of correlators and paths that can be directly related to experimental observables.
\begin{figure}[btp]
	\centering%
	\includegraphics[width=\linewidth]{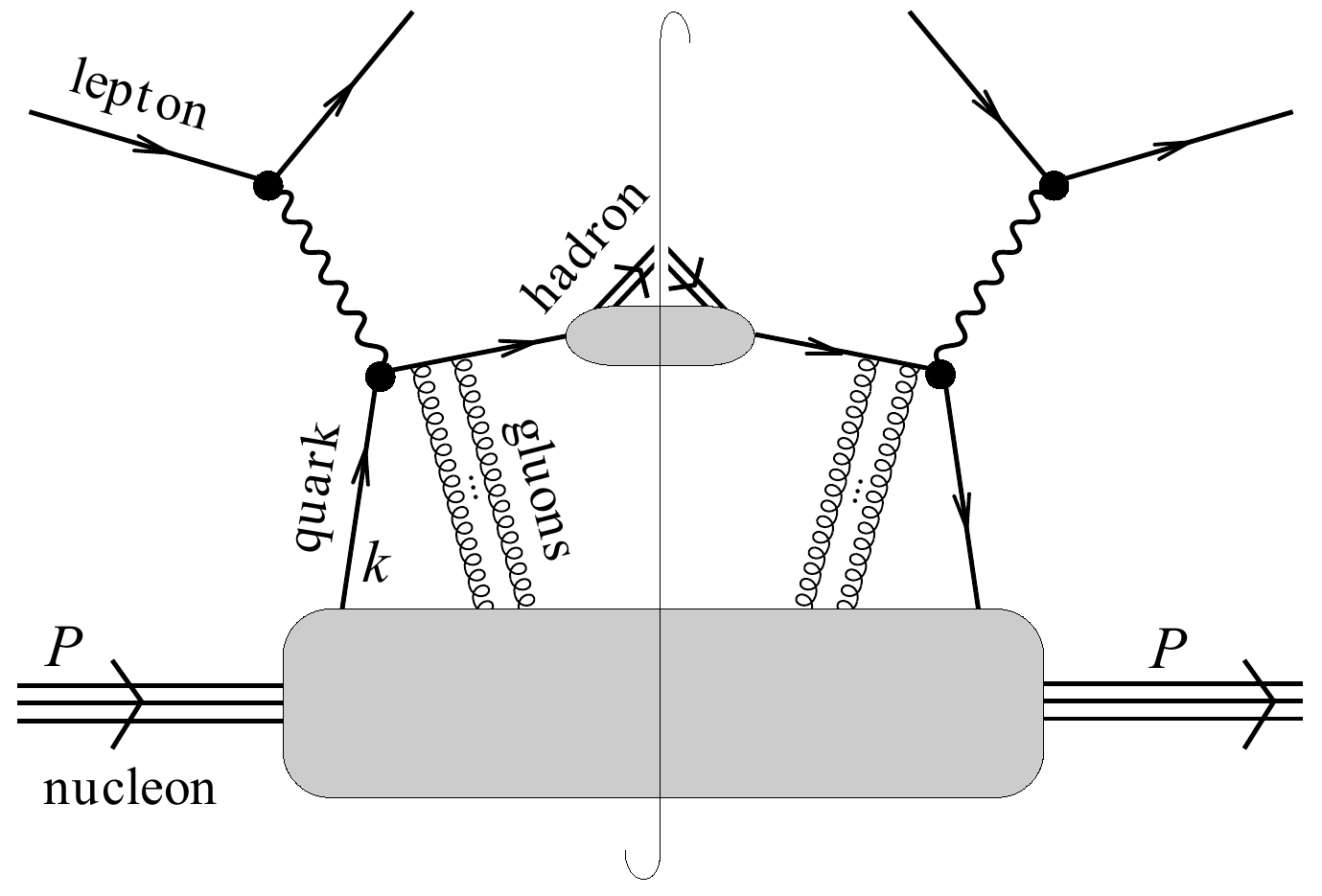}
	\caption{%
		Illustration of the leading contribution to SIDIS.
		\label{fig-SIDISdiagram}
		}
\end{figure}

\begin{figure}[btp]
	\centering%
	\subfloat[][]{%
		\label{fig-link-staple}%
		\includegraphics[]{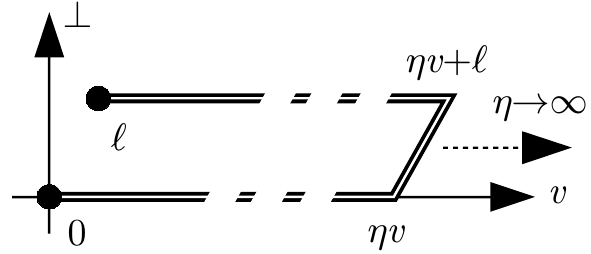}
		}\hfill%
	\subfloat[][]{%
		\label{fig-link-straight}%\
		\includegraphics[trim=0 0 120 0,clip=true]{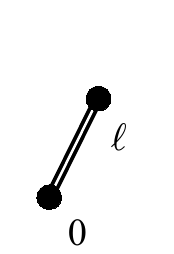}
		}%
	\caption[SIDIS diagram]{%
		\subref{fig-link-staple}\ %
			Staple-shaped gauge link as in SIDIS and DY.\par%
		\subref{fig-link-straight}\ %
			Straight gauge link.\par			
		\label{fig-links}%
		}
\end{figure}

A prominent example is the SIDIS process illustrated in Fig.~\ref{fig-SIDISdiagram}, 
e.g. $n(P)+\gamma^*(q)\rightarrow h(P_h)+X$, in a kinematical region where the photon virtuality is large, $Q^2=-q^2\gg m_N^2$, 
and the measured transverse momentum of the produced hadron is $P_{h\perp}\sim \mathcal{O}(\Lambda_{QCD})$.
In this context,
it is well known that the Wilson line $\WlineC{\mathcal{C}_\elll}$ generically represents gluon mediated interactions of the struck quark
with the nucleon remnants. 
More precisely, in perturbation theory, 
these final state interactions correspond to diagrams where arbitrarily many gluon lines are exchanged, as indicated in the upper part of Fig. \ref{fig-SIDISdiagram}.
From the resummed gluon exchanges (see, e.g., \cite{Bomhof:2006dp}), one obtains at tree-level a 
Wilson line that has the form of a staple of infinite extent, 
as depicted in Fig.~\ref{fig-link-staple}, running along the light-cone to infinity and back. 
With straight Wilson lines denoted by $\Wline{y,z}$,
the staple shaped gauge link is 
given by $\WlineC{\mathcal{C}_\elll^{(\infty v)}} \equiv \Wline{\elll,\infty v + \elll}\Wline{\infty v + \elll,\infty v}\Wline{\infty v, 0}$, where
the direction $v$ is lightlike, $v_\text{SIDIS}=\nminus$.
Importantly, it is not possible to "gauge away" effects of the Wilson lines by choosing, e.g., the light cone gauge $\nminus\cdot A=0$,
since the transverse part of the gauge link, depending on the gauge fields at infinity,
contributes in such gauges \cite{Brodsky:2002cx,Ji:2002aa,Collins:2002kn,Belitsky:2002sm}.
Furthermore, it is essential to note that the form of the path depends on the type of process under consideration.
In particular, it turns out that in the Drell-Yan (DY) process, initial state interactions lead 
to a gauge link that is again staple-like but oriented in the opposite direction, $v_\text{DY}=-v_\text{SIDIS}$, i.e. 
one finds past- in contrast to future-pointing Wilson lines \cite{Collins:2002kn}.
These well known observations clearly show that even in a phenomenological context, already at tree-level in perturbation theory
the form of the Wilson line connecting the quark fields in Eq.~\eqref{eq-corr}, 
and therefore the structure of the correlation function itself, is non-unique.
On the level of the TMDs, the different directions $v$ for SIDIS and DY 
translate for example into a sign change of so-called time reversal odd \TMDs such as the Sivers function, $f_{1T}^\prp(x,\vprp{k}^2;\mathcal{C}^{(\infty n)})=-f_{1T}^\prp(x,\vprp{k}^2;\mathcal{C}^{(-\infty n)})$. 
The important message is that the TMDs can therefore be seen as non-universal objects, 
albeit the "breaking" of universality is exactly calculable, at least in the considered cases.
Another way of formulating these observations is to
consider linear combinations (the sum and difference) of future- and past-pointing Wilson-line operators, leading to "T-even" and "T-odd" correlators that
are separately process-independent. 
The non-universality can then be seen in the fact that there exist two \emph{distinct classes} of TMDs, 
the T-even and T-odd TMDs, which are based on two types of operators with fundamentally different gauge link structures
\cite{Boer:2003cm}.
We note that additional, even more complex gauge-link structures have been found in the framework of tree-level analyses of
$2\rightarrow 2$ hadron scattering processes \cite{Bomhof:2007xt}.
However, a more recent study \cite{Rogers:2010dm} argues that a generalized \TMD factorization of this kind (see also \cite{Bacchetta:2005rm,Bomhof:2006dp}) cannot be achieved for such processes.
The argument is based on a model calculation that gives an explicit example where it is impossible to find standard Wilson line structures that allow factorization.
%A recent study \cite{Rogers:2010dm} argues that for such processes factorization breaks down.
%

In summary, for SIDIS and the Drell-Yan process at tree-level one finds a standard factorization of hard and soft parts, where the latter,
illustrated in the lower part of Fig. \ref{fig-SIDISdiagram}, is
represented by the correlator in Eq.~\eqref{eq-corr}, with a Wilson-line of the form shown in Fig.~\ref{fig-link-staple}. 
This picture changes completely as soon as loop-corrections are taken into account in the lower part of Fig. \ref{fig-SIDISdiagram}.
Already at leading one-loop level, one finds that the lightlike sections of the Wilson lines 
lead to divergences due to light-cone singularities in the additional gluon propagator \cite{Collins:1981uw}.
Hence, to obtain well defined amplitudes, the basic definition in Eq.~\eqref{eq-corr} with the staple-like Wilson line along the light-cone 
has to be modified.
Different improved definitions of TMDs and strategies to remove the divergences have been proposed and discussed in the literature 
\cite{Collins:2003fm,Collins:2004nx,Ji:2004wu,Hautmann:2007uw,Chay:2007ty,Cherednikov:2008ua,Collins:2008ht}.
To illustrate the theoretical status of these issues, we briefly discuss in the following two different approaches.
In \cite{Ji:2004wu}, a QCD factorization theorem for SIDIS has been established at leading 
one-loop level\footnote{The validity with respect to higher order corrections is still under debate.}, 
where the vector $v$ has been taken slightly off the lightcone (i.e., timelike) to
regularize the light-cone divergences. 
This leads to an additional dependence of the correlators on the energy of the incoming hadron, or the variable $\zeta=(2 P \tcdot v)^2/v^2$,
which is described by a known 
evolution equation in certain kinematical regions.
Furthermore, in order to cancel out extra soft contributions from the basic correlator, 
the definition Eq.~\eqref{eq-corr} has to be modified to include appropriate vacuum expectation values of products of Wilson lines.
An important point is, however, that in this approach the light-cone limit, $v^2\rightarrow0$, cannot be taken exactly, and that 
no direct relation to the standard PDFs, e.g. through an integration over $\vprp{k}$, can be established.
This leads clearly to some tension with respect to the increasing number of phenomenological analyses and parametrizations
of SIDIS experiments (e.g., in Ref. \cite{Anselmino:2008sga}), which on the one hand should be based on a QCD factorization theorem, but on the other hand, so far make use of the assumption that the involved TMDs reduce to the PDFs after integration over $\vprp{k}$.

An alternative definition of TMD-correlators 
has been worked out in Ref. \cite{Cherednikov:2008ua}.
It is based on an exactly light-like direction $v$ and a different regularization of the light-cone singularities 
involving certain pole-prescriptions.
In order to remove the prescription dependence at least at one-loop level,
sections of the gauge-link path that run along the transverse direction to infinity, i.e., from  
$(\infty v + 0_\perp)$ to $(\infty v + \infty_\perp )$ and back to $(\infty v + \elll_\perp)$,
have to be explicitly taken into account\footnote{In contrast to the covariant gauge used in Ref. \cite{Ji:2004wu}, the transverse sections
at $\infty v$ cannot be neglected in the light-cone gauge that was employed in this case.}.
In addition, a soft counter term has to be included in the modified definition of the correlation function in Eq.~\eqref{eq-corr}. 
A clear advantage of this approach is that the (dimensionally regularized) $\vprp{k}$-integral of the TMDs defined in this way reproduces
the standard PDFs.
However, it is not known to this date if the TMD-correlator defined in Ref. \cite{Cherednikov:2008ua} is part of any QCD-factorization theorem of a physical
process, which would be a necessary condition for any solid phenomenological analyses.

In summary, the current situation turns out to be quite challenging.
Finding a definition of \TMDs that allows to relate them to the PDFs, and that at the same time is part of a proper factorization theorem
for, e.g., SIDIS, is non-trivial and still a matter of ongoing research.

In view of the issues discussed so far, and the importance of TMDs for our understanding
of hadron structure, we propose to start a program of systematic non-perturbative studies 
of the relevant correlation functions in the framework of lattice QCD, in addition to the ongoing perturbative investigations.
Keeping in mind that the lattice discretization of QCD represents a manifestly gauge-invariant scheme with build-in cutoff, and that
the non-perturbative evaluation of the path integrals doesn't require a fixing of the gauge (which in the perturbative analyses
contributes substantially to the difficulties), the lattice approach has the potential to
provide new insights into the general properties of possible TMD-correlators from a completely different perspective.
The long term plan is to perform non-perturbative studies of matrix elements of manifestly 
non-local operators with different gauge-link structures,  
of potentially relevant soft factors (vacuum expectation values of Wilson-lines and -loops), and to 
get quantitative information from first principles about the $x$-and $\vprp{k}$-dependences of the TMDs.

The direct implementation of non-local operators like $\bar \quark(\elll)\, \GammaOp\ \WlineC{\mathcal{C}_\elll}\ \quark(0)$ on the lattice
is still a novelty.
Therefore, our first steps will be based on simplified operator structures, allowing us to 
establish the basic ideas, formalism and methodology, and to perform first studies of lattice related
issues like the renormalization of potential power-divergences of the Wilson-lines and certain discretization effects.
Specifically, taking into account the fact that there is no straightforward way to realize lightlike gauge links on the lattice, we have performed first investigations with a simple path geometry: We employ a direct, straight Wilson line $\WlineC{\mathcal{C}_\elll^{\text{sW}}}=\Wline{\elll, 0}$, see Fig.~\ref{fig-link-straight}. 
The straight Wilson line (``\sW'') is a process-independent choice that serves us here as a starting point for exploratory calculations.
Note that time reversal odd \TMDs vanish by symmetry for straight Wilson lines, e.g., $f_{1T}^\prp(x,\vprp{k};\mathcal{C}^\text{sW})=0$.
Although our \TMDs defined in this way are thus not directly related
to those defined and used in the literature and for the description of, e.g., SIDIS, they
still can be seen as being elements of the general class of "process-independent, T-even" TMDs, as discussed above.
Although being preliminary, our computations therefore provide some semi-quantitative information about this class of TMDs,
in particular with respect to their signs and (relative) sizes. 
First numerical results have already been presented by us in Ref.~\cite{Hagler:2009mb}, where we observed clear signals
for several TMDs, corresponding to sizable correlations in $\vprp{k}$ and the quark and nucleon spins, $s$ and $S$, 
leading to visibly deformed densities of (polarized) up- and down-quarks in a (polarized) nucleon.
Here, we give a more detailed description of our techniques, and discuss critical issues as well as possible improvements and extensions.

\section{Parametrization in terms of TMDs and invariant amplitudes}
\label{sec-parametrization}

We now come back to the parametrization of the $k^-$-integrated correlator in Eq.~\eqref{eq-corrkminusint} in terms of TMDs.
Following the common conventions in the literature \cite{Ralston:1979ys,Tangerman:1994eh,Mulders:1995dh,Boer:1997nt},
we decompose the correlator for $\Gamma=\gamma^+,\gamma^+\gamma^5,i\sigma^{i+}\gamma^5$ 
into the leading twist-2 \TMDs as follows:
\begin{align}
	\Phi^{[\gamma^+]}(x,\vprp{k}) & = f_1 - \toddmark{\frac{\myeps_{ij}\, \vect{k}_{i}\, \vect{S}_{j}}{m_N}\ f_{1T}^\prp} \label{eq-phigammaplus}\ , \displaybreak[0]  \\
	\Phi^{[\gamma^+\gamma^5]}(x,\vprp{k}) & = \Lambda\, g_{1} + \frac{\vprp{k} \cdot \vprp{S}}{m_N}\ g_{1T} \label{eq-phigammaplusgfive} \ , \displaybreak[0] \\
	\Phi^{[i\sigma^{i+}\gamma^5]}(x,\vprp{k}) & = \vect{S}_i\ h_{1}   +  \frac{(2 \vect{k}_i \vect{k}_j - \vprp{k}^2 \delta_{ij}) \vect{S}_j}{2 m_N^2}\, h_{1T}^\prp \nonumber \\ & + \frac{\Lambda \vect{k}_i}{m_N} h_{1L}^\prp +   \toddmark{\frac{\myeps_{ij} \vect{k}_{j}}{m_N} h_1^\prp}\ .\label{eq-phisigmaplusi} 
\end{align}
% be careful! Goeke,Metz,Schlegel introduce the h's with the oppositie sign! Probably a typo!
Here $i,j=1,2$ are indices denoting transverse directions. The \TMDs in square brackets are odd under time reversal and absent for
our choice of a straight Wilson line. 
For other Dirac structures $\Gamma$, the correlator $\Phi^{[\Gamma]}(x,\vprp{k})$ is suppressed by factors $m_N/P^+$ or $(m_N/P^+)^2$, 
corresponding to contributions of higher twist-3 and twist-4, respectively. 
The parametrizations of the twist-3 correlators are given by \cite{Mulders:1995dh,Boer:1997nt,Goeke:2005hb}
\begin{align}
	\Phi^{[\Eins]}(x,\vprp{k}) & = \frac{m_N}{P^+}\Bigg\{e 
	                        - \toddmark{\frac{\myeps_{ij}\, \vect{k}_{i} \vect{S}_{j}}{m_N} e_T^\prp } \Bigg\}
	\label{eq-phiEins}\ , \displaybreak[0] \\
	\Phi^{[i\gamma^5]}(x,\vprp{k}) & = \frac{m_N}{P^+}\toddmark{\Lambda e_L + \frac{\vprp{k} \cdot \vprp{S}}{m_N}\ e_T }
	\label{eq-phiig5}\ , \displaybreak[0] \\
	\Phi^{[\gamma^i]}(x,\vprp{k}) & = \frac{m_N}{P^+} \Bigg\{ \frac{\vect{k}_i}{m_N} f^\prp 
	+ \toddmark{-\frac{\vect{k}_i\myeps_{jk}\, \vect{k}_j\vect{S}_{k}}{m_N^2}\ f_{T}^{'\prp} \nonumber\\
	             &+ \Lambda \frac{\myeps_{ij}\, \vect{k}_{j}}{m_N} f_{L}^\prp
	             + \frac{\vprp{k} {\cdot} \vprp{S} \myeps_{ij}\, \vect{k}_{j} }{m_N^2}\ f_{T}^\prp }
	\Bigg\} \label{eq-phigammai} , \displaybreak[0] \\
	\Phi^{[\gamma^i\gamma^5]}(x,\vprp{k}) & = \frac{m_N}{P^+}\Bigg\{
	\vect{S}_i g'_T + \frac{\Lambda \vect{k}_i}{m_N} g_{L}^\prp \nonumber\\
	& + \frac{\vprp{k} \cdot \vprp{S} \vect{k}_i }{m_N^2}\ g_{T}^\prp
	  - \toddmark{\frac{\myeps_{ij}\, \vect{k}_{j}}{m_N} g_{}^\prp }
	\Bigg\} \label{eq-phigammaigfive}\ , \displaybreak[0] \\
	\Phi^{[i\sigma^{ij}\gamma^5]}(x,\vprp{k}) & = \frac{m_N}{P^+} \Bigg\{ \frac{\vect{S}_{[i} \vect{k}_{j]}}{m_N}  h_{T}^\prp	 
	 - \toddmark{\myeps_{ij} h} \Bigg\}
\label{eq-phisigmaij} \ , \displaybreak[0]  \\
\Phi^{[i\sigma^{+-}\gamma^5]}(x,\vprp{k}) & =\frac{m_N}{P^+}
  \Bigg\{
	\Lambda h_{L}	+ \frac{\vprp{k} \cdot \vprp{S}}{m_N}\ h_{T}
	\Bigg\}
\label{eq-phisigmapm} \ ,
\end{align}
where square brackets around pairs of indices denote antisymmetrization, $a^{[\mu}b^{\nu]} \equiv a^\mu b^\nu - a^\nu b^\mu$.
Naively, one might ask how the TMDs defined in Eqns.~\eqref{eq-phigammaplus} to \eqref{eq-phisigmapm}, that
are classified according to twist and part of an expansion of correlators in $m_N/P^+$ with large $P^+$,
can ever be accessed in lattice QCD simulations, where the nucleon is at rest or has only a small non-zero three-momentum.
A first step towards the resolution of this potential contradiction is a \emph{frame independent} parametrization of 
$\widetilde \Phi^{[\GammaOp]}_{\quark}(\elll,P,S;\mathcal{C})$ on the right hand side in Eq.~\eqref{eq-corr} 
in terms of Lorentz-invariant amplitudes $\widetilde{A}_i(\elll^2,\elll \tcdot P)$. 
As will be explained in the following sections, the non-local operator technique allows us to evaluate the $\elll$-dependent matrix element $ \widetilde \Phi^{[\GammaOp]}_{\quark}(\elll,P,S;\mathcal{C}) $ directly on the lattice.

Analogous to the procedure outlined in Ref. \cite{Tangerman:1994eh}, we write down all Lorentz-covariant structures compatible with the properties of $ \tilde \Phi^{[\GammaOp]}_{\quark}(\elll,P,S;\mathcal{C}) $ under symmetry transformations, see appendix \ref{sec-symtraf}. For straight gauge links $\mathcal{C}^\text{sW}$, we obtain:
\begin{align}
	\tilde{\Phi}^{[\Eins]} & = 
		2\, m_N\, \tAmp_1 \, ,\displaybreak[0] \nonumber\\
	\tilde{\Phi}^{[\gamma^5]} & = 0 \,, \displaybreak[0] \nonumber\\
	\tilde{\Phi}^{[\gamma^\mu]} & =
		2\,P^\mu\,\tAmp_2 
		+ 2i\,{m_N}^2\,\elll^\mu\,\tAmp_3\,,
		\displaybreak[0] \nonumber\\
	\tilde{\Phi}^{[\gamma^\mu \gamma^5]} & = 
		- 2\, m_N\, S^\mu\, \tAmp_6
		- 2i\,m_N\,P^\mu (\elll \cdot S)\, \tAmp_7 \nonumber\\ & 
		+ 2\,{m_N}^3\,\elll^\mu (\elll \cdot S)\,\tAmp_8 \ ,\displaybreak[0] \nonumber\\
	\tilde{\Phi}^{[i\sigma^{\mu \nu}\gamma^5]} & =
		2\, P^{[\mu} S^{\nu]}\,\tAmp_9 
		+ 2i\,{m_N^2}\, \elll^{[\mu} S^{\nu]}\,  \tAmp_{10}\  \nonumber\\ & 
		+ 2\,m_N^2\,\elll^{[\mu} P^{\nu]} (\elll \cdot S)\tAmp_{11} \,. 
	\label{eq-phitildetraces}
	\end{align}
The structures above can be obtained by replacing $k$ by $i m_N^2 \elll$ in the corresponding structures for the time-reversal-even amplitudes $A_i$ in Ref.~\cite{Goeke:2005hb}.\footnote{We adjust our sign conventions for $\tAmp_{9}$, $\tAmp_{10}$ and $\tAmp_{11}$ as well as the linear combination $\tAmp_{9m}\equiv \widetilde{A}_{9} - \frac{1}{2} m_N^2 \elll^2 \widetilde{A}_{11}$ with respect to previous work \cite{MuschThesis2,Hagler:2009mb} in favor of this simple correspondence.}\setcounter{fnnumberamp}{\thefootnote}
The representation in terms of the $\tAmp_i(\elll^2,\elll \tcdot P)$ is a more convenient choice for our purposes than the conventional parametrization using momentum dependent amplitudes $A_i(k^2, k\tcdot P)$. The $(\elll^2,\elll \tcdot P)$-dependent representation will also be advantageous for the discussion of correlations in the $x$- and $\vprp{k}^2$-dependence of the \TMDs, see section \ref{sec-fac}.
The amplitudes $\widetilde{A}_i$ are complex-valued and fulfill 
\begin{equation}
	\widetilde{A}_i(\elll^2,\elll \tcdot P) = \left[ \widetilde{A}_i(\elll^2,-\elll \tcdot P) \right]^*\ .
	\label{eq-conjugated}
\end{equation}
This property follows from hermiticity and is
analogous to the constraint that the \TMDs and the conventional amplitudes $A_i(k^2,k\tcdot P)$ are real.
Notice that there is in general no one-to-one correspondence between an individual $\tAmp_i(\elll^2,\elll\tcdot P)$ and the Fourier-transform of the analogous $A_i(k^2,k\tcdot P)$. For example, $\tAmp_8$ contributes to $A_6$, $A_7$ and $A_8$ (following the conventions of Ref. \cite{Goeke:2005hb}).

Clearly, the momentum dependent amplitudes $A_i(k^2,k\tcdot P)$, as well as our invariant complex amplitudes $\tAmp_i(\elll^2,\elll \tcdot P)$,
contain information about all leading and higher twist contributions (for the given choice of the Wilson-line path).
To see how the TMDs of different twist can be obtained from the invariant amplitudes, we first note that
combining the definitions \eqref{eq-corr} and \eqref{eq-corrkminusint}, the $k^-$-integral in Eq. \eqref{eq-corrkminusint} translates 
into the constraint $l^+=0$. 
Using $\elll^- = (\elll \tcdot P)/P^+$ for $\elll^+=0$, we obtain 
\begin{align}
	\Phi^{[\Gamma]}(x,\vprp{k};P,S) & = \int \frac{d(\elll \tcdot P)}{(2\pi)}\,e^{-ix(\elll \tcdot P)} \int \frac{d^2 \vprp{l}}{(2\pi)^2}\,e^{i\vprp{\elll}\tcdot\vprp{k}} \nonumber \\ & \times \ \frac{1}{P^+}\ \widetilde{\Phi}^{[\Gamma]}(\elll,P,S) \Big \vert_{\displaystyle \elll^+ = 0} \ .
	\label{eq-tmdopdef}
\end{align}
Inserting the structures in Eq. \eqref{eq-phitildetraces}, the angular part of the $\vprp{l}$-integral can be performed. 
Due to the restriction to $\elll^+=0$, the remaining radial integral can be rewritten as an integral over $\elll^2=-\vprp{\elll}$. 
For the following discussions, it is therefore useful to 
abbreviate the Fourier-transform of amplitudes as
\begin{align}
	\fourint \tAmp_i \equiv & \int \frac{d(\elll \tcdot P)}{(2\pi)} \int \frac{d^2 \vprp{l}}{(2\pi)^2}\,e^{-ix(\elll \tcdot P)+i\vprp{\elll}\tcdot\vprp{k}} \tAmp_i(-\vprp{\elll}^2,\elll \tcdot P) \nonumber \\
	= & \int \frac{d(\elll \tcdot P)}{(2\pi)}\,e^{-ix(\elll \tcdot P)} \nonumber \\ \times & \int_0^\infty \frac{d(-\elll^2)}{2(2\pi)}\ J_0(\sqrt{-\elll^2}\, |\vprp{k}|)\ \tAmp_i(\elll^2, \elll \tcdot P) \, ,
	\label{eq-fourint}
\end{align}
where $J_0$ is a Bessel function. 
Notice that  $x \leftrightarrow (\elll\tcdot P)$  and $\vprp{k}^2 \leftrightarrow \elll^2$ form pairs of conjugate variables with respect to 
the Fourier transform. Notice also that $\elll^2 \leq 0$ in the Fourier-integral above. It turns out that only spacelike and lightlike quark separations $\elll$ occur in the matrix elements needed for \TMDs. In the following, we shall use the abbrevation $|\elll| \equiv \sqrt{-\elll^2}$.
Finally, the TMDs can be identified and extracted from
comparisons of the parametrizations in Eqns. \eqref{eq-phigammaplus}-\eqref{eq-phisigmaplusi} with 
Equations~\eqref{eq-tmdopdef} and \eqref{eq-phitildetraces}, and turn out to be given by
certain linear combinations of ($x$- and $\vprp{k}$-derivatives of) the Fourier-transformed amplitudes.
Specifically, we obtain the twist-2 \TMDs from the amplitudes $\widetilde{A}_{2,6,7,9m,10,11}(\elll^2,\elll \tcdot P)$:
\begin{align}
	 f_1(x,\vprp{k}^2) & = 2 \fourint\ \widetilde{A}_2 \,,\nonumber \displaybreak[0] \\
	 g_1(x,\vprp{k}^2) & = - 2 \fourint\ \widetilde{A}_6 + 2 \partial_x \fourint\ \widetilde{A}_7 \nonumber \,, \displaybreak[0] \\
	 g_{1T}(x,\vprp{k}^2) & = 4 m_N^2 \partial_{\vprp{k}^2} \fourint\ \widetilde{A}_7 \nonumber \,, \displaybreak[0] \\
	 h_{1L}^\prp(x,\vprp{k}^2) & = 4 m_N^2 \partial_{\vprp{k}^2} \left( \fourint \widetilde{A}_{10} 
	                          + \partial_x \fourint\ \widetilde{A}_{11} \right) \nonumber \,, \displaybreak[0] \\
	 h_1(x,\vprp{k}^2) & = - 2 \fourint\ \widetilde{A}_{9m} \nonumber \,, \displaybreak[0] \\
	 h_{1T}^\prp (x,\vprp{k}^2) & = 8 m_N^{4} \left(\partial_{\vprp{k}^2}\right)^2 \fourint\ \widetilde{A}_{11}\,.
	 \label{eq-tmdsfromamps}
\end{align} 
Here $\widetilde{A}_{9m} \equiv \widetilde{A}_{9} - \frac{1}{2} m_N^2 \elll^2 \widetilde{A}_{11}$. 
As an example for corresponding relations at subleading twist, 
we note that the axial-vector TMDs $g_T'$ and $g_T^\perp$ of twist-3 can be obtained from
\begin{align}
	 g'_{T}(x,\vprp{k}^2) & =  - 2 \fourint\ \widetilde{A}_6 
	                           + 4 m_N^2 \partial_{\vprp{k}^2} \fourint\ \widetilde{A}_8 \nonumber \,,\\
	 g_{T}^\prp (x,\vprp{k}^2) & = 8 m_N^4 \left(\partial_{\vprp{k}^2}\right)^2 \fourint\ \widetilde{A}_{8}\ .
	 \label{eq-tmdsfromampsTW3}
\end{align}
Eqns.~\eqref{eq-tmdsfromamps} and \eqref{eq-tmdsfromampsTW3} 
finally show that the specific types of linear combinations and (derivatives) of the involved amplitudes
indeed allow a projection of the invariant $\widetilde{A}_i$ on TMDs of definite twist.

To forestall potential confusion, we also note that the number of independent amplitudes in Eq.~\eqref{eq-phitildetraces} (which is 9)
is already lower than the total number of T-even TMDs of twist-2 and twist-3 TMDs in Eq.~\eqref{eq-tmdsfromamps} and \eqref{eq-tmdsfromampsTW3} 
(which is 16), respectively, leaving aside the contributions of twist-4.
This is a direct consequence of our choice of a straight Wilson-line path,
i.e. the fact that no additional structures depending on a direction vector $v\not\propto\elll$ can appear in 
the parametrization Eq.~\eqref{eq-phitildetraces}.
Accordingly, by a comparison of Eqns.~\eqref{eq-tmdsfromamps} and \eqref{eq-tmdsfromampsTW3} for example, 
it is possible to derive certain relations between (derivatives) of TMDs of twist-2 and twist-3
that are exact for our process-independent choice $\mathcal{C}=\mathcal{C}^\text{sW}$.
Such relations are similar but not identical to the so-called "Lorentz-invariance relations" \cite{Mulders:1995dh,Boer:1997nt,Goeke:2003az}, 
which only hold if the dependence on the direction vector of the staple-like gauge links, i.e. $v=\nminus$ in Fig.~\ref{fig-link-staple}, is neglected.

Integrating Eq.~\eqref{eq-tmdopdef} over $\vprp{k}$, we obtain
\begin{align}
	\Phi^{[\Gamma]}(x;P,S) & \equiv \int \frac{d(\elll \tcdot P)}{(2\pi)P^+}\,e^{-ix(\elll \tcdot P)}  \widetilde{\Phi}^{[\Gamma]}(\elll,P,S) \Big \vert_{\displaystyle \elll^+ {=} \vprp{l} {=} 0} \nonumber \\
	& = \int \frac{d \elll^-}{2(2\pi)}\,e^{-i\elll^- P^+ x}  \nonumber \\ & \times \bra{\nucl{P,S}}\ \bar \quark(\elll^- \nminus)\, \GammaOp\ \WlineC{\mathcal{C}_{\elll^- \nminus}}\ \quark(0)\ \ket{\nucl{P,S}}\ .
	\label{eq-pdfdef}
\end{align}
A parametrization of the above correlator yields the conventional, ``integrated'' PDFs. 
Notice that the staple shaped links of Fig. \ref{fig-link-staple} simplify to a simple 
connecting straight light-like Wilson line in the matrix element above, because the quark fields have no transverse separation.
Due to the perturbative tail of the correlator in Eq.~\eqref{eq-corr} at large transverse momentum,
the $\vprp{k}$-integrations are formally divergent \cite{Bacchetta:2008xw} and require a regularization. 
PDFs are typically introduced directly according to Eq.~\eqref{eq-pdfdef} based on renormalized operators. 
The divergent $\vprp{k}$-integral thus does not appear explicitly. 

\section{Lattice calculations}
\subsection{The discretized non-local operator}
\label{sec-nonloc}

\begin{figure}[tbp]
  \begin{center}
    \includegraphics*{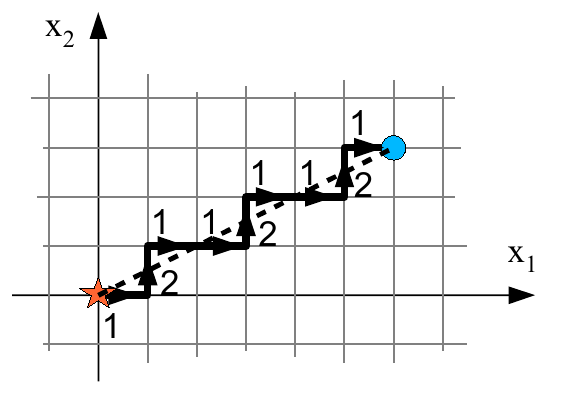}
     \caption{Example of a step-like link path: The straight gauge link in the continuum with $\vect{\elll} = (6,3,0)$ (dashed line) is represented as a product of link variables $U_\mu$ in the directions $\mu = 1, 2, 1, 1, 2, 1, 1, 2, 1$. } \label{fig-steplike}
  \end{center}
\end{figure}

A first important step in the lattice calculation of TMDs is to find a discretized representation of the continuum operator
\begin{equation}
	\mathcal{O}_{\GammaOp,q}[\mathcal{C}_\elll](z) \equiv \bar \quark(\elll+z)\, \GammaOp\ \WlineC{\mathcal{C}_\elll+z}\ \quark(z)\ 	
	\label{eq-contop}
\end{equation}
that appears in the matrix element $\tilde \Phi^{[\Gamma]}$ of Eq.\ \eqref{eq-corr}. Note that we have introduced an overall offset $z$, which does not affect the matrix element: $\tilde \Phi^{[\Gamma]} = \frac{1}{2} \bra{\nucl{P,S}} \mathcal{O}_{\GammaOp,q}[\mathcal{C}_\elll](z) \ket{\nucl{P,S}}$ is independent of $z$.
To implement the non-local operator $\mathcal{O}_{\GammaOp,q}[\mathcal{C}_\elll](z)$ on the lattice, we approximate the Wilson line $\WlineC{\mathcal{C}_\elll+z}$ between the quark fields by a product of connected link variables, as illustrated in Fig. \ref{fig-steplike} and explained in the following. With the notation $U_\mu(x) \equiv U(x,x+a\hat e_\mu)$, $U^\dagger_\mu(x) \equiv U(x+a\hat e_\mu,x)$, the lattice gauge link for a lattice path $\mathcal{C}^\lat_\elll = (x^{(n)},x^{(n-1)},x^{(n-2)},\ldots,x^{(1)},x^{(0)})$ along adjacent lattice sites $x^{(j)}$ is
 \begin{equation}
	\WlineClat{\mathcal{C}^\lat_\elll} \equiv U(x^{(n)},x^{(n-1)})\cdots U(x^{(2)},x^{(1)})\, U(x^{(1)},x^{(0)})\ .
	\label{eq-lat-gaugelink}
\end{equation}
The above expression converges to the Wilson line Eq. \eqref{eq-wlinecont} in the naive continuum limit, provided the  distance of the points $x^{(i)}$ to the continuous path $\mathcal{C}_\elll$ is guaranteed to be of the order of the lattice spacing, see appendix \ref{sec-convproof}.
As a whole, the lattice field combination we employ to probe nucleon structure, 
\begin{align}
	O_{\GammaOp,\quark}^\lat[\mathcal{C}^\lat_\elll](z) \ & \equiv\ \bar \quark(\elll+z)\, \GammaOp\ \WlineClat{\mathcal{C}^\lat_\elll+z}\ \quark(z) \ ,
	\label{eq-olat}
\end{align}
has the same form as the continuum operator in Eq. \eqref{eq-contop}, except for the discretized gauge link along the lattice path $\mathcal{C}^\lat_\elll$ running from the origin, $x^{(0)}= 0$, to $x^{(n)}=\elll$.

If $\elll$ is a multiple of one of the unit vectors $\hat{e}_\mu$, $C^\lat$ 
is a straight path that lies on one of the lattice axes. If $\elll$ is at an oblique angle, we employ a method similar to the Bresenham algorithm \cite{Bresenham:1965} to generate a step-like lattice path close to the 
continuum path, as in the example shown in Fig. \ref{fig-steplike}.

The renormalization of the lattice operators and further properties of the gauge link will be discussed in section \ref{sec-contrenorm}, \ref{sec-rotinv} and \ref{sec-linkren} below.

\subsection{Lattice correlation functions}
\label{sec-correlators}
\begin{figure}[btp]
	\centering%
	\includegraphics[width=0.8\linewidth,clip=true]{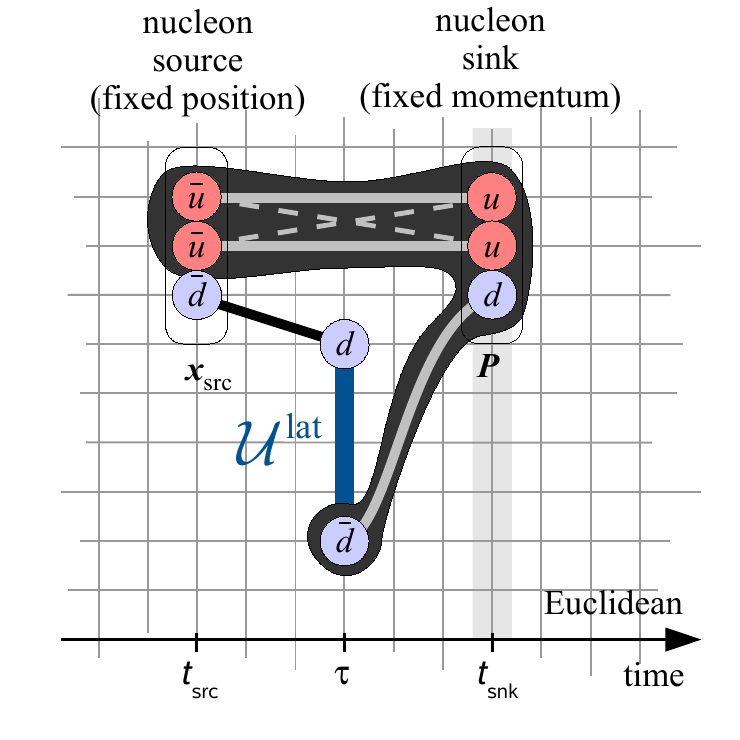}
	\caption{%
		Schematic diagram of a nucleon three-point function on the lattice, here for an operator probing $d$-quarks.
		\label{fig-latpropproduct}%
		}
\end{figure}

Using the discretized non-local operator of the previous section, we extract the invariant amplitudes $\widetilde{A}_i(\elll^2,\elll \tcdot P)$ from lattice three-point correlation functions corresponding to the matrix elements $\tilde \Phi^{[\Gamma]}$. 
A typical lattice three-point-function with a non-local operator insertion at Euclidean time $\tau$ is
illustrated in Fig.~\ref{fig-latpropproduct}, where the nucleon source and sink are placed at $t_\text{src}$ and $t_\text{snk}$, respectively.

The evaluation of three-point functions follows standard techniques \cite{Martinelli:1988rr,Dolgov:2002zm,Hagler:2007xi} which we review very briefly in the following. Only the operators $O_{\GammaOp,\quark}^\lat[\mathcal{C}^\lat_\elll]$ we use to probe the nucleon and the way we interpret the results are specific to our task. 
The purpose of the source and the sink is to create and annihilate states with the quantum numbers of the nucleon.
The nucleon sink has the form 
\begin{align}
	B_\alpha(t,\vect{P})  \equiv &
	\frac{1}{\sqrt{\hat L^3}}\sum_{\vect{x}} e^{-i \vect{P} \cdot \vect{x}} \
	\epsilon_{a b c}\ \times \nonumber \\ & u_{a \alpha} (\vect{x}, t) \
	\Big( u_b^T(\vect{x},t)\ \GammaDiq\ d_c(\vect{x}, t) \Big)\, ,
	\label{eq-nuclsink}
\end{align}
where $a,b,c$ are color indices, $\alpha$ is a Dirac index, $\GammaDiq=\gamma_4 \gamma_2 \gamma_5 (\Eins + \gamma_4)$ and $\vect{P}$ is the three-momentum of the nucleon. An analogous expression $\overline{B}_\alpha(t,\vect{P})$ acts as a nucleon source. To increase the overlap with the nucleon, the quark fields $u$ and $d$ that enter Eq. \eqref{eq-nuclsink} are smeared as described in Ref. \cite{Dolgov:2002zm}. 
We introduce the two-point function by 
\begin{align}
	\latcfn^\text{2pt}(\vect{P}) \equiv & \sum_{\beta\alpha} \GammaTwo_{\beta\alpha}\, \dlangle B_\alpha(t_\text{snk},\vect{P})\, \overline{B}_\beta(t_\text{src},\vect{P}) \drangle \, ,\nonumber
\end{align}
and the three-point function for a general operator $O$ is given by
\begin{align}
	\latcfn^\text{3pt}[O^\lat](\vect{P},\tau) = & \frac{1}{\hat L^3} \sum_{\vect{z}} \sum_{\beta\alpha} \GammaThr_{\beta\alpha}\,\dlangle B_\alpha(t_\text{snk},\vect{P})\ \times\nonumber \\ 
	& O^\lat(\vect{z},\tau)\ \overline{B}_\beta(t_\text{src},\vect{P}) \drangle\ .
	\label{eq-nuclthreepoint}
\end{align}
where $\dlangle \cdots \drangle \equiv \int \mathcal{D}[q,\overline{q},U]\, \cdots \exp( - S^\lat)$ denotes an expectation value 
defined by the lattice path integral, and where $\GammaThr$ is a Dirac matrix projecting out the desired parity and spin polarization of the baryon. 

In order to ensure that the  transfer matrix formalism enables us to rewrite our three-point function in terms of a matrix element $\bra{N(P,S')}O^\lat\ket{N(P,S)}$, we limit ourselves to operators $O_{\GammaOp,\quark}^\lat[\mathcal{C}^\lat_\elll](\vect{z},\tau)$ that do not extend 
in the Euclidean time direction, i.e., the link path is restricted to the spatial plane at $\tau$, and $\elll_4 = \elll^0 = 0$. 
As explained in section \ref{sec-parametrization},
our selection of vectors $\elll$ and $\vect{P}$ on the lattice does not need to correspond to the large momentum frame usually chosen to introduce \TMDs in the context of scattering processes. 
Relevant for the calculation of the \TMDs 
from the amplitudes $\tAmp_i(\elll^2,\elll \tcdot P)$ are only
the Lorentz-invariant quantities formed by the Minkowski four-vectors $\elll$ and $P$, 
which are in the lattice frame given by
$\elll^2 = -\vect{\elll}^2$, or $\sqrt{-\elll^2}= |\vect{\elll}|$, and $\elll \tcdot P = - \vect{\elll} \tcdot \vect{P}$.
Consequently, we will only be able to evaluate the amplitudes $\widetilde A_i(\elll^2,\elll \tcdot P)$ in the range 
\begin{align}
	\elll^2 & \leq 0\, , &
	|\elll \tcdot P| &  \leq |\vect{P}| \sqrt{-\elll^2}\, ,
	\label{eq-lPdomain}
\end{align}  
where $\vect{P}$ is the chosen nucleon momentum on the lattice. 

The transfer matrix formalism shows that the lattice correlation functions decay exponentially in the Euclidean time and the 
energies of the contributing states. If the operator position $\tau$ is far enough away from source $t_\text{src}$ and sink $t_\text{snk}$,  the three-point function is therefore dominated by contributions proportional to nucleon ground state matrix elements $\bra{N(P,S')}\, O^\lat\, \ket{N(P,S)}$.  
The proportionality factors (e.g., overlaps of nucleon source and sink with the nucleon state), the exponential time dependence, as well as part of the statistical noise cancel in the ratio with the two-point function
\begin{align}
	R[O^\lat](\vect{P},\tau) & \equiv  \frac{\latcfn^\text{3pt}[O^\lat](\vect{P},\tau)}{\latcfn^\text{2pt}(\vect{P})}\ .
	\label{eq-ratiodef}
 \end{align} 
If $t_\text{src}$ and $t_\text{snk}$ are far enough apart, we observe a ``plateau'' in a region where the ground state dominates, such that $R[O^\lat](\vect{P},\tau)$ is independent of $\tau$: 
\begin{align}
	R[O^\lat](\vect{P},\tau) & \xrightarrow{ |\tau-t_\text{src}|,|\tau-t_\text{snk}|\gg \Delta E^{-1} } \bar{R}[O^\lat](\vect{P}) \ , \\
	\bar{R}[O](\vect{P}) & \equiv 
	\sum_{S,S'} \frac{\overline{U}(P,S)\ \GammaThr\ U(P,S')}{2E_P\ \Tr_\mathrm{D} \left\lbrace\GammaTwo\, (-i \slashed{P} + m_N)  \right\rbrace} \nonumber\\  & \times   
	\bra{N(P,S')}\, O\, \ket{N(P,S)}\ ,	
	\label{eq-plateaudef}
\end{align}
where $\Delta E=E'-E $ is the difference between the energies of the ground state and the first excited state, and
$U(P,S)$ is the Dirac spinor of a nucleon.
For an appropriately renormalized lattice operator $O^\lat_\ren$, we identify this plateau value with the correspondingly renormalized continuum expression:
\begin{align}
	\bar{R}[O^\lat_\ren](\vect{P}) \xrightarrow{a \rightarrow 0} \bar{R}[O^\ren](\vect{P}) \ .
\end{align}
Thus we finally gain access to the desired continuum matrix elements $\bra{N(P,S')}\, O^\ren\, \ket{N(P,S)}$.
With Equation \eqref{eq-plateaudef} for $\bar{R}[O^{\ren}_{\GammaOp,q}[\mathcal{C}_\elll]](\vect{P})$ and inserting 
(for the case of straight gauge paths $\mathcal{C}_\elll$) our parametrization Eq. \eqref{eq-phitildetraces}, we 
can parametrize the plateau values in terms of the amplitudes $\widetilde A_i$, as given explicitly in Table \ref{tab-ratios} in the appendix.

We now discuss the strategy for evaluating the three-point function $\latcfn^\text{3pt}[O_{\GammaOp,\quark}^\lat[\mathcal{C}^\lat_\elll]](\tau,\vect{P})$.
The average over all offsets $\vect{z}$ in Eq. \eqref{eq-nuclthreepoint} increases statistics and allows us to exploit translation invariance in favor of a fixed source location. 
Integrating out fermions analytically, pairs of quark field variables $u$, $\overline{u}$ and $d$, $\overline{d}$ combine into lattice quark propagators, which we depict as connecting lines between the quark variables in Fig.~\ref{fig-latpropproduct}. Lattice quark propagators are numerically obtained by inversion of the lattice Dirac operator and describe the propagation of a valence quark in a gauge field background, i.e., effects of gluons and sea quarks are included. In principle, all possible contractions of pairs $u$, $\overline{u}$ and $d$, $\overline{d}$ into propagators must be taken into account. In Figure \ref{fig-latpropproduct}, a second diagram, resulting from the permutation of $u$-quarks, is indicated with dashed lines. In practice, however, we neglect here the computationally demanding so called disconnected contributions, where the quark variables of $O_{\GammaOp,\quark}^\lat[\mathcal{C}_\elll]$ contract with each other internally to form a closed quark loop. Disconnected contributions cancel exactly in the isovector case, i.e., for $O_{\GammaOp,u-d}^\lat \equiv O_{\GammaOp,u}^\lat - O_{\GammaOp,d}^\lat$. 

For the numerical calculation, we employ the sequential source technique \cite{Wilcox:1991cq}, which permits us to evaluate the three-point function for arbitrary 
gauge link paths 
using the same given set of point-to-all type lattice propagators. As indicated by the curved gray envelope in Fig.~\ref{fig-latpropproduct}, three of the quark propagators in the diagram can be combined into a single ``sequential propagator'', which can be calculated for fixed $(\vect{x}_\text{src},t_\text{src})$, $t_\text{snk}$ and $\vect{P}$ using a secondary inversion, and which can be used like a backward point-to-all lattice propagator. Finally, the three-point function is evaluated by forming a product of a forward propagator, the link variables and the sequential propagator. 

\subsection{Simulation parameters and computational details}
\label{sec-simpar}

For the purpose of our proof-of-concept calculations, we have chosen existing ensembles and propagators at intermediate pion masses that have already been successfully used in the determination of GPDs \cite{Hagler:2007xi}. The gauge configurations have been generated by the MILC collaboration \cite{Ber01,Aubin:2004wf,Bernard:2007ps}. They feature 2+1 dynamical, improved staggered quarks, with the strange quark mass fixed approximately to the physical value. Employing the ``coarse'' MILC gauge configurations ($a\approx 0.12\units{fm}$), the LHP collaboration has calculated propagators using a domain wall fermion action, where the pion mass has been adjusted to the Goldstone pion mass of the underlying staggered lattice \cite{Hagler:2007xi}. The computationally more expensive domain wall action for the valence quarks exhibits a lattice chiral symmetry, which is in particular advantageous with respect to the operator renormalization. 
Essential ensemble parameters, together with the pion mass determined using the domain wall propagators, are given in  Table~\ref{tab-gaugeconfs}. The MILC collaboration has chosen the strange quark masses $m_{s}$ to correspond roughly to the physical value. 
For our scaling study in Sec. \ref{sec-linkren}, we take advantage of ``fine\nbdash04'', ``superfine\nbdash04'' and ``extracoarse\nbdash04'' gauge configurations that have become available from the MILC collaboration recently. The ensembles listed in the last four lines of Table~\ref{tab-gaugeconfs} all %\mxout{feature}
have the same ratio $\hat m_{u,d}/\hat m_{s}=0.4$, placing them approximately on a line of constant physics, i.e., they feature similar pion and kaon masses. In order to determine the lattice spacing in a uniform way for all six ensembles in the table, we have taken the updated, ``smoothed'' values $r_1/a$ of Ref.~\cite{Bazavov:2009bb}, and $r_1 = 0.3133(26)\units{fm}$ from the recent analysis Ref.~\cite{Davies:2009tsa}.
\footnote{In contrast, Refs.~\cite{Hagler:2007xi,Hagler:2009mb} used $a=0.124\units{fm}$, as determined from the $\Upsilon$ spectrum on the coarse lattices \cite{Wingate:2003gm,Aubin:2004wf}. As a result, numbers in physical units, including the pion masses listed in Table \ref{tab-gaugeconfs}, differ somewhat with respect to these previous references.}\setcounter{fnnumber}{\thefootnote} 

\begin{table*}[tbp]
	\centering
	\renewcommand{\arraystretch}{1.1}
	\begin{tabular}{|l|l|ll|c|c||c|c|c|}
	\hline
	ensemble & $a\units{(fm)}$ & $\hat m_{u,d}$ & $\hat m_{s}$ & $10/g^2$ & $\hat L^3 \times \hat T$\rule{0ex}{1.2em} & $m_\pi^\text{DWF}\units{(MeV)}$ & $m_N^\text{DWF}\units{(GeV)}$ & $\#$conf. \\
	\hline
	coarse\nbdash10       & $0.11664(35)(96)$  & $0.05$   & $0.05$   & $6.85$  & $20^3 \times 64$  & $807.5(16)(92)$ & 1.668(09)(19) & 478 \\
	coarse\nbdash06       & $0.11823(18)(99)$  & $0.03$   & $0.05$   & $6.81$  & $20^3 \times 64$  & $625.4(17)(62)$ & 1.450(11)(15) & 561 \\
	coarse\nbdash04       & $0.11849(14)(99)$  & $0.02$   & $0.05$   & $6.79$  & $20^3 \times 64$  & $519.7(19)(50)$ & 1.355(12)(13) & 425 \\
	fine\nbdash04              & $0.08440(09)(71)$  & $0.0124$ & $0.031$  & $7.11$  & $28^3 \times 96$  & & & \\
	superfine\nbdash04         & $0.05930(08)(50)$ & $0.0072$ & $0.018$  & $7.48$  & $48^3 \times 144$ & & & \\
	extracoarse\nbdash04       & $0.1755(07)(15)$ & $0.0328$ & $0.082$  & $6.485$ & $16^3 \times 48$  & & & \\ \hline
	\end{tabular}\par\vspace{1ex}
	\renewcommand{\arraystretch}{1.0}
	\caption{Lattice parameters of the MILC gauge configurations \cite{Ber01,Aubin:2004wf,Bazavov:2009bb} used in this work. The first error quoted for $a$ estimates statistical errors in $r_1/a$, the second error originates from the uncertainty about $r_1$ in physical units. The sixth and seventh column list the pion and the nucleon masses as determined with the LHPC propagators with domain wall valence fermions \cite{Hagler:2007xi}. The first error is statistical, the second error comes from the conversion to physical units using $a$ as quoted in the table. Note that the masses quoted here in physical units differ slightly from those listed in Ref. \cite{Hagler:2007xi}, because we use a different scheme to fix the lattice spacing, see footnote \thefnnumber. The last column lists the number of configurations used for the calculation of three-point functions.
	\label{tab-gaugeconfs}%
	}
\end{table*}

To reduce computational costs for the production of propagators further, the coarse lattice gauge configurations have been chopped into two halves of temporal extent $\hat T/2 = 32$. Only every sixth trajectory, and alternating temporal halves have been selected, reducing autocorrelations to an undetectable level. Noise has been reduced by application of HYP-smearing \cite{Hasenfratz:2001hp} to the gauge configurations before the propagators have been determined by inversion. Link smearing is an operation in which each link variable is replaced by a unitarized ``average'' of itself and gauge links in the vicinity. In the case of HYP-smearing, only link variables from within the lattice hypercubes adjacent to the original link enter the average, so as to minimize the distortion of physical properties at short distances. An important benefit of HYP-smearing is a reduction of the breaking of rotational symmetry, see also section \ref{sec-rotinv} below. The propagators and sequential propagators provided by LHPC are of the smeared-to-point type, i.e., the quark fields at the source location and the nucleon sink embedded in the sequential propagator are smeared as described in Ref. \cite{Dolgov:2002zm}. Using the smeared-to-point propagator as input, we form a smeared-to-smeared version, in order to be able to compute the appropriate two-point function with smearing both at source and sink. The sequential propagators are available for sink momenta $\vect{P} = 0$ and $\vect{P} = (-1,0,0) \times 2\pi/L$. The latter corresponds to $|\vect{P}|\approx 500\units{MeV}$ and is the lowest non-zero momentum on these lattices. The source-sink separation is fixed to $\hat t_\text{snk} - \hat t_\text{src} = 10 \approx 1.2 \units{fm}$.  

\begin{figure}
	\centering%
	\includegraphics[width=\linewidth]{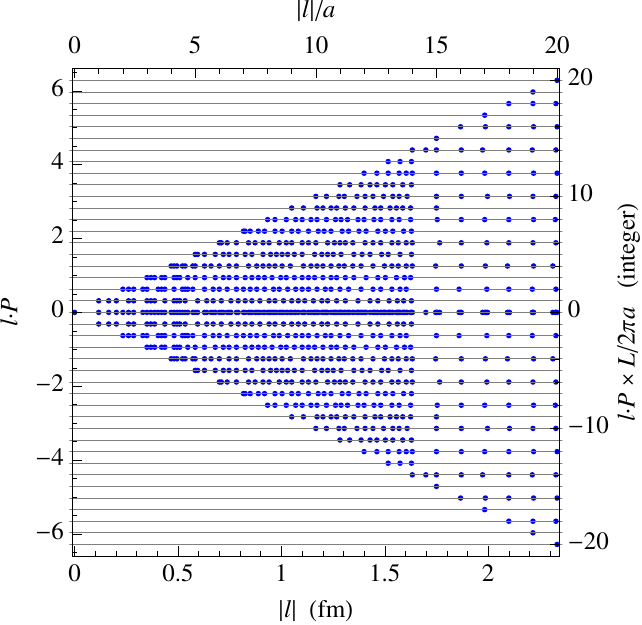}%
	\caption[coverage]{%
		Coverage of the $(\elll^2,\elll \tcdot P)$-plane for our choice of link paths. The scale on top is in lattice units and the scale on the right labels the integer values accessible on the periodic lattice. For the conversion to physical units (scales on the left and bottom axes) we use $L/a = 20$ and $a=0.1166 \units{fm}$, i.e., the values listed in Table \ref{tab-gaugeconfs} for the course\nbdash10 ensemble. Note that $\elll \tcdot P$ is dimensionless in natural units. 
		\label{fig-lPcoverage}
		}
	\end{figure}

For our analysis with lattice nucleon momentum $\vect{P} = (0,0,0)$, we have generated 263 different link paths $\mathcal{C}^\lat_\elll$. We remind the reader that we restrict ourselves to purely spatial extensions of the gauge link. The quark separations $\vect{\elll}$ cover the three lattice axes up to a link length $|\elll|=20a$, three quadrants in the $(\vect{\elll}_1,\vect{\elll}_2)$- and $(\vect{\elll}_1,\vect{\elll}_3)$-planes for $|\elll|\leq 8a$ and a choice of additional links with $|\elll|\leq15a$ in the first octant. For the analysis with $\vect{P} = (-1,0,0)\times 2\pi/L$, we choose 743 further vectors $\vect{\elll}$  from the two octants with $\vect{\elll}_2 \geq 0$, $\vect{\elll}_3 \geq 0$ such that the $(|\elll|,\elll \tcdot P)$-plane is densely covered in the range accessible on the lattice, see Fig. \ref{fig-lPcoverage}. 

\begin{figure}
	\centering%
	\label{fig-plateauPzero}%
	\includegraphics{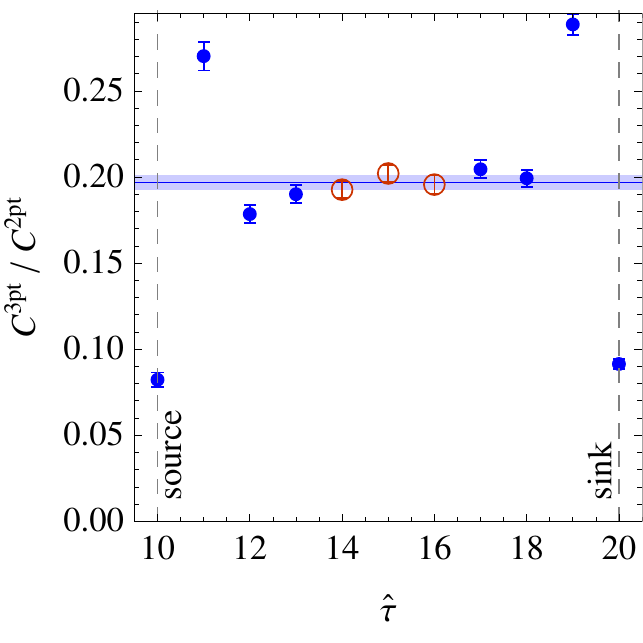}%
	\caption[plateaus]{%
		Plateau plot for the real part of the ratio $R[O^\lat_{\gamma_4,u-d}[\mathcal{C}^\lat_\elll]](\vect{P},\tau)$ as a function of $\tau$ for the HYP-smeared coarse\nbdash10 lattice, for $\hat{\vect{P}} = 0$ and the link path $\mathcal{C}^\lat_\elll$ depicted in Fig.~\ref{fig-steplike}. The plateau value $\overline{R}[O^\lat_{\gamma_4,u-d}[\mathcal{C}^\lat_\elll]](\vect{P})$ is extracted from the three encircled points and is displayed as a horizontal error band. 
		\label{fig-plateaus}
		}
	\end{figure}

In Figure \ref{fig-plateaus}, we show an example plot of the ratio $R[\mathcal{O}^\lat_{\GammaOp,q}[\mathcal{C}^\lat_\elll]](\tau,\vect{P})$ 
as a function of $\tau$ between $t_\text{src}$ and $t_\text{snk}$. Even for the rather long link path of Fig. \ref{fig-steplike} with $|\vect{\elll}| \approx 0.8\units{fm}$, the signal to noise ratio is good. We follow the strategy of Ref. \cite{Hagler:2007xi} and take the average of the three data points at $\tau-t_\text{src} = 4, 5, 6$ as an estimate of the plateau value $\overline{R}[\mathcal{O}^\lat_{\GammaOp,q}[\mathcal{C}^\lat_\elll]](\vect{P})$ defined in Eq. \eqref{eq-ratiodef}. Potential contaminations from excited states can be neglected at our present level of accuracy. 

In order to estimate statistical uncertainties, we consistently employ the Jackknife method \cite{Quenouille49,Quenouille56,tukey1958bac}. 
For fits to lattice data, we minimize for each configuration $j$
\begin{equation}
	\chi^2 \equiv \sum_{i}  \left[ f_i(p_1^{(j)},\ldots,p_n^{(j)}) - y^{(j)}_i \right]^2 \Delta y_i^{-2}\, .
	\label{eq-chisqr}
\end{equation}
Here $f_i$ denotes the fit function evaluated at location $i$, the lattice data at this location are given by Jackknife samples $y_i^{(j)}$, and the Jackknife error is $\Delta y_i$. 
The parameter estimates $p_1^{(j)}, \ldots, p_n^{(j)}$ thus obtained are again Jackknife samples.
The functional form of Eq. \eqref{eq-chisqr} does not reflect correlations among the data points by means of the covariance matrix. Nevertheless, the least squares fit using $\chi^2$ as given above implements a consistent estimator \cite{EDJS71} for the Jackknife samples $p_1^{(j)},\ldots,p_{n}^{(j)}$.
Hence, the Jackknife errors that are finally obtained for the parameters and for functions of the parameters adequately include correlations. 

\subsection{Renormalization of the non-local operators}
\label{sec-contrenorm}
The renormalization properties of the continuum operator $\mathcal{O}_{\GammaOp,q}[\mathcal{C}_\elll]$ have been studied with the help of an auxiliary field technique (``$z$-field'') in Refs. \cite{Dotsenko:1979wb,Arefeva:1980zd,Craigie:1980qs,Dorn:1986dt} and independently in leading order perturbative QCD in Ref. \cite{Stefanis:1983ke}. For a smooth open path $\mathcal{C}_\elll$, the renormalized Wilson line has the form 
\begin{equation}
	\Wlineren{\mathcal{C}_\elll} = \renZ_z^{-1} e^{ -\delta m\, \len[\mathcal{C}_\elll] } \Wline{\mathcal{C}_\elll}	\, ,	
	\label{eq-linkrencont}
\end{equation} 
where $\len[\mathcal{C}_\elll]$ is the total length of the path. The length dependent, exponential factor corresponds to the self-energy of the Wilson line. The dimensionful renormalization constant $\delta m$ removes a divergence linear in the cutoff scale (i.e., $a^{-1}$ on the lattice). In dimensional regularization, $\delta m$ vanishes, but renormalon ambiguities appear, see e.g., Ref. \cite{Pineda:2005kv}. The renormalization factor $ \renZ_z^{-1}$ can be associated with the end points of the gauge link and does not appear in a Wilson loop. For a piecewise smooth gauge link, we would have to add an angle-dependent renormalization factor for each corner point. For the composite operator $\mathcal{O}_{\GammaOp,q}[\mathcal{C}_\elll]$, we get an additional renormalization factor $\renZ_\psi^{-1}$ for the quark field renormalization and a factor $\renZ_{(\psi z)}^2$ for the quark -- gauge link vertices:
\begin{equation}
	\mathcal{O}^\ren_{\GammaOp,q}[\mathcal{C}_\elll] = \underbrace{\renZ_\psi^{-1}\, \renZ_{(\psi z)}^2\, \renZ_z^{-1}}_{\displaystyle \renZ^{-1}_{\psi,z}}\ e^{ -\delta m\, \len[\mathcal{C}_\elll]}\ \mathcal{O}_{\GammaOp,q}[\mathcal{C}_\elll]\ . 
	\label{eq-opren}
\end{equation}
Note that the renormalization constants do not depend on $\Gamma$. This is in contrast to the renormalization of \emph{local} operators of the form $\overline{q}(0) \Gamma q(0)$, $\overline{q}(0) \Gamma D_{\mu} q(0)$,  $\overline{q}(0) \Gamma D_{\mu} D_{\nu} q(0)$, $\ldots$ as they are used, e.g., in the calculation of moments of GPDs. The basic explanation for the $\Gamma$-independent renormalization of the non-local object is that the spatially separated quark fields are renormalized individually. However, the precise relation between the derivative operators $\overline{q}(0) \Gamma D_{\mu} D_{\nu} \cdots q(0)$ and the non-local operator $\mathcal{O}^\ren_{\GammaOp,q}[\mathcal{C}_\elll]$ remains to be studied further. The interested reader is referred to appendix \ref{sec-localops}, where we rewrite the non-local lattice operator $O_{\GammaOp,\quark}^\lat[\mathcal{C}^\lat_\elll]$ explicitly as a weighted sum of derivative operators. The main purpose of appendix \ref{sec-localops} is to address the question whether and how mixing among local operators affects the non-local object.  

As discussed in section \ref{sec-linkren}, it is known how to renormalize straight Wilson lines on the lattice that run along the lattice axes. It turns out that renormalization in this case is also of the form Eq. \eqref{eq-linkrencont}.  

Fundamental for the remainder of this work, we will make the assumption that the \emph{discretized} operator $O_{\GammaOp,\quark}^\lat[\mathcal{C}^\lat_\elll]$ has the same renormalization properties as the continuum operator. Specifically, we will employ Eq. \eqref{eq-opren} to renormalize our lattice operator $O_{\GammaOp,\quark}^\lat[\mathcal{C}^\lat_\elll]$.  This assumption relies on the physical argument that, for a given discretization prescription of the gauge link, the operator $O_{\GammaOp,\quark}^\lat[\mathcal{C}^\lat_\elll]$ becomes an approximate representation of the continuum operator $O_{\GammaOp,\quark}[\mathcal{C}_\elll]$ as soon as the length of the gauge link is large compared to the lattice spacing $a$.  Note that the numerical values of the renormalization constants we obtain for given renormalization conditions depend on the lattice action used and on the details of implementation of the discretized operator. 
Numerical checks of these assumptions and the non-perturbative methods that are employed to determine the renormalization constants for given lattice action, lattice spacing and renormalization conditions will be discussed in 
sections \ref{sec-rotinv} and \ref{sec-linkren}.
We point out that more detailed work on the renormalization of the general, step-like non-local lattice operator could benefit from the method of constructing symmetry improved operators as described in appendix \ref{sec-symimpop}. This is to be expected because at the level of local operators, increased symmetry reduces complications caused by mixing.

\section{Numerical results}
\label{sec-firstobs}

\subsection{Mapping out the \texorpdfstring{$(\elll^2,\elll \tcdot P)$}{(l.l,l.P)}-plane}
\label{sec-lsqrldpplane}

\begin{figure}[tbp]
	\centering
 	\subfloat[][]{%
 		\label{fig-lpsurface-re}%
 		\includegraphics[width=\linewidth,trim=0 200 0 50,clip=true]{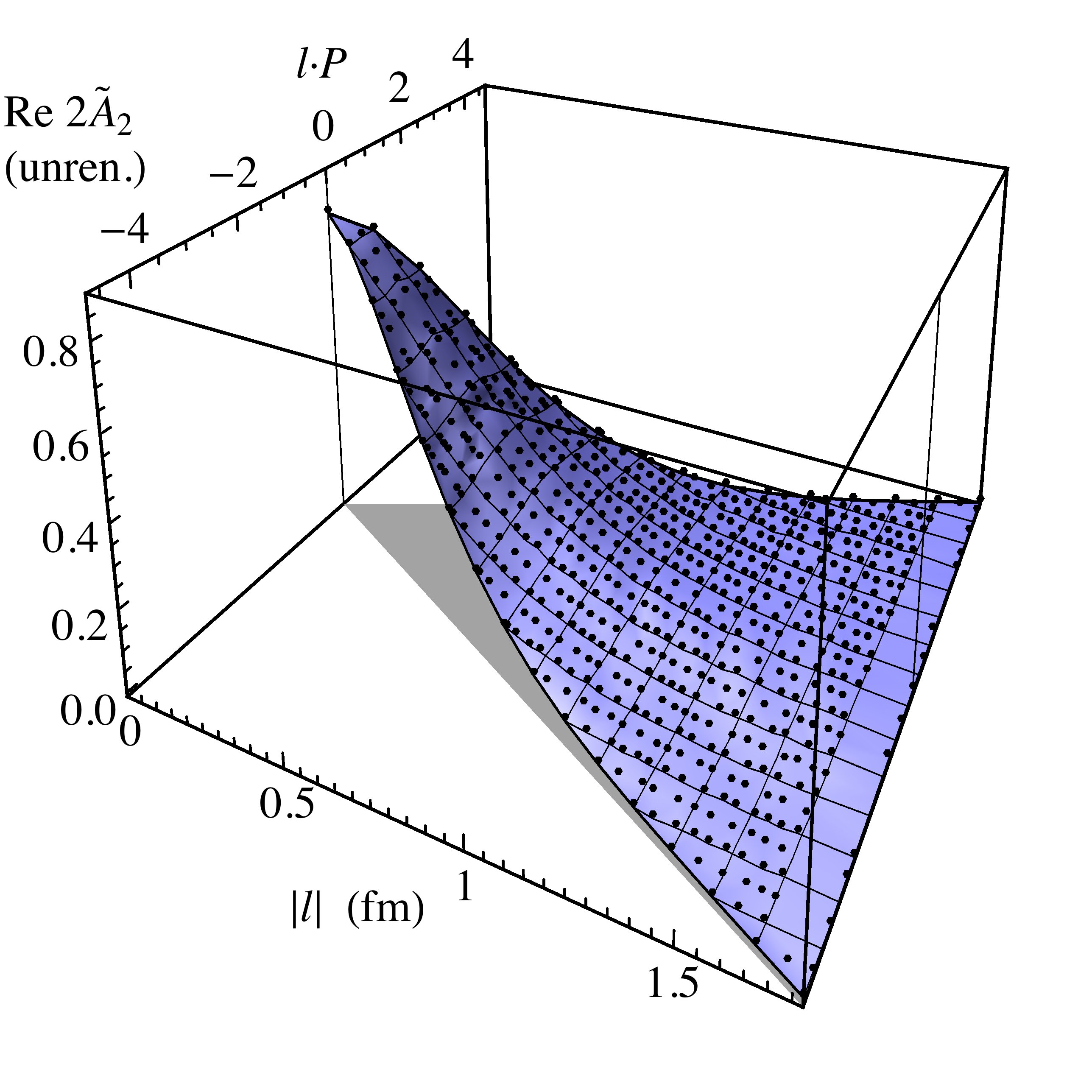}%
 		}\\%
 	\subfloat[][]{%
 		\label{fig-lpsurface-im}%
 		\includegraphics[width=\linewidth,trim=0 200 0 50,clip=true]{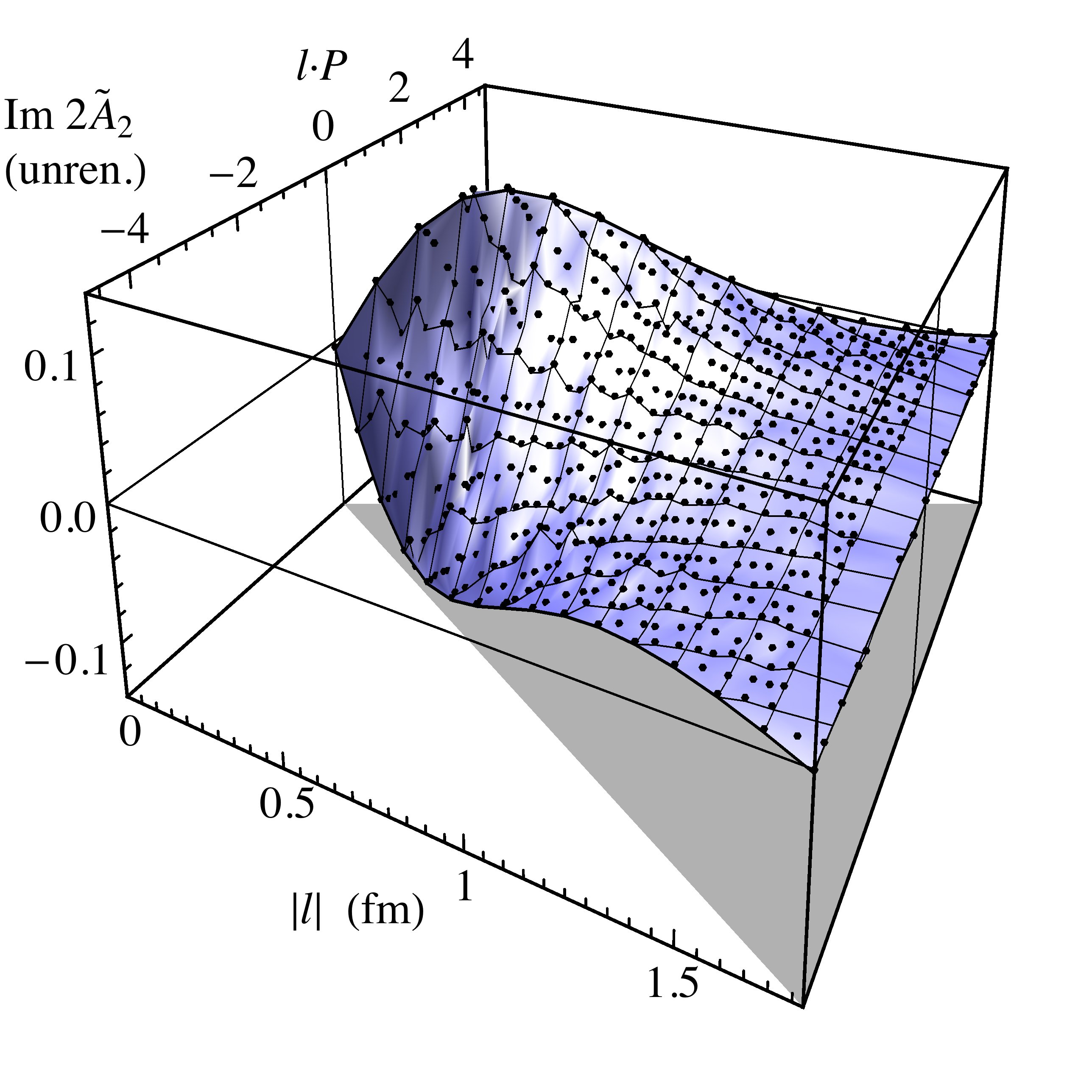}%
 		}\par%
	\caption[3D plots]{%
		The unrenormalized amplitude $\tAmp_{2}^\unren(\elll^2,\elll\tcdot P)$ obtained directly from the ratio $\bar R[O^\lat_{\gamma_\Eu{4},u-d}[\mathcal{C}^\lat_\elll]](\vect{P})$ using the sequential propagators with $\vect{P}=(-1,0,0)\times2\pi/L$ on the coarse\nbdash10 ensemble and applying HYP smearing to the gauge fields.
		\subref{fig-lpsurface-re}~real part,
		\subref{fig-lpsurface-im}~imaginary part.%
		\label{fig-lpsurface}
		}
\end{figure}
Following the methods outlined in \ref{sec-correlators}, we have computed the invariant amplitudes $\tAmp_{i}^\unren(\elll^2,\elll\tcdot P)$ for the coarse\nbdash10 ensemble, using unrenormalized operators. According to Table \ref{tab-ratios} in the appendix, a straight link calculation with the operator $O_{\Gamma}[C_\elll]$ for $\Gamma=\gamma_4$ gives us access to $\tilde A_2(\elll^2,\elll \tcdot P)$. Results for $\tAmp_{2}^\unren(\elll^2,\elll\tcdot P)$, in the domain we can reach with the available lattice nucleon momenta $\vect{P}$, are displayed in Fig. \ref{fig-lpsurface}.
The accessible ``kinematical'' domain is characterized by a triangle with an opening angle given by the largest nucleon momentum $|\vect{P}|$ available in the calculation, see Eq. \eqref{eq-lPdomain}.  At $\elll^2 = 0$, all amplitudes can only be extracted 
for the single data point $\elll\tcdot P = 0$. 
The $\elll\tcdot P$ dependence can thus only be studied at non-vanishing values of $\elll^2$. Therefore, the $x$-dependence of PDFs cannot be obtained from a direct evaluation of Eq. \eqref{eq-pdfdef} on the lattice, in accordance with the common
knowledge that the lightlike gauge links in the gauge invariant definition of PDFs cannot be realized on an Euclidean space-time lattice. Nevertheless, we will be able to discuss the $\vprp{k}$-dependence of the lowest $x$-moment of \TMDs, and, beyond that, to draw some conclusions about the $x$-dependence from data at non-zero $\elll^2$.

Coming back to the amplitude in Fig.~\ref{fig-lpsurface}, we note that the real part  $\myRe\,\tAmp_{2}^\unren$ is dominated by a Gaussian-like drop with $|\elll|$, while the dependence on $\elll\tcdot P$ at constant $|\elll|$ features only a slight curvature.  
Our results for the imaginary part $\myIm\,\tAmp_{2}^\unren$ in 
Fig.~\ref{fig-lpsurface-im} form a surface twisted around the $|\elll|$-axis at $\elll\tcdot P{=}0$, where the amplitude must vanish, cf. Eq.~\eqref{eq-conjugated}. 
The slope of the surface flattens out towards larger $|\elll|$. 
 We will investigate this behavior in Section \ref{sec-fac}.

\subsection{A study of rotational symmetry}
\label{sec-rotinv}
We now study the amplitude $\tilde A_2(\elll^2,\elll \tcdot P)$ in Fig.~\ref{fig-lpsurface-re} 
in greater detail for $\elll^0=0$, $\vect{P} = 0$. In this case, $\elll \tcdot P=0$, i.e., the amplitude only depends on the (Euclidean) length of the gauge link $|\elll|$. 
Carrying out the calculation with an unrenormalized lattice operator $O_{\gamma^4}^\lat[C^\lat_\elll]$, we obtain an unrenormalized amplitude $\tilde A_2^\unren$. Renormalization will eventually be based on Eq. \eqref{eq-opren}. However, it is not a priori clear to what extent $\delta m$ should be independent of the direction of the vector $\vect{\elll}$ of the link path on the lattice, since the discretization prescription for the gauge link is not (and cannot be) rotationally invariant. 
Consider the set of plateau values $\bar R[\mathcal{O}^\lat_{\gamma_4,u-d}[\mathcal{C}^\lat_\elll]](\vect{P}{=}0)$ obtained from our selection of link paths $\mathcal{C}^\lat_\elll$. The lattice action is invariant under reflections and permutations of the lattice axes, i.e., under symmetry transformations of the $\mathrm{H}(4)$ group. We have checked that the plateau values are indeed numerically equal within statistics for link paths $\mathcal{C}^\lat_\elll$ that are equivalent up to reflections and permutations of the (spatial) axes. Next, we ask how severely continuous rotational symmetry is broken. 
\begin{figure}[btp]
	\centering%
	\subfloat[][]{%
		\label{fig-A2unren-sm}%
		\includegraphics[width=0.98\linewidth]{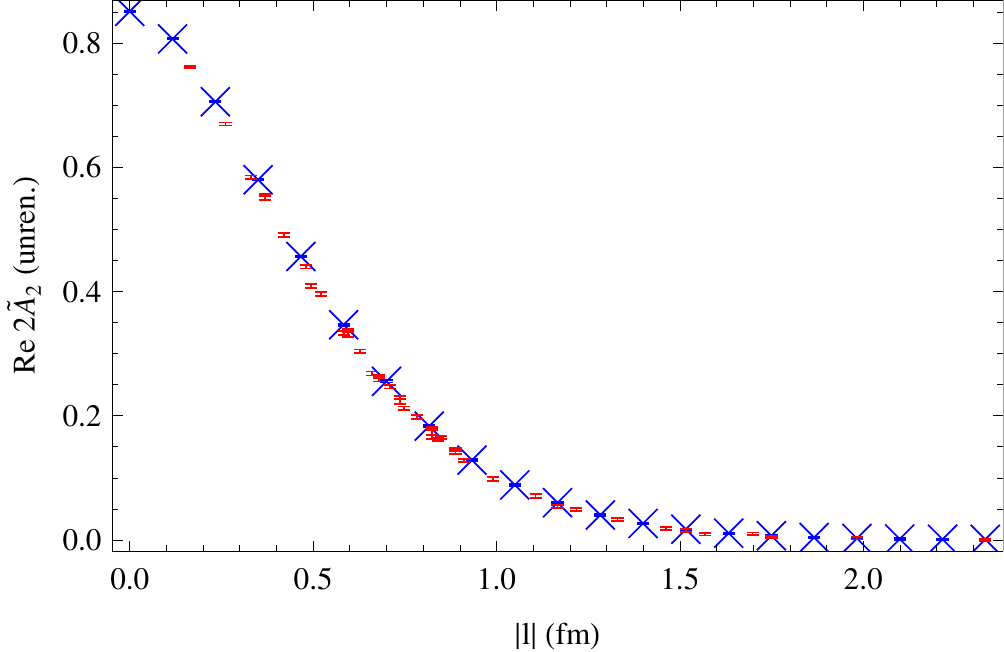}
		}\\%
	\subfloat[][]{%
		\label{fig-A2unren-unsm}%\
		\includegraphics[width=0.98\linewidth]{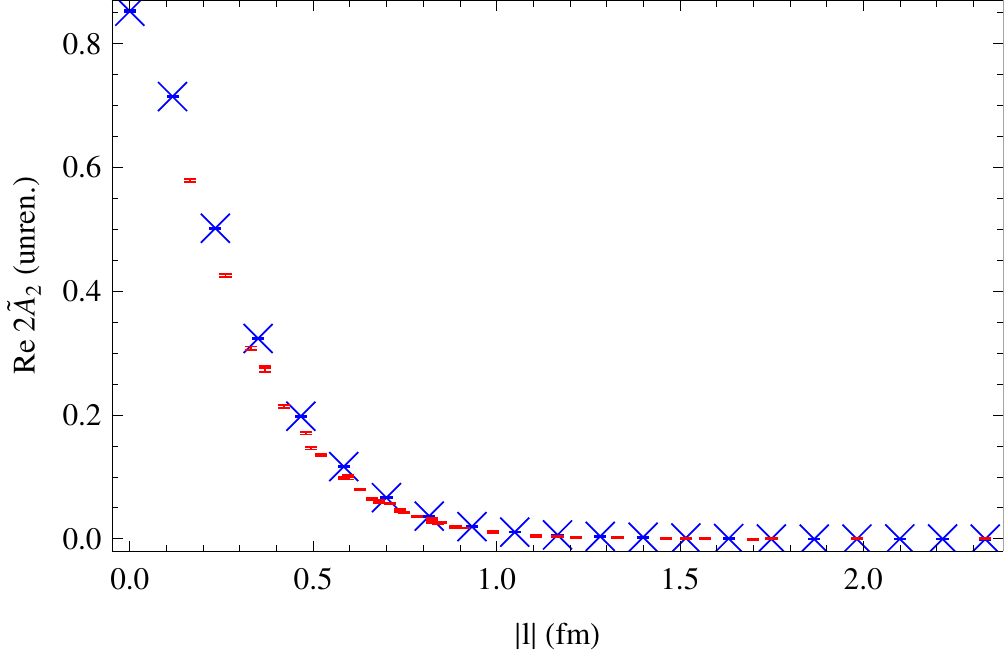}
		}%
	\caption[A2 unren]{%
		Unrenormalized data obtained for the amplitude $2 \tilde A_{2,u-d}(\elll^2,\elll \tcdot P{=}0)$ using the lattice operator $O_{\gamma^4}^\lat[C^\lat_\elll]$ and a nucleon momentum $\vect{P}=0$ on the coarse\nbdash10 lattice. Link paths coinciding with the lattice axes are marked with  a blue cross, the red error bars belong to link paths at oblique angles. The gauge path was constructed
		\subref{fig-A2unren-sm}\ %
			on HYP smeared gauge configurations,\ %
		\subref{fig-A2unren-unsm}\ %
			on unsmeared gauge configurations.			
		\label{fig-A2unren}%
		}
\end{figure}
In Fig. \ref{fig-A2unren} we plot the plateau values as a function of the quark separation $|\elll|$. To avoid a cluttered plot, we have taken the averages over link paths equivalent under $\mathrm{H}(4)$ transformations. 
In Fig. \ref{fig-A2unren-sm} the operator has been evaluated on the HYP smeared gauge configurations. Here the results from step-like link paths and results from gauge links on the axes agree very well,  and may be described by a smooth, $|\elll|$-dependent function. A distance $|\elll|$ where we have a results both from paths along the axes and from step-like paths can be found, for example, at $|\elll|=a\sqrt{4_x^2+3_y^2}=a\sqrt{5^2}=5a=0.58\units{fm}$.
We find a relative difference of $4\pm1$ percent between the two results.

In the unsmeared case, Fig. \ref{fig-A2unren-unsm}, data points from step-like links are visibly and systematically lower than data points from links along the axes. (At $|\elll|=5a$, the discrepancy amounts to $17\pm2$ percent.)
We found a very similar picture when we studied the breaking of rotational invariance of the vacuum expectation value of the gauge link $\dlangle  \Tr_c\ \WlineClat{\mathcal{C}^\lat_\elll} \drangle$ on a Landau gauge fixed ensemble.
As a side remark, we note that a simple correction model, the ``taxi driver correction'', reduces the deviations particularly well in the unsmeared case \cite{MuschThesis2}. 
As a whole, we conclude that rotational symmetry is only weakly broken, especially if the gauge link is smeared. We rate this as an important indication that the discretized operator does indeed approximate the continuum operator.
In the following, we will analyze nucleon structure with the smeared gauge link, and acknowledge a systematic discretization error of the order of four percent associated with the violation of rotational symmetry.
Last but not least, we notice an overall faster drop-off of the data with $|\elll|$ in the unsmeared case, Fig. \ref{fig-A2unren-unsm}, than in the smeared case Fig. \ref{fig-A2unren-sm}. This can be explained by the fact that two different values $\delta m$ are needed to renormalize the smeared and unsmeared case.

\subsection{Link renormalization}
\label{sec-linkren}

\subsubsection{Method}

In lattice QCD, we work in a cutoff scheme that depends on the lattice action, with a UV cutoff of the order of $1/a$. In order to be able to present results for amplitudes that have a well-defined continuum limit, and that are independent of the lattice spacing and action, we need to renormalize our operator, in particular with respect to the self-energy of the gauge link, as discussed in Section \ref{sec-contrenorm}. 
The crucial question is how to determine $\delta m$ in Equations \eqref{eq-linkrencont} and \eqref{eq-opren}. Since we observe approximate rotational invariance for our operator on the smeared lattices, we can restrict ourselves to the determination of $\delta m$ for straight gauge links along one of the lattice axes. The renormalization of the Wilson line on the lattice has a long history in the context of heavy quark propagators, where it has been found that the respective power divergence requires a non-perturbative subtraction \cite{Maiani:1991az}. Calculations in lattice perturbation theory \cite{Eichten:1989kb,Boucaud:1989ga,Martinelli:1998vt} confirm that the gauge link can be renormalized by a factor $\exp(-\delta m\, L)$, but will not serve us here to determine an accurate value for $\delta m$. Instead, we turn to non-perturbative methods. We choose a gauge invariant procedure based on the static quark potential that has been applied in the literature for the renormalization of the Polyakov loop \cite{Fodor:2004ft,Aoki:2006br,Cheng:2007jq,Bazavov:2009zn}. 
Here we outline the basic idea. Implementation details are given in appendix \ref{sec-renimpl}. The static potential $V(R)$ for a system 
of a heavy quark and antiquark with relative distance $R$ can be obtained from the asymptotic behavior of the expectation value of a rectangular Wilson loop $W(R,T)$ 
\begin{equation}
	W(R,T) = c(R) e^{ - V(R)\, T} + \text{higher excitations}\, ,
	\label{eq-potfromloop}
\end{equation}
where the contributions from higher excitations are exponentially suppressed for large $T$.
The Wilson loop is renormalized according to
\begin{equation}
	W^\ren(R,T) = e^{- \delta m\, (2R + 2T) - 4 \nu(90^\circ)} W(R,T)\, ,
\end{equation}
where $\nu(90^\circ)$ is the renormalization constant corresponding to the $90^\circ$ corners of the loop. Inserting this form into Eq. \eqref{eq-potfromloop} shows that the renormalized static quark potential
\begin{equation}
	V^\ren(R) = V(R) + 2\,\delta m 
\end{equation}
obtains a constant offset caused by the self-energy of the gauge links in $T$-direction. 
Note that we must ensure that the loop's gauge links in $T$-direction are implemented the same way as those we use as part of our non-local operator. Smearing of the gauge configurations, for example, affects $\delta m$.

A simple renormalization condition that fixes $\delta m$ would be to demand $V^\ren(R_0) = 0$ at some $R_0$, which has to have a fixed value in physical units, see, e.g., Ref. \cite{Fodor:2004ft,Aoki:2006br}. 
An alternative idea \cite{Cheng:2007jq,Bazavov:2009zn} makes use of the fact that the lattice data is quite well approximated by the string potential \cite{Luscher:1980fr}
\begin{equation}
	V_\text{string}(R) = \sigma R - \frac{\pi}{12 R} + C^\ren
	\label{eq-stringpot}
\end{equation}
for not too small quark distances $R$. Matching this form to lattice data\footnote{introducing only a weak dependence on a matching point, chosen here to be $1.5 r_0$, in terms of the Sommer scale $r_0$ \cite{So93}} and demanding $C^\ren = 0$ fixes $\delta m$ and avoids introduction of another dimensionful constant. By setting $C^\ren = 0$, we have introduced a renormalization condition. In simple terms, it can be understood as the asymptotic condition $V^\ren(R) - \sigma R \rightarrow 0$ for large $R$. 

Applying renormalization with $\delta m$ obtained in this way, we 
eliminate the lattice cutoff dependence of our 
gauge links in favor of a reproducible, non-perturbative renormalization condition. 
A future challenge is to find the connection of our renormalization condition with the scale dependence of \TMDs, see also the discussion at the end of section \ref{sec-Gauss}.

\subsubsection{Numerical results}

Table \ref{tab-renconst}
\begin{table}[tb]
	{\centering
	\small
	\renewcommand{\arraystretch}{1.1}
	\begin{tabular}{|c|lc||r|l|}
	\hline
	$\hat m_{u,d}/\hat m_s$ & ensemble \rule{0ex}{1.2em} & & 
	$\hat T_\mmin$ & $-\delta \hat m$ \\
	\hline
	\hline
	1.0 & coarse\nbdash10    & smeared & 6 & \textbf{0.1440(37)}   \\
	\hline
	0.6 & coarse\nbdash06    & smeared & 6 & \textbf{0.1491(31)}  \\
	\hline
	0.4 & extracoarse\nbdash04      & smeared & 4 & 0.1043(94)  \\ 
	0.4 & coarse\nbdash04    & smeared &  6 & \textbf{0.1554(45)} \\
	0.4 & fine\nbdash04           & smeared &  8 & 0.1639(35)     \\
	0.4 & superfine\nbdash04     & smeared & 10 & 0.1578(17)  \\
	\hline \hline                                                       
	1.0 & coarse\nbdash10    & & 5 & 0.4239(89)  \\
	\hline	
	0.4 & extracoarse\nbdash04      &  & 3 & 0.361(60)  \\ 
	0.4 & coarse\nbdash04    &  & 4 & 0.397(35)  \\
	0.4 & fine\nbdash04                   &  & 4 & 0.382(10)   \\
	0.4 & superfine\nbdash04         &  & 5 & 0.361(11)   \\
	\hline                     
	\end{tabular}\par\vspace{1ex}
	\renewcommand{\arraystretch}{1.0}}
	\caption{Renormalization constant $\delta \hat m$ from the static quark potential. Errors in brackets are statistical.
	}
	\label{tab-renconst}%
\end{table}
lists our numerical results for $\delta \hat m = \delta m / a$ based on matching to the string potential with $C^\ren = 0$. 
We have fit the exponential form Eq. \eqref{eq-potfromloop} to Wilson loops, where the minimal temporal extent that was taken into account is given by $\hat T_\mmin$ (in lattice units). 
Most important for the following analysis of the invariant amplitudes are the smeared coarse lattices, 
where the full set of available gauge configurations has been used.
The corresponding numbers are shown in bold letters.
The other lattices serve us to convince ourselves that the method works, but do not enter our results on \TMDs and could be improved with full statistics and larger values of $\hat T_\mmin$. 
In particular, the extracoarse\nbdash04 lattice may exhibit strong discretization errors, and the rather low values $\delta \hat m$ obtained with $\hat T_\mmin = 3,4$ may not be reliable.
Note that our values $\delta \hat m$ correspond to $-C(\beta)/2$ in the notation of Ref. \cite{Bazavov:2009zn}.

\begin{figure}[tb]
	\centering%
	\includegraphics[width=\linewidth]{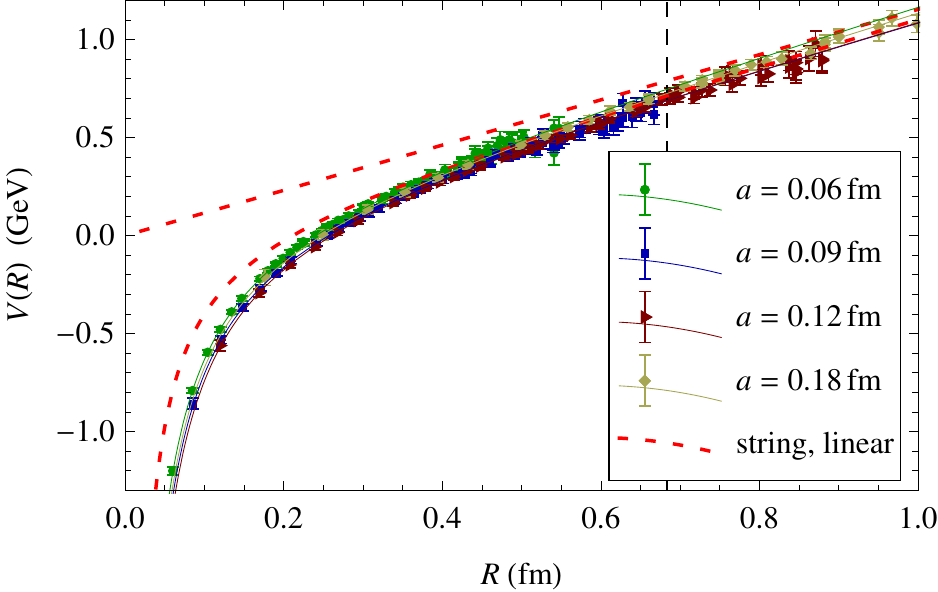}%
	\caption{Renormalized potential for the four smeared lattices with $m_{u,d} = 0.4 m_s$. }
	\label{fig-renpot}
\end{figure}

Figure \ref{fig-renpot} 
displays the renormalized potential for four lattices with different lattice spacings $a$ but equal ratio of quark masses $m_{u,d}/m_s = 0.4$.
The data points have been corrected for known discretization errors by adding $\lambda ( V_\text{pert}^\lat(\vect{r}) - 1/R )$ as in Eq. \eqref{eq-potfitfn} in the appendix, and the solid lines have been obtained from the model function $\hat{V}(R)$ in that same equation.
The curved dashed line shows the string potential Eq. \eqref{eq-stringpot}, plotted for an average $\sigma$. The vertical dashed line indicates the matching point. The string potential approaches asymptotically a straight line through the origin, which we show as a straight 
dashed line in the figure. 
We see that the method yields a renormalized potential that agrees on several ensembles of very different lattice spacings.

\subsubsection{Cross-check with open gauge links}
\label{sec-openlinks}

\begin{figure}[btp]
	\centering%
	\subfloat[][]{%
		\label{fig-Yline}%
		\includegraphics[width=0.98\linewidth]{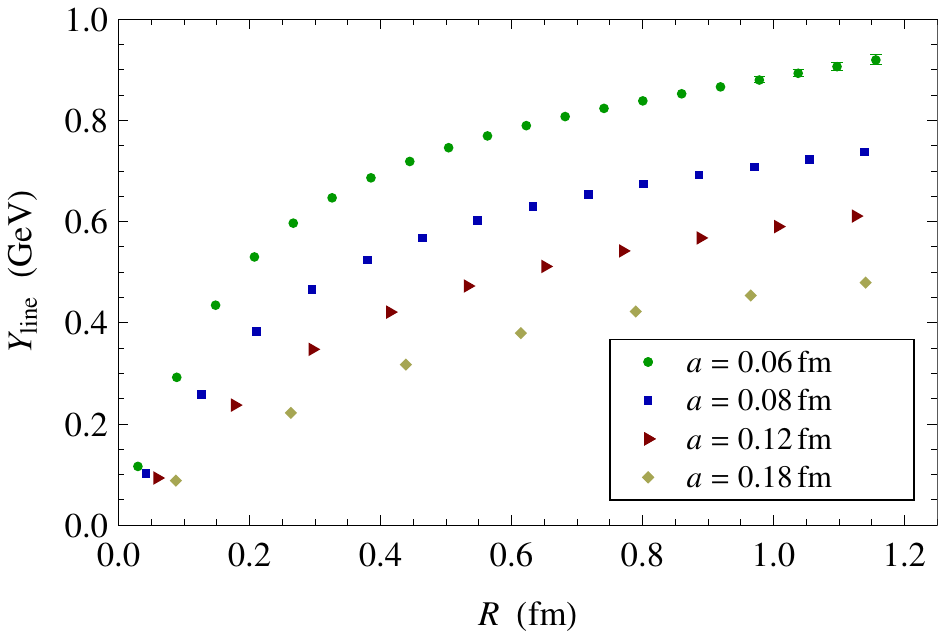}
		}\\%
	\subfloat[][]{%
		\label{fig-YlineBazRen}%\
		\includegraphics[width=0.98\linewidth,clip=true]{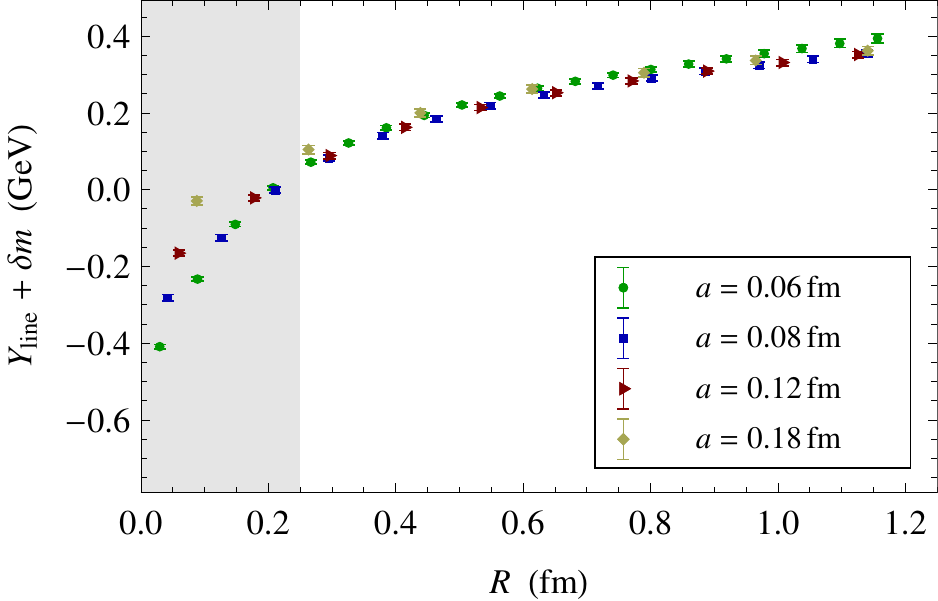}
		}%
	\caption[Yline]{%
		\subref{fig-Yline}\ %
			$Y_\text{line}(R)$ evaluated on the smeared gauge configurations for the four lattices with $m_{u,d} = 0.4 m_s$. \par%
		\subref{fig-YlineBazRen}\ %
			$Y_\text{line}^\ren(R)$, renormalized using $\delta m$ determined from the static quark potential. The gray background highlights a region of link lengths $R$ in which lattice cutoff effects lead to visible discrepancies between the different ensembles. \par			
		\label{fig-SIDISdetails}%
		}
\end{figure}

To convince ourselves that the renormalization constant $\delta m$ obtained from the static quark potential renormalizes straight gauge links in general, we study expectation values of straight gauge links on Landau gauge-fixed ensembles. A convenient quantity to analyze is
\begin{equation}
	Y_\text{line}(R) \equiv -\frac{1}{a} \ln \frac{\dlangle \Tr_c\ \WlineClat{\mathcal{C}_{\elll'}}\drangle_\text{Landau-gauge}}{\dlangle \Tr_c\ \WlineClat{\mathcal{C}_{\elll}}\drangle_\text{Landau-gauge}}\ ,	
	\label{eq-Ylinedef}
\end{equation}
where $\mathcal{C}_{\elll'}$ and $\mathcal{C}_\elll$ are straight link paths of lengths $R+a/2$ and $R-a/2$, respectively. Note that the expectation values of open gauge links are not meaningful quantities without gauge fixing.
The renormalization constants $Z_z$ cancel in the ratio of gauge links, so that the renormalized quantity is $Y_\text{line}^\ren(R) = Y_\text{line}(R) + \delta m$. Indeed, unrenormalized lattice results for $Y_\text{line}(R)$ at different lattice spacings exhibit visible offsets, see Fig. \ref{fig-Yline}. 
It is encouraging to see that the offsets nearly disappear in Fig. \ref{fig-YlineBazRen}, 
where we have renormalized with the values $\delta m$ determined from the static quark potential.
Except in a region roughly below $R<0.25\units{fm}$, we find in fact a very reasonable agreement of the lattice results for $Y_\text{line}^\ren(R)$ between the different ensembles. 
We conclude that lattice cutoff effects become strong for gauge links shorter than about three lattice spacings. Therefore, in the following, we will exclude data points with $R<0.25\units{fm}$ from our analysis. 

A quantitative comparison of $Y_\text{line}(R)$ at different lengths $R$, different lattice spacings $a$ and an extrap\nolinebreak olation to the continuum $a=0$ can provide a rough estimate of the size of discretization errors.
We perform such an extrapolation in appendix \ref{sec-discerr}. The resulting number $\Delta[\delta \hat m]_\text{dis} = 0.0194$ for the coarse\nbdash04 ensemble can be effectively treated as an uncertainty in the renormalization constant $\delta m$.

\section{The lowest \texorpdfstring{$x$}{x}-moment of \TMDs with straight gauge links}
\label{sec-lowmom}

\subsection{The \texorpdfstring{$x$}{x}-integrated correlator and \TMDs}

We already stated in section \ref{sec-lsqrldpplane} that the restriction to the triangle shaped domain in the $(|\elll|,\elll \tcdot P)$-plane given in Eq. \eqref{eq-lPdomain} precludes us from performing the full Fourier transform Eq. \eqref{eq-fourint}. However, within our approach, we do have access to the $x$-integral of the correlator Eq. \eqref{eq-tmdopdef}, i.e., to the lowest $x$-moment :
\begin{align}
	\int_{-1}^{1} dx\ \Phi^{[\Gamma]}(x,\vprp{k};P,S) & = \int \frac{d^2 \vprp{l}}{(2\pi)^2}\,e^{i\vprp{\elll}\tcdot\vprp{k}} \nonumber \\ & \times \ \frac{1}{P^+}\ \widetilde{\Phi}^{[\Gamma]}(\elll,P,S) \Big \vert_{\displaystyle \elll^+{=}l^-{=}0} \ .
	\label{eq-corrxmom}
\end{align}
The above correlator can be parametrized in terms of the lowest $x$-moments of \TMDs, cf. Eqns.~\eqref{eq-phigammaplus} to \eqref{eq-phisigmapm}. 
As an example, consider the case of $f_1$, where we define (see also Eq.~\eqref{eq-tmdsfromamps}) 
\begin{equation}
	f_1^\xmom{1}(\vprp{k}^2) \equiv \int_{-1}^{1} dx\ f_1(x,\vprp{k}^2) = 2 \xfourint \tAmp_2
  \label{eq-xintf1}
\end{equation} 
with 
\begin{align}
	\xfourint \tAmp_i & = \int_0^\infty \frac{d(-\elll^2)}{2(2\pi)}\ J_0(\sqrt{-\elll^2}\, |\vprp{k}|)\ \tAmp_i(\elll^2, 0) \, .
  \label{eq-xintFT}
\end{align} 
Expressions for the lowest $x$-moments of other \TMDs are obtained analogously, in accordance with Eq. \eqref{eq-tmdsfromamps}. 
Lattice data for the amplitudes at $\elll \tcdot P = 0$ are available, e.g., from simulations with the nucleon at rest on the lattice, $\vect{P}=0$. 

The $x$-integral in Eq. \eqref{eq-corrxmom} is taken over the whole support of $\Phi^{[\Gamma]}(x,\vprp{k};P,S)$. The contributions from the integration region with $x<0$ can be related to anti-quark distributions using the correlator $\Phi^c$ defined with charge conjugated fields, see Ref. \cite{Mulders:1995dh} and relation Eq. \eqref{eq-Ccorr} in the appendix. For straight link paths $\mathcal{C}^{\text{sW}}$ as well as staple shaped gauge links $\mathcal{C}^{(v)}$, we can decompose the $x$-integrated correlator as
\begin{align}
	\int_{\mathrlap{-1}}^{\mathrlap{1}} dx\ \Phi^{[\GammaOp]}(x,\vprp{k};P,S;\mathcal{C}) & = \int_{\mathrlap{0}}^{\mathrlap{1}} dx\ \Phi^{[\Gamma]}(x,\vprp{k};P,S;\mathcal{C}) \nonumber \\
	 & + \int_{\mathrlap{0}}^{\mathrlap{1}} dx\ \Phi^{c[\GammaOp^c]}(x,-\vprp{k};P,S;\mathcal{C})\, ,
	\label{eq-corrxmomqantiq}
\end{align}
where $\GammaOp^c = - \gamma^0 \gamma^2 \GammaOp^\transp \gamma^2 \gamma^0$. For $\GammaOp = \Eins$, $\gamma^\mu \gamma^5$ and $\gamma^5$, one finds, $\GammaOp^c = \GammaOp$, while for $\GammaOp = \gamma^\mu$ and $i \sigma^{\mu\nu}\gamma^5$, the sign changes, $\GammaOp^c = -\GammaOp$. 
For the lowest $x$-moment of \TMDs, this translates into, e.g.,
\begin{equation}
	f_1^\xmom{1}(\vprp{k}^2) = \int_{0}^{1} dx\ f_1(x,\vprp{k}^2) - \int_{0}^{1} dx\ \bar{f}_1(x,\vprp{k}^2)\, ,
	\label{eq-f1xmomdecomp}
\end{equation}
where $\bar{f}_1$ is the anti-quark \TMD defined with respect to $\Phi^c$. 
Analogously, $g_{1T}^\xmom{1}$, $h_1^\xmom{1}$, and $h_{1T}^{\prp \xmom{1}}$ are differences of quark- and anti-quark \TMDs. On the other hand, $f_{1T}^{\prp\xmom{1}}$, $g_1^\xmom{1}$, $h_{1L}^{\prp\xmom{1}}$ and $h_{1}^{\prp \xmom{1}}$ are the \emph{sum} of quark and anti-quark \TMDs.

\subsection{Gaussian fits and renormalized data}
\label{sec-Gauss}

To be able to perform the Fourier transforms Eq. \eqref{eq-xintFT} and to renormalize our amplitudes according to Eq. \eqref{eq-opren},  we follow a simple scheme  (This approach circumvents potential problems with divergences of the amplitudes at $|\elll|=0$ in the continuum limit, see section \ref{sec-xktmom}. Limitations of our approach will be discussed later):
%To  renormalize our amplitudes according to Eq. \eqref{eq-linkrencont}, and to circumvent the problems with the divergence of the amplitudes at $|\elll|=0$ in the continuum limit described in the previous section, we follow a simple scheme (limitations of our approach will be discussed later):
\begin{enumerate}
  \item We multiply our unrenormalized data $\tAmp_i^\unren(\elll^2,0)$ by the length dependent renormalization factor  $\exp(-\delta m |l|)$, using the renormalization constant from Table \ref{tab-renconst} \footnote{We remind the reader that the renormalization procedure involves a renormalization condition. In our case, we have chosen a condition based on the static quark potential. Changing this condition would modify the renormalized data for the amplitudes significantly.}. 
  \item We parametrize the resulting data points in terms of Gaussian functions, 
  \begin{equation}
  e^{-\delta m |l|} \times  \tAmp^\unren_{i,q}(\elll^2,0) \xrightarrow{\text{fit}}
  \frac{1}{2} c^{\unren}_{i,q} e^{- \frac{|\elll|^2}{\sigma_{i,q}^2}}  \, ,
  \end{equation}
  where the parameters $c_{i,q}^\unren$, 
  $\sigma_{i,q}$ are obtained from fits to the lattice data points.
  In the fit, we only include data points with $|\elll| > 0.25\units{fm}$, to avoid sensitivity to lattice cutoff effects. It turns out that the Gaussian ansatz fits our data reasonably well in this range. 
  An exception is the amplitude $\tAmp_1$, which appears at subleading twist only.
  \item We determine the multiplicative renormalization constant $Z^{-1}_{\Psi,z}$ 
by demanding that
    \begin{equation}
  \int_{-1}^{1}dx\int d^2\vprp{k}\ f_{1,_\quark}(x,\vprp{k}^2)=2 \tAmp_{2,q}(0,0)=g_{V,\quark}\stackrel{!}{=}n_\quark \, , \nonumber
  \end{equation}
  where $n_\quark$ is the number of valence quarks (quarks minus anti-quarks). 
  After substitution of the renormalized fit expression for $2 \tAmp_2(0,0)$, 
 the equation above reads $g_V =  Z^{-1}_{\Psi,z}\, c_2^\unren$ .
  Since the isovector channel is free of contributions from disconnected diagrams, we fix
   $Z^{-1}_{\Psi,z}$ numerically by setting
  \begin{equation}
   Z^{-1}_{\Psi,z} := \frac{ n_{u-d} }{ c^\unren_{2,u-d} } \, ,
  \end{equation} 
  where $n_{u-d}=1$ and where $c^\unren_{2,u-d}$ is directly determined from a Gaussian fit to data for the isovector amplitude $\tAmp_{2,u-d}$.
  \item The renormalization constant $Z^{-1}_{\Psi,z}$ thus extracted from the long-range behavior of $\tAmp_{2,u-d}$ is applied to all amplitudes: We obtain renormalized data points from
    \begin{equation}
 \tAmp_{i,q}(\elll^2,0) =  Z^{-1}_{\Psi,z} \, e^{-\delta m |l|} \tAmp_{i,q}^\unren(\elll^2,0) \, ,
  \end{equation} 
  as well as renormalized fit functions 
\begin{equation}
	\tAmp^\text{Gauss}_{i,q}(|\elll|) = \frac{1}{2} c_{i,q} e^{ - |\elll|^2/\sigma_{i,q}^2 }
	\label{eq-rengauss}
\end{equation}  
with $c_{i,q} \equiv Z^{-1}_{\Psi,z} c_{i,q}^\unren$.
\end{enumerate}
   
The prescription above is designed to provide lattice scheme and lattice spacing independent results for the long-range behavior of the amplitudes $\tAmp_i$. Qualitatively, the large-$|\elll|$-behavior of our amplitudes is linked by a Fourier-transform to the small-$|\vprp{k}|$-behavior of the corresponding \TMDs, cf. Eq.~\eqref{eq-fourint} to \eqref{eq-tmdsfromampsTW3}. Since we can successfully fit (most of) our data with Gaussians for $|\elll| > 0.25\units{fm}$, we expect to obtain a reasonable description of the corresponding \TMDs at small $|\vprp{k}|$, $|\vprp{k}| \lesssim 1/0.25\units{fm} \approx 0.8\units{GeV}$. 

By restricting the fit to $|\elll| \ge 0.25\units{fm}$ 
and using (smooth) Gaussians to bridge the gap between $|\elll|=0.25\units{fm}$ and $|\elll|=0$, 
we effectively regularize any potential continuum divergence at $|\elll|=0$, albeit in a parametrization dependent way. 
This will be important for the definition and interpretation of $(\vprp{k})^{n}$-weighted integrals of the TMDs below in section \ref{sec-densities}.

\begin{figure*}
	\centering
	\includegraphics{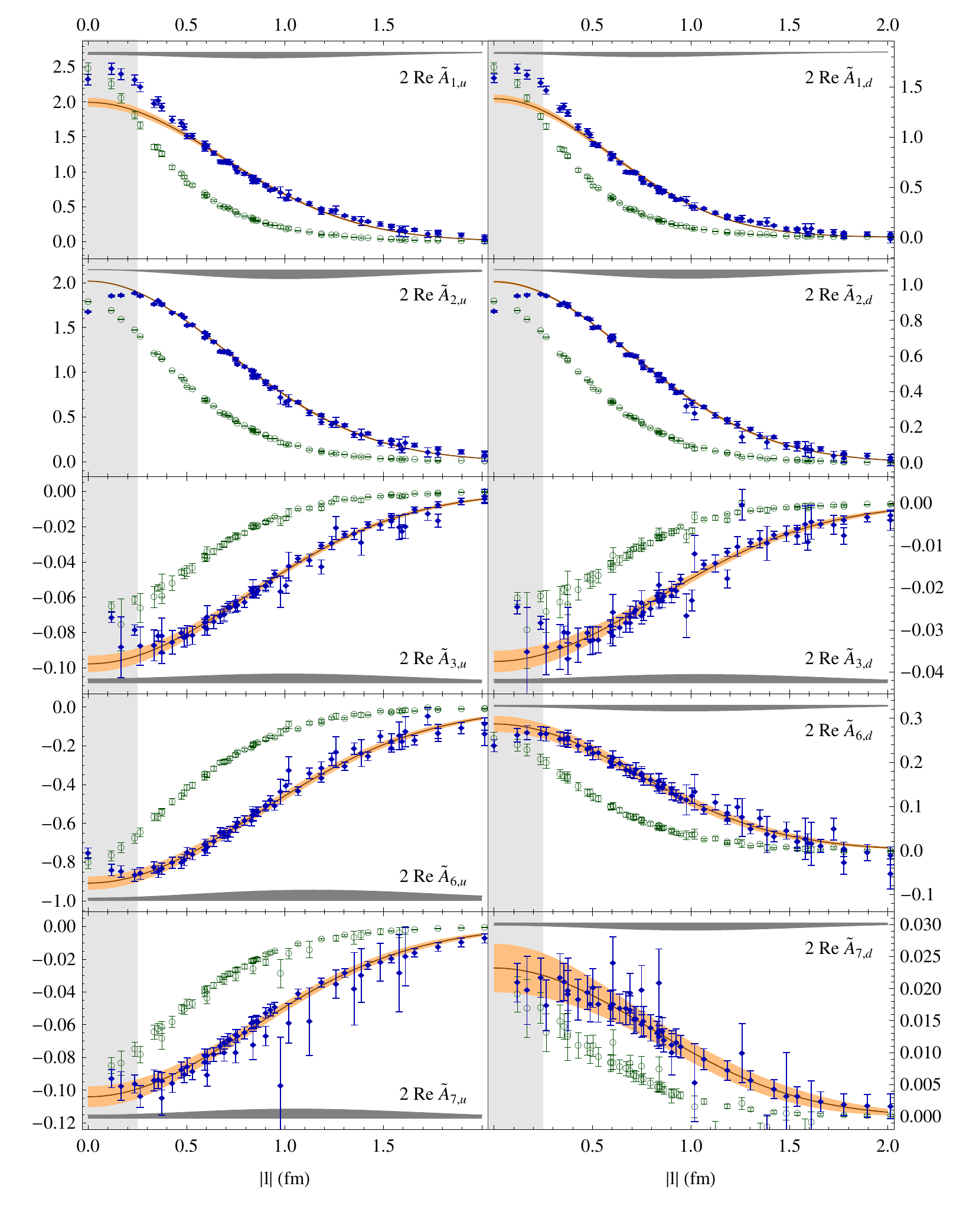}
	\caption{Amplitudes on the coarse\nbdash04 ensemble at $m_\pi \approx 500\units{MeV}$. We show the unrenormalized data (open symbols), renormalized data (full symbols) and Gaussian fits. The uncertainties combined in $\Delta[\delta\hat m]$, are given by the shaded horizontal bands.\label{fig-amps}}
\end{figure*}

\begin{figure*}
	\centering
	\includegraphics{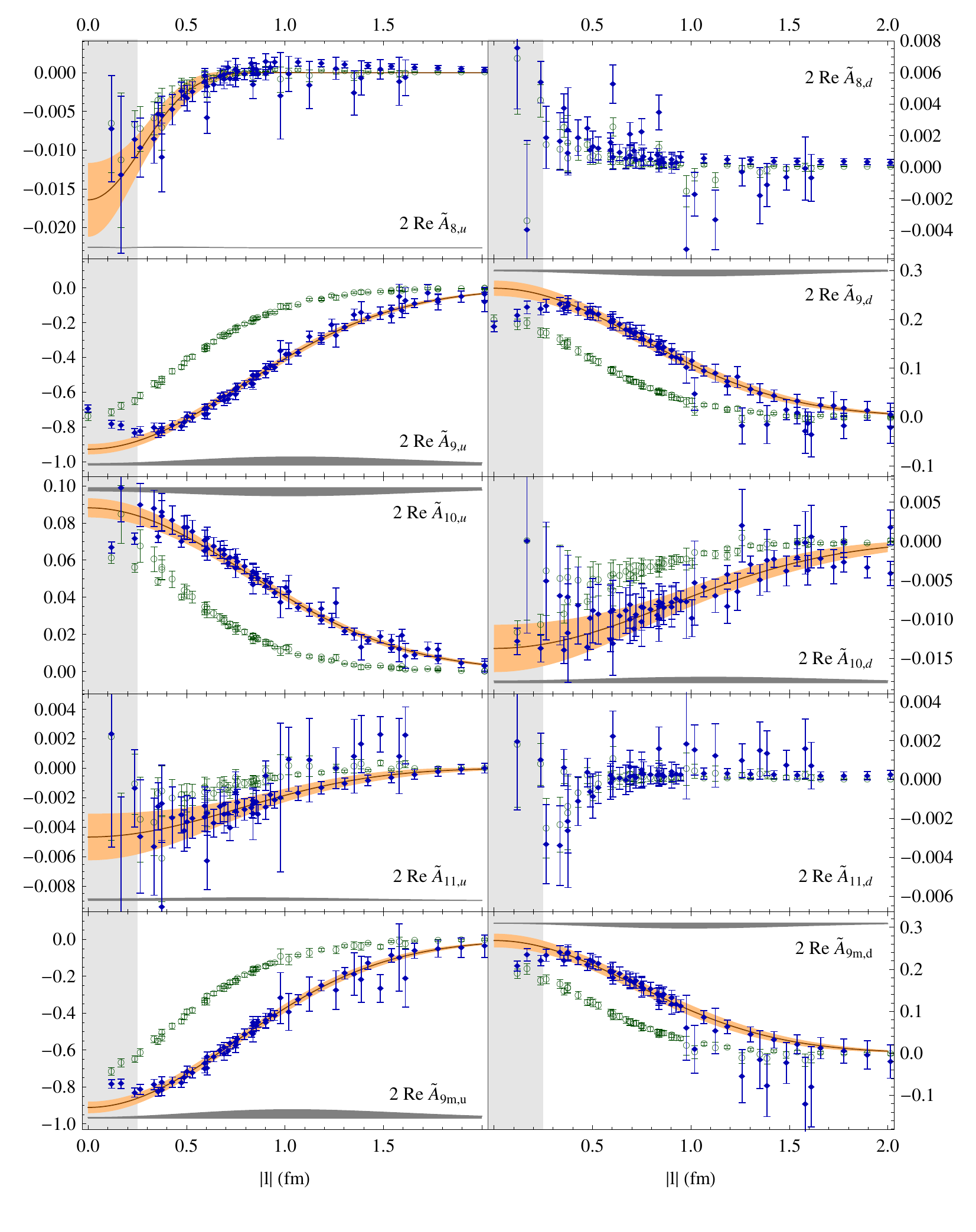}
	\caption{Amplitudes on the coarse\nbdash04 ensemble, continued. For convenience, we have introduced a combined amplitude $\tAmp_{9m}$, which is associated with the \TMD $h_1$.  The uncertainties combined in $\Delta[\delta\hat m]$, are given by the shaded horizontal bands.\label{fig-amps2}}
\end{figure*}

We now discuss results for the coarse\nbdash04 ensemble, with a pion mass of about $500\units{MeV}$. 
In Figs. \ref{fig-amps} and \ref{fig-amps2}, the open data points show the unrenormalized amplitudes obtained at $\elll \tcdot P = 0$.
From the Gaussian fit to $\tAmp_2$, we determine $Z^{-1}_{\Psi,z} = 0.938 \pm 0.005_\text{stat.} \pm 0.042_{\Delta[\delta \hat m]}$, where the second error is associated with the combined uncertainly $\Delta[\delta \hat m]$ that will be specified in 
the paragraph below. 
The fully renormalized data points are shown as solid symbols in Figs. \ref{fig-amps} and \ref{fig-amps2}. The curves and error bands correspond to the Gaussian fits after renormalization with $Z^{-1}_{\Psi,z}$. Data points inside the gray shaded area below $0.25\units{fm}$ have been excluded from the fits. 
The uncertainty obtained from $\Delta[\delta\hat m]$ in Eq.~\eqref{eq-deltamerr} (see the following paragraph) 
is given by the shaded horizontal bands.
The fit parameters obtained for the various amplitudes are listed in Table \ref{tab-gaussparam}. 
\begin{table}
	{\centering\small
	\renewcommand{\arraystretch}{1.2}
	\begin{tabular}{|l|c|c|}
	\hline
	$\tAmp_i$ & $c_{i}$ & $\sigma_{i}\ (\mathrm{fm})$ \\
	\hline\hline
% insert mathematica stuff here
$\tAmp_{2,u}$ &
  $\phantom{-}2.0186 \tpm 0.0063 \tpm 0.0008$ &
  $1.001 \tpm 0.010 \tpm 0.068$\\
$\tAmp_{2,d}$ &
  $\phantom{-}1.0171 \tpm 0.0064 \tpm 0.0005$ &
  $0.975 \tpm 0.012 \tpm 0.063$\\
$\tAmp_{2,u-d}$ &
  $\phantom{-}1.0000 \phantom{{}\tpm 0.0001 \tpm 0.0001}$ &
  $1.029 \tpm 0.018 \tpm 0.073$\\
$\tAmp_{3,u}$ &
  $-0.0978 \tpm 0.0047 \tpm 0.0024$ &
  $1.136 \tpm 0.032 \tpm 0.066$\\
$\tAmp_{3,d}$ &
  $-0.0375 \tpm 0.0026 \tpm 0.0009$ &
  $1.159 \tpm 0.047 \tpm 0.071$\\
$\tAmp_{3,u-d}$ &
  $-0.0599 \tpm 0.0037 \tpm 0.0014$ &
  $1.125 \tpm 0.044 \tpm 0.065$\\
$\tAmp_{6,u}$ &
  $-0.9080 \tpm 0.035\phantom{0} \tpm 0.015\phantom{0}$ &
  $1.207 \tpm 0.036 \tpm 0.089$\\
$\tAmp_{6,d}$ &
  $\phantom{-}0.2870 \tpm 0.019\phantom{0} \tpm 0.0033$ &
  $1.023 \tpm 0.048 \tpm 0.059$\\
$\tAmp_{6,u-d}$ &
  $-1.1920 \tpm 0.037\phantom{0} \tpm 0.019\phantom{0}$ &
  $1.164 \tpm 0.026 \tpm 0.080$\\
$\tAmp_{7,u}$ &
  $-0.1041 \tpm 0.0064 \tpm 0.0021$ &
  $1.151 \tpm 0.047 \tpm 0.074$\\
$\tAmp_{7,d}$ &
  $\phantom{-}0.0232 \tpm 0.0038 \tpm 0.0004$ &
  $1.079 \tpm 0.12\phantom{0} \tpm 0.063$\\
$\tAmp_{7,u-d}$ &
  $-0.1278 \tpm 0.0063 \tpm 0.0025$ &
  $1.140 \tpm 0.037 \tpm 0.073$\\
$\tAmp_{8,u}$ &
  $-0.0164 \tpm 0.0048 \tpm 0.0001$ &
  $0.359 \tpm 0.058 \tpm 0.004$\\
$\tAmp_{8,u-d}$ &
  $-0.0178 \tpm 0.0035 \tpm 0.0001$ &
  $0.433 \tpm 0.047 \tpm 0.007$\\
$\tAmp_{9,u}$ &
  $-0.9268 \tpm 0.030\phantom{0} \tpm 0.011\phantom{0}$ &
  $1.101 \tpm 0.028 \tpm 0.073$\\
$\tAmp_{9,d}$ &
  $\phantom{-}0.2636 \tpm 0.016\phantom{0} \tpm 0.0027$ &
  $1.057 \tpm 0.051 \tpm 0.066$\\
$\tAmp_{9,u-d}$ &
  $-1.1944 \tpm 0.034\phantom{0} \tpm 0.015\phantom{0}$ &
  $1.089 \tpm 0.023 \tpm 0.070$\\
$\tAmp_{10,u}$ &
  $\phantom{-}0.0881 \tpm 0.0052 \tpm 0.0020$ &
  $1.134 \tpm 0.036 \tpm 0.067$\\
$\tAmp_{10,d}$ &
  $-0.0137 \tpm 0.0031 \tpm 0.0003$ &
  $1.188 \tpm 0.18\phantom{0} \tpm 0.076$\\
$\tAmp_{10,u-d}$ &
  $\phantom{-}0.1024 \tpm 0.0054 \tpm 0.0024$ &
  $1.139 \tpm 0.033 \tpm 0.067$\\
$\tAmp_{11,u}$ &
  $-0.0047 \tpm 0.0016 \tpm 0.0002$ &
  $0.986 \tpm 0.16\phantom{0} \tpm 0.041$\\
$\tAmp_{11,u-d}$ &
  $-0.0045 \tpm 0.0015 \tpm 0.0002$ &
  $1.102 \tpm 0.19\phantom{0} \tpm 0.053$\\
\hline
$\tAmp_{\text{9m},u}$ &
  $-0.9110 \tpm 0.032\phantom{0} \tpm 0.0053$ &
  $1.058 \tpm 0.035 \tpm 0.072$\\
$\tAmp_{\text{9m},d}$ &
  $\phantom{-}0.2683 \tpm 0.017\phantom{0} \tpm 0.0015$ &
  $1.013 \tpm 0.062 \tpm 0.064$\\
$\tAmp_{\text{9m},u-d}$ &
  $-1.1822 \tpm 0.034\phantom{0} \tpm 0.0077$ &
  $1.046 \tpm 0.027 \tpm 0.069$\\
$\tAmp_{2{+}6,u}$ &
  $\phantom{-}1.1206 \tpm 0.035\phantom{0} \tpm 0.0054$ &
  $0.851 \tpm 0.021 \tpm 0.039$\\
$\tAmp_{2{+}6,d}$ &
  $\phantom{-}1.2962 \tpm 0.021\phantom{0} \tpm 0.0088$ &
  $0.989 \tpm 0.015 \tpm 0.058$\\
$\tAmp_{2{+}6,u-d}$ &
  $-0.2451 \tpm 0.034\phantom{0} \tpm 0.0064$ &
  $1.622 \tpm 0.18\phantom{0} \tpm 0.17\phantom{0}$\\
$\tAmp_{2{-}6,u}$ &
  $\phantom{-}2.8989 \tpm 0.035\phantom{0} \tpm 0.023\phantom{0}$ &
  $1.066 \tpm 0.014 \tpm 0.071$\\
$\tAmp_{2{-}6,d}$ &
  $\phantom{-}0.7265 \tpm 0.020\phantom{0} \tpm 0.0041$ &
  $0.956 \tpm 0.025 \tpm 0.054$\\
$\tAmp_{2{-}6,u-d}$ &
  $\phantom{-}2.1756 \tpm 0.036\phantom{0} \tpm 0.022\phantom{0}$ &
  $1.104 \tpm 0.019 \tpm 0.075$\\
$\tAmp_{2{+}\text{9m},u}$ &
  $\phantom{-}1.0969 \tpm 0.032\phantom{0} \tpm 0.0031$ &
  $0.956 \tpm 0.029 \tpm 0.058$\\
$\tAmp_{2{+}\text{9m},d}$ &
  $\phantom{-}1.2805 \tpm 0.019\phantom{0} \tpm 0.0039$ &
  $0.986 \tpm 0.017 \tpm 0.062$\\
$\tAmp_{2{+}\text{9m},u-d}$ &
  $-0.1980 \tpm 0.034\phantom{0} \tpm 0.0020$ &
  $1.068 \tpm 0.13\phantom{0} \tpm 0.066$\\
$\tAmp_{2{-}\text{9m},u}$ &
  $\phantom{-}2.9113 \tpm 0.032\phantom{0} \tpm 0.011\phantom{0}$ &
  $1.024 \tpm 0.015 \tpm 0.069$\\
$\tAmp_{2{-}\text{9m},d}$ &
  $\phantom{-}0.7483 \tpm 0.018\phantom{0} \tpm 0.0015$ &
  $0.958 \tpm 0.029 \tpm 0.059$\\
$\tAmp_{2{-}\text{9m},u-d}$ &
  $\phantom{-}2.1673 \tpm 0.034\phantom{0} \tpm 0.011\phantom{0}$ &
  $1.044 \tpm 0.019 \tpm 0.071$\\
% end of mathematica stuff
\hline
  \end{tabular}
 	 \renewcommand{\arraystretch}{1}
	 }
	\caption{Results from Gaussian fits on the coarse\nbdash04 ensemble at $m_\pi \approx 500\units{MeV}$. The first error is statistical. The second error includes the statistical uncertainty in $\delta m$ and an estimate of discretization uncertainties, as given in Eq.~\eqref{eq-deltamerr}. The values for $u-d$-quarks have been obtained directly from Gaussian fits to the $u-d$ data. Note that we have performed the conversion to physical units using the values for the lattice spacing $a$ given in Table \ref{tab-gaugeconfs}. See also footnotes \thefnnumberamp\  and \thefnnumber.\label{tab-gaussparam}}
\end{table}
Most importantly, we find clearly non-zero signals for all amplitudes, even at larger distances, except for $\tAmp_{8}$ and $\tAmp_{11}$.
Furthermore, the lattice data points show a high degree of consistency within the (in many cases encouragingly small) statistical and systematic uncertainties. 
These results already point towards rather non-trivial correlations between momentum and spin degrees of freedom inside the nucleon.
In case of the ``unpolarized'' amplitude $\tAmp_{2,u}$, our data have very small statistical errors, and we obtain a comparatively large value of $3.9$ for $\chi^2$ per degree of freedom.\footnote{Strictly speaking, we cannot make strong probabilistic arguments based on our values of $\chi^2/\text{p.d.o.f}$, because we do not treat potential correlations explicitly in Eq. \eqref{eq-chisqr}.}
In a fit that excludes step-like link paths, $\chi^2/\text{p.d.o.f}$ is reduced to $2.0$, indicating that the the small violation of rotational symmetry present in our calculation is to a large degree responsible for the high $\chi^2$-value. 
In the case of the twist-4 amplitude $\tAmp_{1,u}$, we obtain an even larger value, $\chi^2/\text{p.d.o.f} = 4.8$ both with and without step-like paths. In contrast to the case of $\tAmp_{2,u}$, the data points visually follow a different curve that deviates from the Gaussian fit function. The same is true for $\tAmp_{1,d}$. We conclude that the Gaussian model does not adequately describe amplitude $\tAmp_1$. Statistical fluctuations are still too large to obtain stable fits to $\tAmp_{8,d}$ and $\tAmp_{11,d}$. 
The meaning of the amplitudes $\tAmp_{2\pm6}$ and $\tAmp_{2\pm\text{9m}}$ will be discussed in section \ref{sec-A2A6}.

In order to get an estimate for systematic errors, we combine the statistical error $\Delta[\delta \hat m]_\text{stat}$ and the estimate of discretization uncertainties $\Delta[\delta \hat m]_\text{dis}$ of appendix \ref{sec-discerr}:
\begin{equation}
	\Delta[\delta \hat m] \equiv \sqrt{ \Delta[\delta \hat m]_\text{stat}^2 + \Delta[\delta \hat m]_\text{dis}^2 }
	\label{eq-deltamerr}
\end{equation}
and find that $ \Delta[\delta \hat m]_\text{dis}^2 $ dominates.
It turns out that $\Delta[\delta \hat m]$ mainly affects the widths of the renormalized Gaussians, not so much the renormalized $c_{i,q}$, because variations in the $c_{i,q}^\text{unren}$ largely cancel in the process of renormalization with $Z^{-1}_{\Psi,z}$. Next, we estimate discretization errors associated with the breaking of rotational invariance. We compare two different Gaussian fits to the self-energy-renormalized data for $\tAmp_{2,u}$. In one fit, we use all the data points above $|\elll|\geq 0.25\units{fm}$, in another fit we restrict ourselves to data points from straight link paths on the axes. On the coarse\nbdash04 ensemble, the relative difference in $c^\unren_{2,u}$ is just $0.6\%$, and the relative difference in $\sigma_{2,u}$ is $1.5\%$.  Analogous to the case of $\Delta[\delta \hat m]$, the effect on the renormalized parameters $c_{i,q}$ is expected to be even smaller. We assume that our estimate is also valid for the other amplitudes, where it is more difficult to make such a comparison due to larger statistical errors. In the following, we do not show uncertainties from violation of rotational invariance, because they are negligible compared to statistical uncertainties and uncertainties accounted for in $\Delta[\delta \hat m]$. Quantities given in physical units are also affected by the uncertainty in the lattice spacing $a$, which is not included in the errors we quote.
It can, however, easily be obtained by adding a relative uncertainty of $|d| \Delta a/a$ to any quantity given in units $\mathrm{GeV}^d$ or $\mathrm{fm}^d$. Other sources of errors we do not treat here include contributions from excited states in the three-point function and the static quark potential, contributions from disconnected diagrams, and effects of the finite lattice volume. 
Finally, in order to obtain results at the physical point, the lattice results as functions of the pion mass have to be extrapolated to $m_\pi^{\text{phys}}$.
Although we have already performed some preliminary studies with respect to the above mentioned issues, 
they are beyond the scope of this initial investigation and will have to be left for future work.

A remaining challenge within our procedure is to associate a 
renormalization scale with the self-energy renormalization condition we employ. Especially the widths of our amplitudes and of the resulting $x$-integrated \TMDs are very sensitive to $\delta m$, and thus to the employed renormalization condition. We remark that the issue of gauge link self-energy appears for any link geometry that contains space-like sections. 
Of great interest for future lattice studies in particular is the development of theoretically more accurate
definitions of the correlator Eq. \eqref{eq-corr} as discussed in the introduction \ref{sec-intro}.
For our purposes, it would be important to have subtraction and/or soft factors
included that cancel the gauge-link self-energies right from the start, as discussed already in Ref. \cite{Collins:2008ht}.

\subsection{Interpretation of the lattice results in terms of transverse momentum dependent distributions and quark densities}
\label{sec-densities}

\begin{figure}[tb]
	\centering%
	\includegraphics[width=\linewidth]{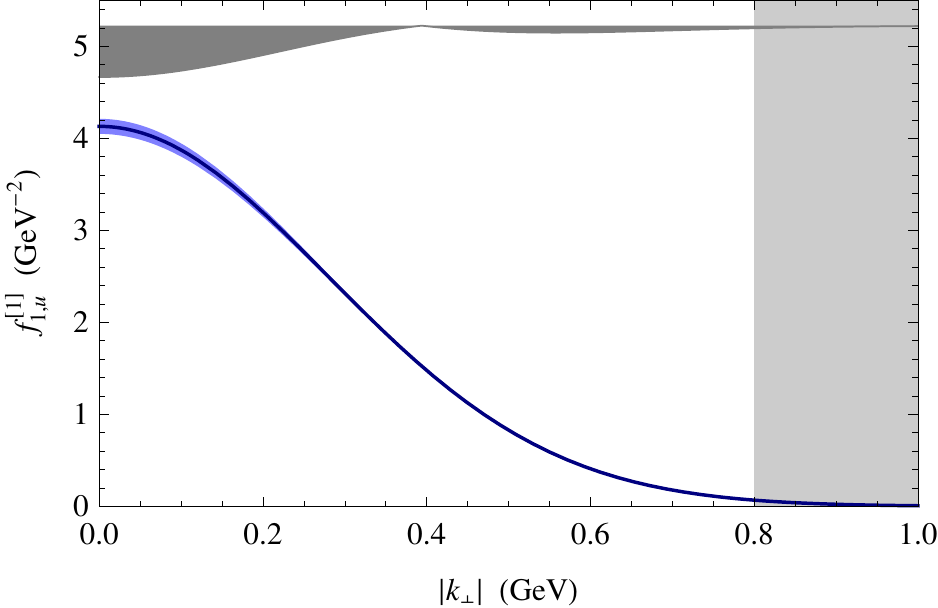}
	\caption{$f_1^\xmom{1}(\vprp{k}^2)$ for up-quarks obtained using the Gaussian parametrization at a pion mass $m_\pi\approx 500 \units{MeV}$. The solid curve and the statistical error band in blue have been obtained from a Gaussian fit to the amplitude $\tAmp_2$, as shown in Fig. \ref{fig-amps}. The gray band on the top indicates uncertainties that can effectively be expressed as an error in $\delta m$. The gray region at large $|\vprp{k}|$ indicates that we qualitatively expect strong parametrization dependence to set in at $|\vprp{k}|\gtrsim 1/0.25 \units{fm} \approx 0.8\units{GeV}$. }
	\label{fig-f1}
\end{figure}

Using Eqns. \eqref{eq-xintf1}, \eqref{eq-xintFT} and analogous Fourier-trans\-forms for the other \TMDs, we can now determine $x$-integrated \TMDs from the Gaussian fits to the amplitudes discussed in the previous section. 
As an example, for the unpolarized distribution $f_1$ we obtain from Eqns.~\eqref{eq-xintf1}, \eqref{eq-xintFT} 
\begin{equation}
  f_1^\xmom{1}(\vprp{k}^2) = \frac{c_2\,\sigma_2^2}{4\pi} e^{-\frac{\vprp{k}^2}{(2 / \sigma_2)^2}}\, .
  \label{eq-f1xmom}
\end{equation}
The result for up-quarks is shown in Fig. \ref{fig-f1}. 
Using the $x$-integral of Eq. \eqref{eq-tmdsfromamps}, it is easy to express all $x$-integrated \TMDs in terms of the parameters $c_i$, $\sigma_i$
provided in Table \ref{tab-gaussparam}. 
Note that we have chosen to determine $c_{9m}$, $\sigma_{9m}$ directly from Gaussian fits to the combined amplitude $\tAmp_{9m}$.
This way, all resulting expressions for the leading twist \TMDs are again single Gaussians of the form $\tilde{c} \exp(- \vprp{k}^2/\tilde{\sigma}^2)$. 
For convenience, we list the numerical results for $\tilde{c}$ and $\tilde{\sigma}$ in Table \ref{tab-gausstmds}.
\begin{table*}
  \centering 
  \renewcommand{\arraystretch}{1.4} 
  \begin{tabular}{|l|rc|rc|}
    \hline
    & \multicolumn{2}{c|}{$\tilde{c}\ (\mathrm{GeV}^{-2})$} 
    & \multicolumn{2}{c|}{$\tilde{\sigma}\ (\mathrm{GeV})$} \\ \hline\hline
    % insert Mathematica results here
$f_{1,u}^\xmom{1}$ &
  $c_2\, \sigma_2^2 / (4\pi) = $ &
  $\phantom{-}4.13 \pm 0.09 \pm 0.56$ &
  $2 / \sigma_2 = $ &
  $0.394 \pm 0.004 \pm 0.027$\\
$f_{1,d}^\xmom{1}$ &
  $c_2\, \sigma_2^2 / (4\pi) = $ &
  $\phantom{-}1.98 \pm 0.05 \pm 0.26$ &
  $2 / \sigma_2 = $ &
  $0.405 \pm 0.005 \pm 0.027$\\
$g_{1,u}^\xmom{1}$ &
  $-c_6\, \sigma_6^2 / (4\pi) = $ &
  $\phantom{-}2.70 \pm 0.17 \pm 0.44$ &
  $2 / \sigma_6 = $ &
  $0.327 \pm 0.010 \pm 0.025$\\
$g_{1,d}^\xmom{1}$ &
  $-c_6\, \sigma_6^2 / (4\pi) = $ &
  $-0.61 \pm 0.07 \pm 0.08$ &
  $2 / \sigma_6 = $ &
  $0.385 \pm 0.018 \pm 0.023$\\
$f_{1,u}^\xmom{1} + g_{1,u}^\xmom{1}$ &
  $c_{2{-}6}\, \sigma_{2{-}6}^2 / (4\pi) = $ &
  $\phantom{-}6.73 \pm 0.21 \pm 0.94$ &
  $2 / \sigma_{2{-}6} = $ &
  $0.370 \pm 0.005 \pm 0.025$\\
$f_{1,d}^\xmom{1} + g_{1,d}^\xmom{1}$ &
  $c_{2{-}6}\, \sigma_{2{-}6}^2 / (4\pi) = $ &
  $\phantom{-}1.36 \pm 0.08 \pm 0.17$ &
  $2 / \sigma_{2{-}6} = $ &
  $0.413 \pm 0.011 \pm 0.024$\\
$f_{1,u}^\xmom{1} - g_{1,u}^\xmom{1}$ &
  $c_{2{+}6}\, \sigma_{2{+}6}^2 / (4\pi) = $ &
  $\phantom{-}1.66 \pm 0.09 \pm 0.16$ &
  $2 / \sigma_{2{+}6} = $ &
  $0.463 \pm 0.011 \pm 0.022$\\
$f_{1,d}^\xmom{1} - g_{1,d}^\xmom{1}$ &
  $c_{2{+}6}\, \sigma_{2{+}6}^2 / (4\pi) = $ &
  $\phantom{-}2.59 \pm 0.08 \pm 0.33$ &
  $2 / \sigma_{2{+}6} = $ &
  $0.399 \pm 0.006 \pm 0.024$\\
$h_{1,u}^\xmom{1}$ &
  $-c_{\text{9m}}\, \sigma_{\text{9m}}^2 / (4\pi) = $ &
  $\phantom{-}2.08 \pm 0.15 \pm 0.30$ &
  $2 / \sigma_{\text{9m}} = $ &
  $0.373 \pm 0.013 \pm 0.026$\\
$h_{1,d}^\xmom{1}$ &
  $-c_{\text{9m}}\, \sigma_{\text{9m}}^2 / (4\pi) = $ &
  $-0.56 \pm 0.08 \pm 0.08$ &
  $2 / \sigma_{\text{9m}} = $ &
  $0.388 \pm 0.024 \pm 0.025$\\
$f_{1,u}^\xmom{1} + h_{1,u}^\xmom{1}$ &
  $c_{2{-}\text{9m}}\, \sigma_{2{-}\text{9m}}^2 / (4\pi) = $ &
  $\phantom{-}6.24 \pm 0.19 \pm 0.86$ &
  $2 / \sigma_{2{-}\text{9m}} = $ &
  $0.385 \pm 0.006 \pm 0.026$\\
$f_{1,d}^\xmom{1} + h_{1,d}^\xmom{1}$ &
  $c_{2{-}\text{9m}}\, \sigma_{2{-}\text{9m}}^2 / (4\pi) = $ &
  $\phantom{-}1.40 \pm 0.09 \pm 0.18$ &
  $2 / \sigma_{2{-}\text{9m}} = $ &
  $0.412 \pm 0.013 \pm 0.026$\\
$f_{1,u}^\xmom{1} - h_{1,u}^\xmom{1}$ &
  $c_{2{+}\text{9m}}\, \sigma_{2{+}\text{9m}}^2 / (4\pi) = $ &
  $\phantom{-}2.05 \pm 0.13 \pm 0.26$ &
  $2 / \sigma_{2{+}\text{9m}} = $ &
  $0.412 \pm 0.013 \pm 0.025$\\
$f_{1,d}^\xmom{1} - h_{1,d}^\xmom{1}$ &
  $c_{2{+}\text{9m}}\, \sigma_{2{+}\text{9m}}^2 / (4\pi) = $ &
  $\phantom{-}2.54 \pm 0.09 \pm 0.33$ &
  $2 / \sigma_{2{+}\text{9m}} = $ &
  $0.400 \pm 0.007 \pm 0.026$\\
$g_{1T,u}^\xmom{1}$ &
  $- m_N^2 c_7\, \sigma_7^4 / (8\pi) = $ &
  $\phantom{-}8.72 \pm 1.3\phantom{0} \pm 2.4\phantom{0}$ &
  $2 / \sigma_7 = $ &
  $0.342 \pm 0.014 \pm 0.022$\\
$g_{1T,d}^\xmom{1}$ &
  $- m_N^2 c_7\, \sigma_7^4 / (8\pi) = $ &
  $-1.46 \pm 0.59 \pm 0.35$ &
  $2 / \sigma_7 = $ &
  $0.362 \pm 0.039 \pm 0.022$\\
$h_{1L,u}^{\prp\xmom{1}}$ &
  $- m_N^2 c_{10}\, \sigma_{10}^4 / (8\pi) = $ &
  $-6.96 \pm 0.82 \pm 1.8\phantom{0}$ &
  $2 / \sigma_{10} = $ &
  $0.348 \pm 0.012 \pm 0.021$\\
$h_{1L,d}^{\prp\xmom{1}}$ &
  $- m_N^2 c_{10}\, \sigma_{10}^4 / (8\pi) = $ &
  $\phantom{-}1.24 \pm 0.71 \pm 0.31$ &
  $2 / \sigma_{10} = $ &
  $0.325 \pm 0.047 \pm 0.023$\\
$h_{1T,u}^{\prp\xmom{1}}$ &
  $m_N^4 c_{11}\, \sigma_{11}^6 / (16\pi) = $ &
  $-3.77 \pm 4.6\phantom{0} \pm 0.76$ &
  $2 / \sigma_{11} = $ &
  $0.348 \pm 0.012 \pm 0.021$\\    
  % end of Mathematica results
   \hline
  \end{tabular}
  \renewcommand{\arraystretch}{1}
  \caption{Numerical results for $x$-integrated leading twist \TMDs parametrized in terms of Gaussians 
  of the form $\tilde{c} \exp(- \vprp{k}^2/\tilde{\sigma}^2)$, for a pion mass of $m_\pi\approx 500 \units{MeV}$, straight gauge links, and a renormalization condition based on the static quark potential. 
  We also include results for linear combinations of \TMDs corresponding to an alternative Gaussian parametrization, see section \ref{sec-A2A6}. 
  The first error is statistical. The second error includes the statistical uncertainty in $\delta m$ and an estimate of discretization uncertainties, as given in Eq.~\eqref{eq-deltamerr}. 
  Note that we have performed the conversion to physical units using the values for the lattice spacing $a$ given in Table \ref{tab-gaugeconfs}, see also footnote \thefnnumber.\label{tab-gausstmds}}
\end{table*}
\begin{figure}[p]
	\centering%
	\subfloat[][]{%
		\label{fig-f1uof1d}%
		\includegraphics[width=0.96\linewidth]{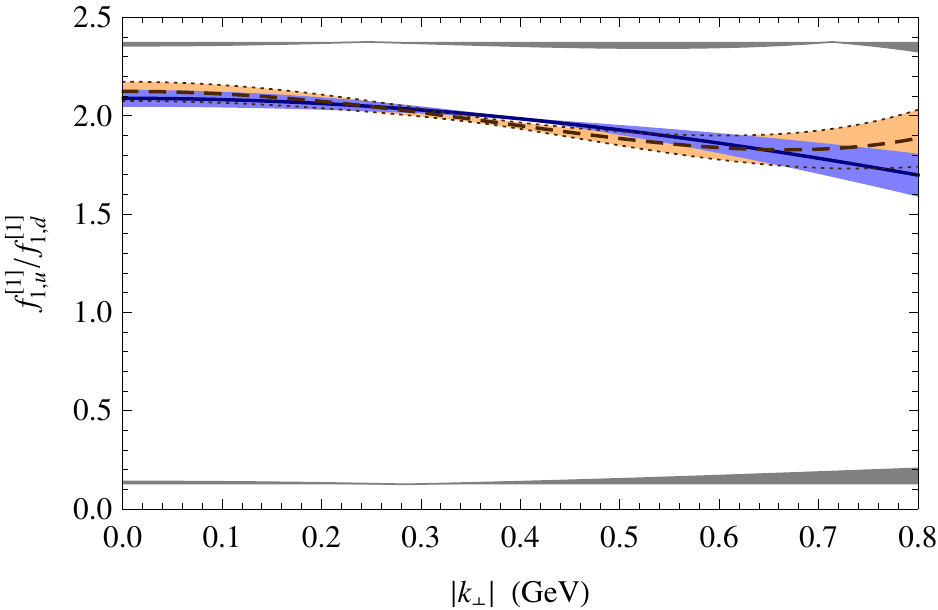}
		}\\%
	\subfloat[][]{%
		\label{fig-g1uog1d}%\
		\includegraphics[width=0.96\linewidth,clip=true]{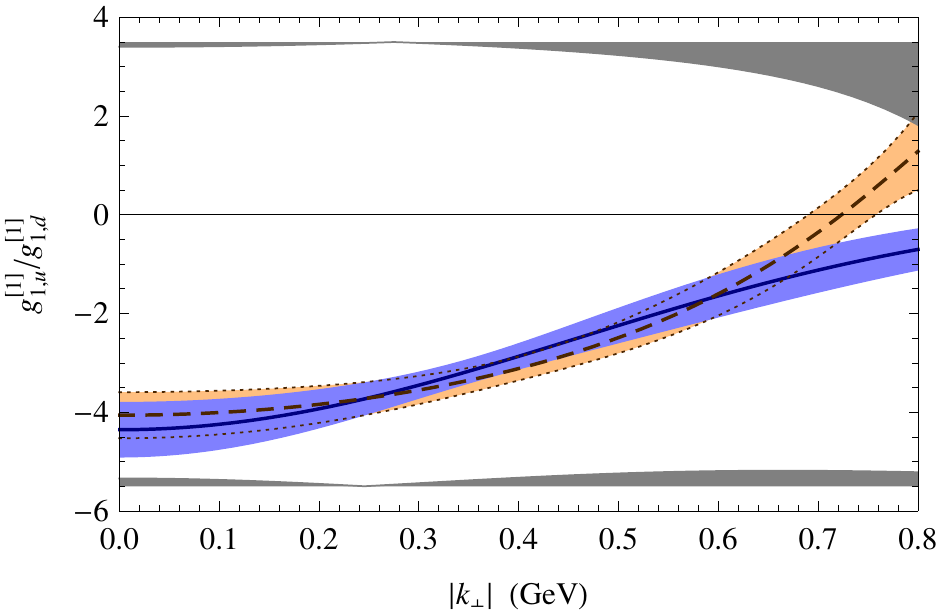}
		}\\%
	\subfloat[][]{%
		\label{fig-h1uoh1d}%\
		\includegraphics[width=0.96\linewidth,clip=true]{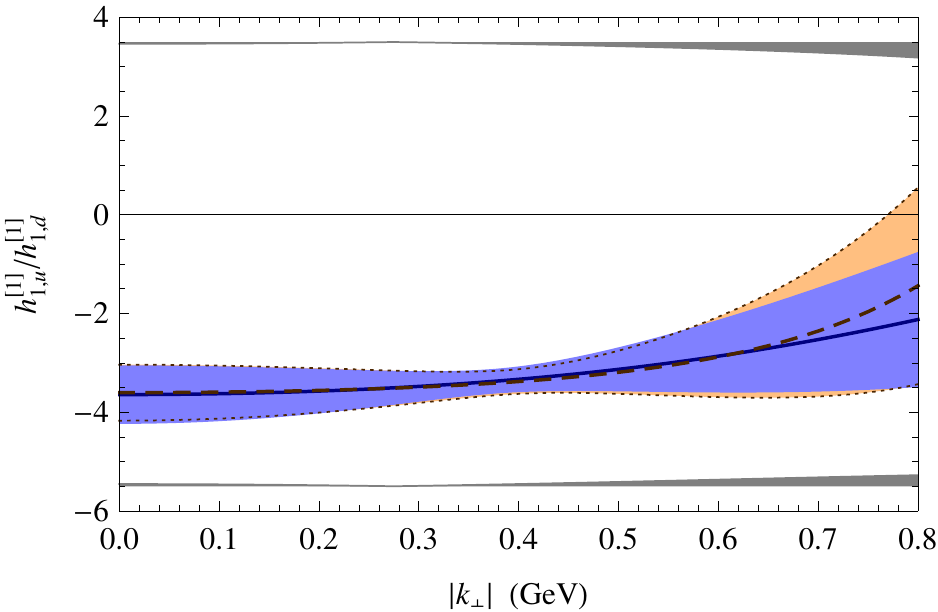}
		}%		
	\caption[Yline]{%
	Flavor-ratios at a pion mass $m_\pi\approx 500 \units{MeV}$. The solid curve and the statistical error band in blue have been obtained from the Gaussian fits displayed in Fig. \ref{fig-amps} and \ref{fig-amps2}. The corresponding errors associated with $\Delta[\delta m]$ are shown as a gray band at the bottom. For the dashed curve and the band in orange we have used alternative Gaussian parametrizations as discussed in section \ref{sec-A2A6}. The respective  uncertainties from $\Delta[\delta m]$ are shown at the top of each plot.  \\
		\subref{fig-f1uof1d}\ 
			$f_{1,u}^\xmom{1}(\vprp{k}^2)/f_{1,d}^\xmom{1}(\vprp{k}^2)$ from $\tAmp_{2}$ (solid) and $\tAmp_{2\pm6}$ (dashed)\\
		\subref{fig-g1uog1d}\ 
			$g_{1,u}^\xmom{1}(\vprp{k}^2)/g_{1,d}^\xmom{1}(\vprp{k}^2)$ from $\tAmp_{6}$ (solid) and $\tAmp_{2\pm6}$ (dashed)\\
		\subref{fig-h1uoh1d}\ 
			$h_{1,u}^\xmom{1}(\vprp{k}^2)/h_{1,d}^\xmom{1}(\vprp{k}^2)$ from $\tAmp_{\text{9m}}$ (solid) and $\tAmp_{2\pm\text{9m}}$ (dashed)\\
			\par			
		\label{fig-flavorratios}%
		}
\end{figure}

In most cases, the widths $\tilde \sigma$ turn out to be fairly similar. Correspondingly, flavor ratios  $f_{1,u}^\xmom{1}(\vprp{k}^2)/f_{1,d}^\xmom{1}(\vprp{k}^2)$ and $h_{1,u}^\xmom{1}(\vprp{k}^2)/h_{1,d}^\xmom{1}(\vprp{k}^2)$ 
shown in Figs. \ref{fig-f1uof1d} and \ref{fig-h1uoh1d}, respectively, are relatively flat functions of $\vprp{k}$. 
In contrast, the width of $g_{1,u}^\xmom{1}$ is significantly lower than that of $g_{1,d}^\xmom{1}$, resulting in a 
clearly visible slope of the flavor ratio in Fig. \ref{fig-g1uog1d}. 
By and large, it is interesting to see that the $\vprp{k}$-distribution for the down-quarks appear in all three cases to be broader
than for the up-quarks.
In qualitative agreement with our findings, experimental results by the CLAS collaboration \cite{Avakian:2010ae} analyzed using the approach of Ref. \cite{Anselmino:2006yc} favor a reduced width of $g_1$ as compared to $f_1$.
Note that the plots also show results obtained for the same quantities with an alternative Gaussian parametrization which will be discussed in section \ref{sec-A2A6}.

It is natural to think of \TMDs as functions that characterize probability densities of partons in the nucleon. Although the probability interpretation is not rigorous, see, e.g., Ref. \cite{Collins:2003fm}, 
we provide an interpretation of our results in this fashion for the sake of an intuitive picture. 
Transverse momentum dependent quark densities are introduced as 
%\muline{(see also \cite{Diehl:2005jf})}
\begin{align}
	\MoveEqLeft
  \rho_q(x,\vprp{k};\lambda,\vprp{s},\Lambda,\vprp{S}) \nonumber \\
	& \equiv \Phi_q^{[(\gamma^+ + \lambda \gamma^+\gamma^5 - s^j i \sigma^{+j} \gamma^5)/2]}(x,\vprp{k};P,S)\, , 
	\label{eq-densdef}
\end{align}
Here the choice of the matrix $\GammaOp = {\scriptstyle\frac{1}{2}} ( \gamma^+ + \lambda \gamma^+\gamma^5 - s^j i \sigma^{+j} \gamma^5 )$ ensures projection on the 
``good'' spinor components \cite{Kogut:1969xa,Burkardt:1995ct} and, simultaneously, on the desired light cone quark helicity $\lambda$ and transverse quark polarization $\vprp{s}$ \cite{Soper:1972xc,Diehl:2005jf}. We introduce the following special cases of densities:
\begin{align}
	\rho_{UU,q} & \equiv \frac{1}{2} \sum_{\lambda, \Lambda = \pm 1} \rho_q(x,\vprp{k};\lambda,0,\Lambda,0) = f_{1,q}\, , \label{eq-rhoUU} \displaybreak[0] \\
	\rho_{TU,q} & \equiv \sum_{\lambda = \pm 1} \rho_q(x,\vprp{k};\lambda,0,0,\vprp{S}) \nonumber \\ & = f_{1,q} + \toddmark{ \frac{\vect{S}_j \vect{\epsilon}_{ji} \vect{k}_i}{m_N}\, f_{1T,q}^\perp }\, , \label{eq-rhoTU} \displaybreak[0] \\
	\rho_{UT,q} & \equiv \frac{1}{2}\sum_{\Lambda = \pm 1} \rho_q(x,\vprp{k};0,\vprp{s},\Lambda,0) \nonumber  \\
	& = \frac{1}{2}\Big(f_{1,q} +  \toddmark{ \frac{\vect{s}_j\vect{\epsilon}_{ji} \vect{k}_i}{m_N}\, h_{1,q}^\perp} \Big)\, , \label{eq-rhoUT} \displaybreak[0] \\
	\rho_{LL,q} & \equiv \rho_q(x,\vprp{k};\lambda,0,\Lambda,0) = \frac{1}{2}\Big( f_{1,q} + \lambda \Lambda g_{1,q} \Big)\, , \label{eq-rhoLL} \displaybreak[0] \\
	\rho_{TL,q} & \equiv \rho_q(x,\vprp{k};\lambda,0,0,\vprp{S}) \nonumber \\
		& = \frac{1}{2}\Big( f_{1,q} + \lambda\frac{\vprp{k} \cdot \vprp{S}}{m_N} g_{1T,q} + \toddmark{ \frac{\vect{S}_j \vect{\epsilon}_{ji} \vect{k}_i}{m_N}\, f_{1T,q}^\perp }\Big)\, , 
		\label{eq-rhoTL} \displaybreak[0] \\%
	\rho_{LT,q} & \equiv \rho_q(x,\vprp{k};0,\vprp{s},\Lambda,0) \nonumber \\
		& = \frac{1}{2}\Big( f_{1,q} + \Lambda\frac{\vprp{k} \cdot \vprp{s}}{m_N} h_{1L,q}^\perp + \toddmark{ \frac{\vect{s}_j\vect{\epsilon}_{ji} \vect{k}_i}{m_N}\, h_{1,q}^\perp} \Big)\, , \label{eq-rhoLT}  \displaybreak[0] \\		
	\rho_{TT,q} & \equiv \rho_q(x,\vprp{k};0,\vprp{s},0,\vprp{S}) = \frac{1}{2}\Big( f_{1,q} \nonumber \\ & +  \vprp{s}\cdot \vprp{S} h_{1,q} + \frac{\vect{s}_j(2 \vect{k}_j \vect{k}_i - \vprp{k}^2 \delta_{ji}) \vect{S}_i}{2m_N^2}\, h_{1T,q}^\perp \nonumber \\
		& + \toddmark{ \frac{\vect{s}_j\vect{\epsilon}_{ji} \vect{k}_i}{m_N}\, h_{1,q}^\perp} \Big) 	\, ,
		\label{eq-rhoTT}
\end{align}
where the first and the second index of $\rho$ indicates the nucleon and quark polarization, respectively.

From the $x$-moments of amplitudes $\tAmp_i$ obtained on the lattice, we can construct $x$-integrated densities $\rho^\xmom{1}_q$, and decompose them in analogy to Eq. \eqref{eq-corrxmomqantiq} as
\begin{align}
	\MoveEqLeft \rho_q^\xmom{1}(\vprp{k};\lambda,\vprp{s},\Lambda,\vprp{S}) \nonumber \\
	& \equiv \int_{-1}^1 dx\,  \rho_q(x,\vprp{k};\lambda,\vprp{s},\Lambda,\vprp{S}) \nonumber \\
	& =  \int_0^1 dx\,  \rho_q(x,\vprp{k};\lambda,\vprp{s},\Lambda,\vprp{S}) \nonumber \\
	& -   \int_0^1 dx\,  \rho_{\bar q}(x,-\vprp{k};-\lambda,\vprp{s},\Lambda,\vprp{S}) \, .
	\label{eq-xintdensdecomp}
\end{align}
where the anti-quark density $\rho_{\bar q}$ is defined as in Eq. \eqref{eq-densdef} but using the correlator $\Phi^c_q$ of Eq.~\eqref{eq-Ccorr} in the appendix. 
Here the appearance of minus signs in front of $\rho_{\bar q}$
and $\lambda$ accommodates the sign changes in the Dirac matrix $\Gamma$ after charge conjugation, i.e., $\Gamma^c = - {\scriptstyle\frac{1}{2}} ( \gamma^+ - \lambda \gamma^+\gamma^5 - s^j i \sigma^{+j} \gamma^5 )$.
We conclude that the $x$-integrated densities $\rho^\xmom{1}_q$ are differences of quark densities $\rho_q$ and anti-quark densities $\rho_{\bar q}$ of \par
\vbox{
\begin{itemize*}
	\item opposite transverse momentum $-\vprp{k}$,
	\item opposite light cone helicity $-\lambda$, 
	\item same transverse polarization $\vprp{s}$.
\end{itemize*}
}
Strictly speaking, the densities that are integrated over $x$ from $-1$ to $+1$ are thus not densities themselves and can, at least in principle, become negative.

With the Gaussian $x$-moments of \TMDs from Table \ref{tab-gausstmds} as input, we are in a position to draw plots of the $x$-integrated transverse momentum dependent densities of quarks in the nucleon. 
Two particularly interesting and statistically well-determined $x$-integrated densities are $\rho_{LT}^\xmom{1}$ and $\rho_{TL}^\xmom{1}$. They feature significant dipole deformations due 
to correlations in the transverse spins and intrinsic transverse momentum, as can be seen from 
the terms proportional to $g_{1T}$ and $h_{1L}^\prp$ in Eqns.~\eqref{eq-rhoTL} and \eqref{eq-rhoLT}, in combination with our non-zero 
results for the relevant amplitudes $\tAmp_7$ and $\tAmp_{10}$, see Eq.~\eqref{eq-tmdsfromamps}. 
For corresponding density plots and their interpretation, we refer to our previous publication Ref.~\cite{Hagler:2009mb}.
The dipole deformations can be characterized by average transverse momentum shifts
of the quarks, denoted by $\langle \vect{k}_x \rangle_{TL}$ and $\langle \vect{k}_x \rangle_{LT}$.
These are defined by ratios of specific moments in $x$- \emph{and} $\vprp{k}$ of the densities, as we will discuss in the following section. 

\begin{figure}[tb]
	\centering%
	\subfloat[][]{%
		\label{fig-f1pg1uof1pg1d}%
		\includegraphics[width=0.98\linewidth]{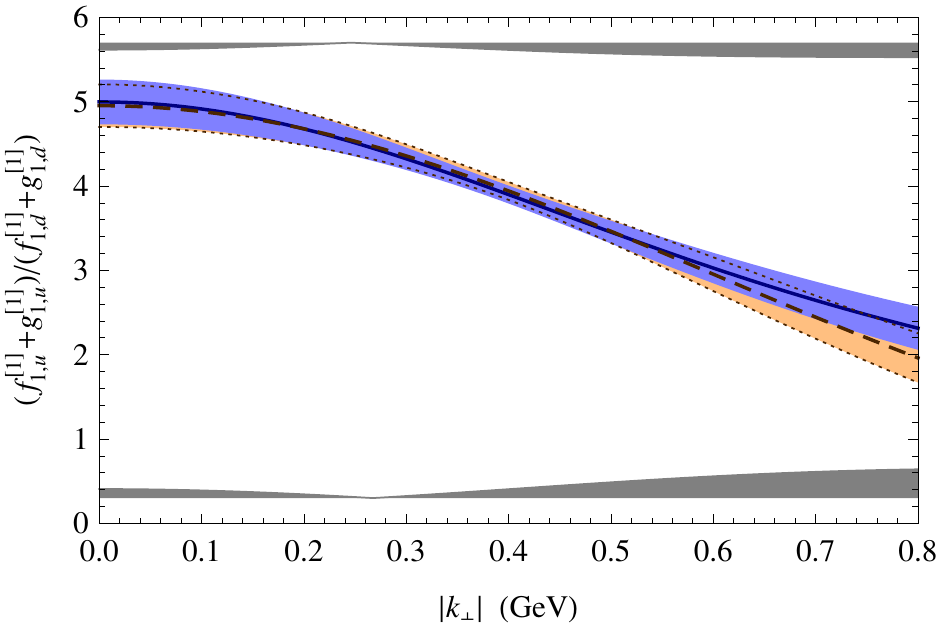}
		}\\%
	\subfloat[][]{%
		\label{fig-f1ph1uof1ph1d}%\
		\includegraphics[width=0.98\linewidth,clip=true]{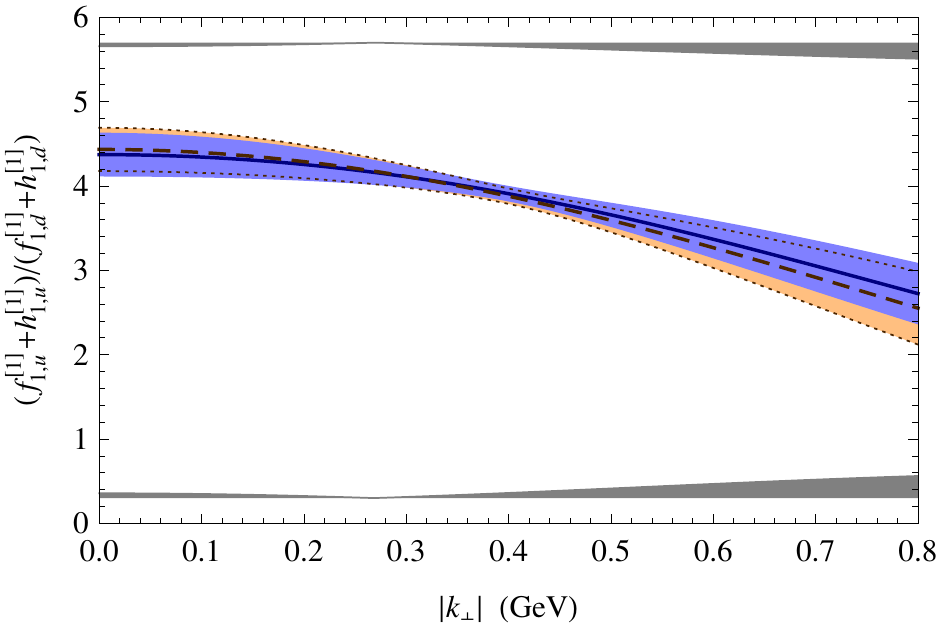}
		}\\%
	\caption[flavordensratios]{%
	Flavor ratios of $x$-integrated densities at a pion mass $m_\pi\approx 500 \units{MeV}$. The solid curve and the statistical error band in blue have been obtained from the Gaussian fits displayed in Fig. \ref{fig-amps} and \ref{fig-amps2}. The corresponding errors associated with $\Delta[\delta m]$ are shown as a gray band at the top. For the dashed curve and the band in orange we have used alternative Gaussian parametrizations as discussed in section \ref{sec-A2A6}. The respective errors from $\Delta[\delta m]$ are shown at the bottom of each plot.  We show up vs. down ratios \\
		\subref{fig-f1pg1uof1pg1d}\ 
			of $f_1^\xmom{1}+g_1^\xmom{1}$ from $\tAmp_{2}$, $\tAmp_{6}$ (solid) and $\tAmp_{2\pm6}$ (dashed), and\\
		\subref{fig-f1ph1uof1ph1d}\ 
			of $f_1^\xmom{1}+h_1^\xmom{1}$ from $\tAmp_{2}$, $\tAmp_{\text{9m}}$ (solid) and $\tAmp_{2\pm\text{9m}}$ (dashed)\\
			\par			
		\label{fig-flavordensratios}%
		}
\end{figure}

The density interpretation also guides us in our qualitative understanding of the flavor ratio $f_{1,u}^\xmom{1}/f_{1,d}^\xmom{1}$. 
According to Eq. \eqref{eq-f1xmomdecomp}, we can decompose this ratio as
\begin{equation}
	\frac{f_{1,u}^\xmom{1}(\vprp{k}^2)}{f_{1,d}^\xmom{1}(\vprp{k}^2)} =
	\frac{\int_0^1 dx\, f_{1,u}(x,\vprp{k}^2) - \int_0^1 dx\, \bar{f}_{1,u}(x,\vprp{k}^2)}{\int_0^1 dx\, f_{1,d}(x,\vprp{k}^2) - \int_0^1 dx\, \bar{f}_{1,d}(x,\vprp{k}^2)}
\end{equation}
where, according to Eq. \eqref{eq-rhoUU}, each of the four terms on the right hand side has an interpretation as a $\vprp{k}$-dependent density of unpolarized quarks/antiquarks.
Integrating numerator and denominator individually with respect to $\vprp{k}$ yields the flavor ratio of valence quarks $n_u / n_d = 2$ in the proton. If $f_{1,u}^\xmom{1}(\vprp{k}^2)/f_{1,d}^\xmom{1}(\vprp{k}^2)$ were constant, we would thus expect to find a value of $2$.
Indeed, our result shown in Fig. \ref{fig-f1uof1d} is quite close to $2$. At low $|\vprp{k}|$, the ratio $f_{1,u}^\xmom{1}/f_{1,d}^\xmom{1}$ is slightly higher than 2, for large $|\vprp{k}|$ it drops below $2$. 
According to the equation above, the larger ratio at low $|\vprp{k}|$ could be attributed, for example, to an enhancement of the density of up-quarks $\int_0^1 dx\, f_{1,u}(\vprp{k}^2)$ at low $|\vprp{k}|$, to a depletion of up-antiquarks $\int_0^1 dx\, \bar{f}_{1,u}(\vprp{k}^2)$ at low $|\vprp{k}|$, or to converse effects with regard to the down-flavor densities in the denominator. 
The flavor ratio for $f_1^\xmom{1}+g_1^\xmom{1}$ shown in Fig. \ref{fig-f1pg1uof1pg1d} corresponds to the $x$-integral of $\rho_{LL}$ for $\lambda=\Lambda$, i.e. the density of quarks with the same helicity as the nucleon, minus an antiquark contribution of opposite helicity, see Eqns. \eqref{eq-rhoLL} and \eqref{eq-xintdensdecomp}. In this spin-polarized channel, we see a strong excess of the $x$-integrated up quark density as compared to the $x$-integrated down quark density. It is well known that up quarks tend to be aligned with the proton helicity, while down quarks exhibit the opposite behavior. It is therefore not surprising to find a flavor ratio larger than $2$ in this channel. However, it is interesting to observe that this effect occurs mainly at low transverse momentum, as suggested by the notable decline of the flavor ratio with $|\vprp{k}|$.
Since the Boer-Mulders function $h_{1}^\prp$ vanishes in the straight link case, the combination $f_1+h_1$ involving the transversity distribution corresponds to the density $\rho_{TT}$ when $\vprp{s}=\vprp{S}$ and $(\vprp{k} \tcdot \vprp{s})^2 = \vprp{k}^2/2$, i.e., on the lines where $\vprp{k}$ is at an angle of $45^\circ$ with the transverse spin vectors of proton and quark. The flavor ratio for this combination is displayed in Fig. \ref{fig-f1ph1uof1ph1d}, 
where we observe a similar but somewhat less pronounced effect compared to the longitudinally polarized case in Fig. \ref{fig-f1pg1uof1pg1d}.

\subsection{Combined \texorpdfstring{$x$-$\vprp{k}$}{x-kt}-moments of \TMDs and densities}
\label{sec-xktmom}

In the following, we denote the combined $x$-$\vprp{k}$-mo\-ments of \TMDs as
\begin{align}
	f_{1}^\xktmom{n}{m} & = \int dx\ x^{n-1} \int d^2 \vprp{k}\ \left(\frac{\vprp{k}^2}{2 m_N^2}\right)^m f_{1}(x,\vprp{k})\, ,
\end{align}
and analogously for the other \TMDs $g_1$, $g_{1T}$, $\ldots$ .

As has already been mentioned before, $\vprp{k}$-integrals of TMDs taken over the full range of $\vprp{k}$ 
are in general not well defined due to their asymptotic $\vprp{k}$-dependence.
Perturbative calculations show that, e.g., $f_1(x,\vprp{k}) \sim 1/\vprp{k}^2$ for large $\vprp{k}$, leading to a logarithmically divergent $\vprp{k}$-integral, see e.g., Ref. \cite{Bacchetta:2008xw}.
Correspondingly, in the continuum, the amplitude $2 \tAmp_2(\elll^2,\elll \tcdot P)$ is expected to diverge for $|\elll| \rightarrow 0$.
The required (systematic) regularization of these potential divergencies will in general introduce a dependence on a regularization scheme and parameter, e.g. a UV cut-off scale $\lambda$.
Here, we follow a simpler, more practical approach and employ the Gaussian parametrizations of the amplitudes as discussed in Section \ref{sec-Gauss},
which allowed us to perform the necessary extrapolation in $|\elll|$ to $|\elll| = 0$,
and which in turn lead to Gaussian (i.e. exponential) fall-offs of the \TMDs as $\vprp{k} \rightarrow \infty$.
With this provisional Gaussian regularization in mind, we can now define 
a number of ratios of $\vprp{k}$-moments of \TMDs and densities that have clear and interesting physical interpretations:
\begin{align}
	\frac{g_A}{g_V} & = \frac{g_{1}^\xktmom{1}{0}}{f_1^\xktmom{1}{0}}%\nonumber\\
	 %& = \frac{ \int d^2\vprp{k}\  \sum_{\lambda=\pm 1} \lambda\, \rho_{LL}^\xmom{1}(\vprp{k},\lambda,\Lambda{=}1)  }{  \int d^2\vprp{k}\ \rho_{UU}^\xmom{1}(\vprp{k}) } \nonumber\\ &
	  \mathop{=}^{\text{sW}} \frac{-\tAmp_6(0,0)}{\tAmp_2(0,0)} \, , \label{eq-gA} \\
	\frac{g_T}{g_V} & = \frac{h_{1}^\xktmom{1}{0}}{f_1^\xktmom{1}{0}} %\nonumber \\ 
	%& = \frac{ \int d^2\vprp{k}\  \sum_{\mu=\pm 1} \mu\,\rho_{TT}^\xmom{1}(\vprp{k},\vprp{S}{=}\mu \vprp{s}{=}(1,0)) }{  \int d^2\vprp{k}\ \rho_{UU}^\xmom{1}(\vprp{k}) }  \nonumber\\ &
	 \mathop{=}^{\text{sW}} \frac{-\tAmp_{9m}(0,0)}{\tAmp_2(0,0)}\, , \label{eq-gT}
\end{align}
%\todo{Using integrals of densities to define these basic quantities is maybe a bit of an overkill; Maybe just leave the ratio of
%moments of the TMDs?}
giving the well-known axial vector and tensor charges, respectively, and
\begin{align}
	\langle \vect{k}_x \rangle_{TL} & \equiv \frac{ \int d^2\vprp{k}\ \vect{k}_x\ \rho_{TL}^\xmom{1}(\vprp{k},\lambda{=}1,\vprp{S}{=}(1,0)) }{  \int d^2\vprp{k}\  \rho_{TL}^\xmom{1}(\vprp{k},\lambda{=}1,\vprp{S}{=}(1,0)) } \nonumber \\
	& = m_N \frac{g_{1T}^\xktmom{1}{1}}{f_1^\xktmom{1}{0}} \mathop{=}^{\text{sW}} -m_N \frac{\tAmp_7(0,0)}{\tAmp_2(0,0)}\, ,  \label{eq-shiftTL} \displaybreak[0] \\
	\langle \vect{k}_x \rangle_{LT} & \equiv \frac{ \int d^2\vprp{k}\ \vect{k}_x\ \rho_{LT}^\xmom{1}(\vprp{k},\vprp{s}{=}(1,0),\Lambda{=}1) }{  \int d^2\vprp{k}\ \rho_{LT}^\xmom{1}(\vprp{k},\vprp{s}{=}(1,0),\Lambda{=}1) } \nonumber \\
	& = m_N \frac{h_{1L}^{\prp\xktmom{1}{1}}}{f_1^\xktmom{1}{0}} \mathop{=}^\text{sW} m_N \frac{-\tAmp_{10}(0,0)}{\tAmp_2(0,0)} \, ,\label{eq-shiftLT} \displaybreak[0] \\ 
	\langle \vect{k}_y \rangle_{TU} & \equiv \frac{ \int d^2\vprp{k}\ \vect{k}_y\ \rho_{TU}^\xmom{1}(\vprp{k},\vprp{S}{=}(1,0)) }{  \int d^2\vprp{k}\ \rho_{TU}^\xmom{1}(\vprp{k},\vprp{S}{=}(1,0)) } \nonumber \\
	& = m_N \frac{f_{1T}^{\prp\xktmom{1}{1}}}{f_1^\xktmom{1}{0}} \, .\label{eq-shiftTU} 
\end{align}
The first two are the above mentioned transverse momentum shifts for longitudinally polarized quarks in a transversely polarized nucleon ($TL$)
and vice-versa. 
For later discussions, we have also introduced the transverse momentum shift perpendicular to the transverse nucleon spin for unpolarized quarks ($TU$), which is given by the Sivers function $f_{1T}^\prp$ and thus vanishes for straight gauge links.
We note that the quantities above can be expressed in terms of simple ratios of amplitudes, as shown in Eqns. \eqref{eq-gA} - \eqref{eq-shiftLT} for the case of straight Wilson lines (``sW'').
A noteworthy advantage of such \emph{ratios of amplitudes} compared to individual amplitudes is that
they in general need no renormalization with respect to the self-energy of the gauge link and the multiplicative renormalization factor $Z^{-1}_{\Psi,z}$ in Eq.~\eqref{eq-opren}, i.e., 
\begin{equation}
  \frac{\tAmp_i(\elll^2,\ldots)}{\tAmp_j(\elll^2,\ldots)} = 
  \frac{\tAmp_i^\unren(\elll^2,\ldots)}{\tAmp_j^\unren(\elll^2,\ldots)} \, ,
\end{equation}
due to cancellations of the factors in the numerator and denominator.
We have to keep in mind, however, that we do not evaluate the amplitudes directly at small $|\elll| <0.25\units{fm}$, but rather use
the Gaussian parametrizations to perform an extrapolation to $|\elll| = 0$ .
Therefore, our results can have a residual dependence on $\delta m$, and thus on the employed renormalization condition, i.e. $C^{\text{ren}}=0$.
Numerically, it turns out that this dependence is weak.
It is important to note that apart from $g^{u-d}_A$, the tensor charge $g_T$ as well as the 
transverse momentum shifts are generically scale and scheme dependent quantities, due to
to the required regularization of the potential singularities at very short distances, i.e. 
the renormalization properties of the underlying local operators.
At this point, we are unfortunately not able to relate our simple Gaussian regularization 
to a standard scheme like the $\MSbar$-scheme at a certain scale $\mu$.
This most likely requires a detailed theoretical understanding of the behavior of the lattice 
amplitudes at small $|\elll|$, which may be obtained for example using lattice perturbation theory.
We plan to address this issue in future works.

The numerical values for the observables given in Eqns.~\eqref{eq-gA} to \eqref{eq-shiftLT} are listed in Table \ref{tab-xktmoms} for different flavor combinations.
\begin{table}
  \centering
  \renewcommand{\arraystretch}{1.2}  
  \begin{tabular}{|l|c|rl|}
    \hline
    observable & flavor & \multicolumn{2}{c|}{ value } \\
    \hline \hline
    % insert Mathematica results here
$ g_A/g_V $ & u &
  $\phantom{-}0.450 \pm 0.018 \pm 0.008$ & \\
$ g_A/g_V\ [\tAmp_{2{\pm}6}] $ & u &
  $\phantom{-}0.442 \pm 0.017 \pm 0.002$ & \\
$ g_A/g_V $ & d &
  $-0.282 \pm 0.018 \pm 0.004$ & \\
$ g_A/g_V\ [\tAmp_{2{\pm}6}] $ & d &
  $-0.282 \pm 0.018 \pm 0.001$ & \\
$ g_A/g_V $ & u-d &
  $\phantom{-}1.192 \pm 0.037 \pm 0.019$ & \\
$ g_A/g_V\ [\tAmp_{2{\pm}6}] $ & u-d &
  $\phantom{-}1.254 \pm 0.036 \pm 0.005$ & \\
$ g_T/g_V $ & u &
  $\phantom{-}0.451 \pm 0.016 \pm 0.003$ & \\
$ g_T/g_V\ [\tAmp_{2{\pm}\text{9m}}] $ & u &
  $\phantom{-}0.453 \pm 0.016 \pm 0.001$ & \\
$ g_T/g_V $ & d &
  $-0.264 \pm 0.017 \pm 0.002$ & \\
$ g_T/g_V\ [\tAmp_{2{\pm}\text{9m}}] $ & d &
  $-0.262 \pm 0.017 \pm 0.001$ & \\
$ g_T/g_V $ & u-d &
  $\phantom{-}1.182 \pm 0.034 \pm 0.008$ & \\
$ g_T/g_V\ [\tAmp_{2{\pm}\text{9m}}] $ & u-d &
  $\phantom{-}1.201 \pm 0.034 \pm 0.002$ & \\
$ \langle \vect{k}_x \rangle_{TL} $ & u &
  $\phantom{-}69.7 \pm \phantom{0}\phantom{0}4.3 \pm \phantom{0}\phantom{0}1.4$ & MeV\\
$ \langle \vect{k}_x \rangle_{TL} $ & d &
  $-30.9 \pm \phantom{0}\phantom{0}5.1 \pm \phantom{0}\phantom{0}0.6$ & MeV\\
$ \langle \vect{k}_x \rangle_{TL} $ & u-d &
  $\phantom{-}172.8 \pm \phantom{0}\phantom{0}8.5 \pm \phantom{0}\phantom{0}3.3$ & MeV\\
$ \langle \vect{k}_x \rangle_{LT} $ & u &
  $-59.1 \pm \phantom{0}\phantom{0}3.5 \pm \phantom{0}\phantom{0}1.4$ & MeV\\
$ \langle \vect{k}_x \rangle_{LT} $ & d &
  $\phantom{-}18.3 \pm \phantom{0}\phantom{0}4.1 \pm \phantom{0}\phantom{0}0.4$ & MeV\\
$ \langle \vect{k}_x \rangle_{LT} $ & u-d &
  $-138.5 \pm \phantom{0}\phantom{0}7.4 \pm \phantom{0}\phantom{0}3.2$ & MeV\\  
  % end of Mathematica results
  \hline 
  \end{tabular}
  \renewcommand{\arraystretch}{1}
  \caption{Numerical results for $x$-$\vprp{k}$-moments of \TMDs obtained using the Gaussian amplitudes at a pion mass $m_\pi\approx 500 \units{MeV}$. 
  We also include results corresponding to an alternative Gaussian parametrization based on linear combinations of amplitudes, as indicated in square brackets, see section \ref{sec-A2A6}.
  The first error is statistical.  The second error includes the statistical uncertainty in $\delta m$ and an estimate of discretization uncertainties, as given in Eq.~\eqref{eq-deltamerr}. 
  The values for $u-d$-quarks have been obtained directly from Gaussian fits to the $u-d$ data. Note that we have performed the conversion to physical units using the values for the lattice spacing $a$ given in Table \ref{tab-gaugeconfs}, see also footnote \thefnnumber.  \label{tab-xktmoms}}
\end{table}
We note that the value we obtain for the isovector axial vector coupling $g_A^{u-d} = g^{u-d}_A/g^{u-d}_V=1.192 \pm 0.037 \pm 0.019$ agrees within statistics with the value $1.173\pm 0.029$ of Ref. \cite{Edwards:2005ym}, obtained using conventional, 
local operators on the same ensemble, and is
also reasonably close to the experimental result $g_A^{u-d} = 1.2694(28)$ \cite{Nakamura:2010zzi}.
Our result for the isovector tensor charge $g_T^{u-d} = g^{u-d}_T/g^{u-d}_V=1.182 \pm 0.034 \pm 0.008$
turns out to be $\approx10\%$ larger than the value $g_T^{u-d}\simeq1.06\pm 0.02$ from Ref.~\cite{Edwards:2006qx}
obtained for the same ensemble using local operators\footnote{in the $\MSbar$-scheme at $\mu^2=4\text{GeV}^2$}.
This may be related to the fact that the Gaussian parametrization of the corresponding amplitude
$\tAmp_{9m}$ in Fig.~\ref{fig-amps2} in fact overshoots the lattice data points at small values of  $|\elll| \sim 0.25\text{fm}$
by $\approx7-10\%$, in contrast to the case of the amplitude $\tAmp_{6}$ in Fig.~\ref{fig-amps} that gives $g_A$.
A more sophisticated parametrization of the $|\elll|$-dependency of the lattice data for the amplitudes could help
to resolve this issue.
In any case, we interpret the outcome of these comparisons as a first non-trivial, successful consistency check of our method.

As we have already discussed in Ref. \cite{Hagler:2009mb}, the average transverse momentum shifts, $\langle \vect{k}_x \rangle_{TL}$ and $\langle \vect{k}_x \rangle_{LT}$ (cf. Table \ref{tab-xktmoms}) turn out to be sizeable and of opposite sign for up- and for down-quarks.
Moreover, as has been observed in Ref. \cite{Pasquini:2009eb}, our values are 
quite similar to the results from a light-cone constituent quark model calculation \cite{Pasquini:2008ax}.
This is remarkable, not only because the quark masses employed in the lattice calculation are still unphysically large,
but also because possible dependencies on the UV-cutoff scale have neither been investigated by us nor in the model calculation.
As discussed earlier, these dependencies may be weak in particular 
for quantities like $\langle \vect{k}_x \rangle_{TL}$ and $\langle \vect{k}_x \rangle_{LT}$ that can be expressed as ratios of amplitudes.
It is also interesting to note that the gauge link and its geometry do not enter explicitly in the calculation of time-reversal even \TMDs within the aforementioned constituent quark model.

Finally, we note that as an alternative to the Gaussian approach, it is conceivable to regularize the quantities defined in Eqns. \eqref{eq-gA}-\eqref{eq-shiftLT} by evaluating the ratio at a small but nonzero $|\elll|$:
\begin{equation}
  \left[ \frac{\tAmp_i(0,0)}{\tAmp_j(0,0)} \right]^\text{reg} \equiv
  \frac{\tAmp_i(\elll_\mmin^2,0)}{\tAmp_j(\elll_\mmin^2,0)} \ .
  \label{eq-regratio}
\end{equation}
For a direct calculation on the lattice, $|\elll_\mmin|$ would have to be chosen large enough compared to the lattice spacing $a$ to avoid significant discretization errors.

\subsection{Parametrization dependence using the Gaussian prescription}
\label{sec-A2A6}

\newcommand{\prmtr}{parametrization\xspace}

The simple Gaussian ansatz for the $\vprp{k}$-dependence of TMDs is very successful at parametrizing experimental data \cite{Schweitzer:2010tt,D'Alesio:2004up,Anselmino:2005nn,Collins:2005ie}.
It also describes the $\elll^2$-dependence of our lattice data for the invariant amplitudes at $\elll \tcdot P=0$ 
quite well and enables us to perform the Fourier-transform to obtain $x$-moments of \TMDs in a simple way.
However, this ansatz clearly introduces additional \prmtr uncertainties. 

\begin{figure}[tb]
	\centering%
	\includegraphics[width=\linewidth]{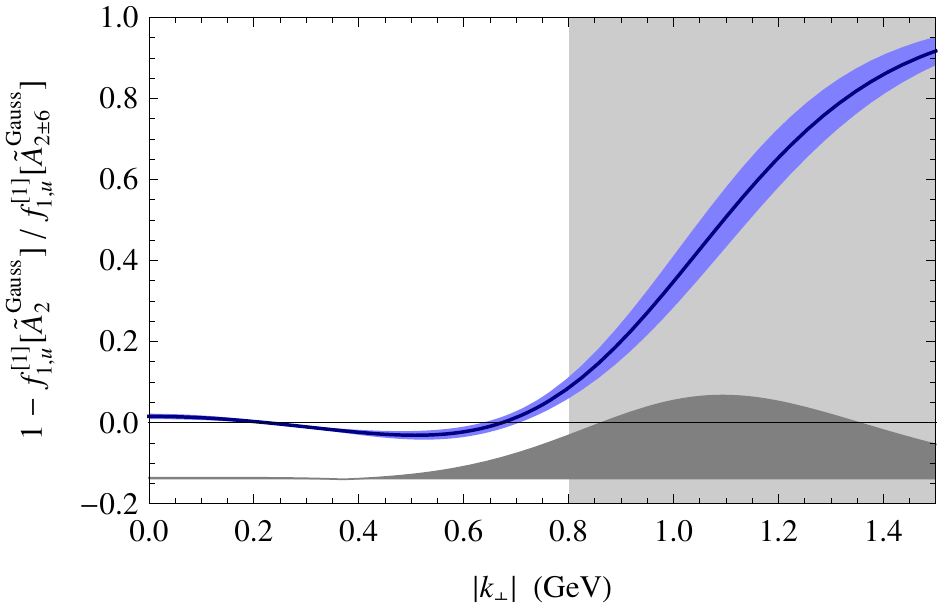}
	\caption{The solid curve and error band in blue give the relative difference between two different parametrizations of $f_1^\xmom{1}$ for up-quarks at a pion mass $m_\pi \approx 500\units{MeV}$.  The gray band at the bottom indicates uncertainties that can effectively be expressed as an error in $\delta m$. The gray region at large $|\vprp{k}|$ indicates the scale where we qualitatively expect strong parametrization uncertainties to set in.}
	\label{fig-f1-compare}
\end{figure}

In the following case study of \prmtr uncertainties we compare two different ways to use Gaussians for the parametrization of our data.
Consider $x$-integrated densities of longitudinally polarized quarks in the longitudinally polarized nucleon
\begin{align}
	\rho^{\pm\xmom{1}}(\vprp{k}^2) & \equiv \rho_{LL}^\xmom{1}(\vprp{k};\lambda{=}{\pm}1,\Lambda{=}{+}1) \nonumber \\ & = \frac{1}{2}\left( f_1^\xmom{1}(\vprp{k}^2) \pm g_1^\xmom{1}(\vprp{k}^2) \right) \nonumber \\
	& = \xfourint \left( \tAmp_2 \mp \tAmp_6 \right) \equiv \xfourint \tAmp_{2 \mp 6} \ .
	\label{eq-combinedA2A6}
\end{align}
In the previous sections, we have discussed individual Gaussian fits 
to $\tAmp_2$ and $\tAmp_6$. This translates into a Gaussian parametrization of $f_1^\xmom{1}$ and $g_1^\xmom{1}$ with the help of Eq. \eqref{eq-tmdsfromamps}. Let us label the corresponding results $f_1^\xmom{1}[\tAmp_2^\text{Gauss}]$, etc. An alternative is to fit Gaussians to each of the combined amplitudes $\tAmp_{2\pm 6} \equiv \tAmp_2 \pm \tAmp_6$. This translates directly into a Gaussian parametrization of $\rho^{\pm\xmom{1}}$, while $f_1^\xmom{1}$ and $g_1^\xmom{1}$ now need to be expressed as linear combinations of Gaussians. Specifically, we obtain
\begin{align}
	& f_1^\xmom{1}[\tAmp_{2\pm 6}^\text{Gauss}](\vprp{k}^2) = \frac{1}{2}\left( \rho^{+\xmom{1}}(\vprp{k}^2) + \rho^{-\xmom{1}}(\vprp{k}^2) \right) \nonumber \\
	= & \frac{1}{2} \left( \frac{c_{2-6} \sigma_{2-6}^2 }{4\pi} e^{-\frac{|\vprp{k}^2|}{(2/\sigma_{2-6})^2}} + \frac{c_{2+6} \sigma_{2+6}^2 }{4\pi} e^{-\frac{-|\vprp{k}^2}{(2/\sigma_{2+6})^2}} \right)\, .
\end{align}
Note that a single Gaussian function does not change sign. Therefore, the alternative parametrization in terms of $\tAmp_{2\pm6}$ is 
in this sense physically 
better motivated, 
since the quantities $\rho^{\pm\xmom{1}}(\vprp{k}^2)$ have an interpretation as densities of longitudinally polarized quarks, and should be positive \cite{Bacchetta:1999kz}, as long as we ignore the (small) contribution from anti-quarks, cf. section \ref{sec-densities}. 
The Gaussian fits to data for $\tAmp_{2+6}$ and $\tAmp_{2-6}$ are of similar quality as those for $\tAmp_2$ and $\tAmp_6$. In Figure \ref{fig-f1-compare}, we plot for $f_1^\xmom{1}$ the relative difference between the two parametrizations, namely $1-f_1^\xmom{1}[\tAmp_{2}^\text{Gauss}]/f_1^\xmom{1}[\tAmp_{2\pm 6}^\text{Gauss}]$, as a function of $|\vprp{k}|$. The difference between the two parametrizations 
stays below $5\%$ for $|\vprp{k}|\lesssim 0.7\units{GeV}$ , then it rises to an asymptotic value of $100\%$ at large $|\vprp{k}|$. This picture is compatible with our qualitative expectations of large parametrization dependence beyond $|\vprp{k}|\gtrsim 1/0.25 \units{fm} \approx 0.8\units{GeV}$.

\begin{figure}[tb]
	\centering%
	\includegraphics[width=\linewidth]{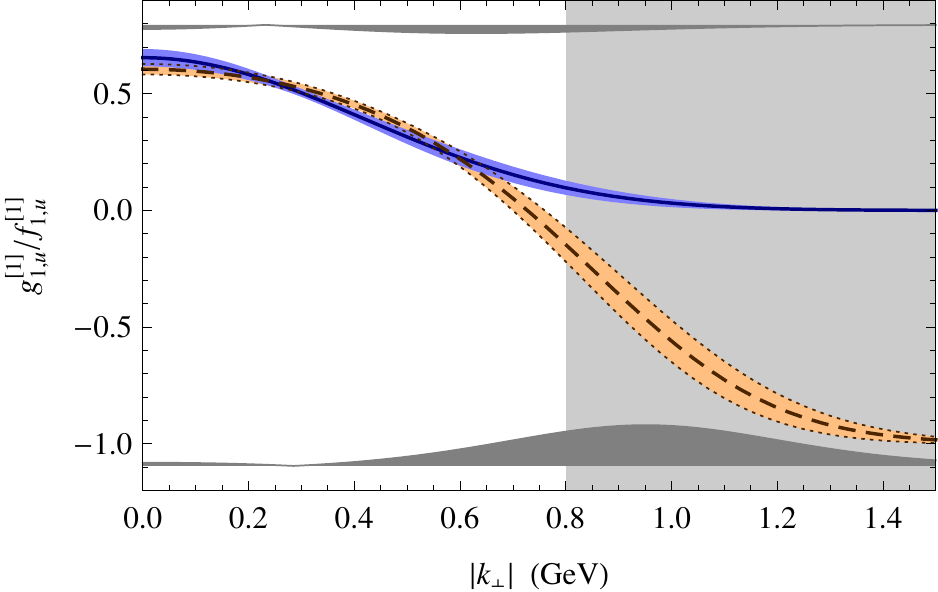}
	\caption{$g_1^\xmom{1}(\vprp{k}^2)/f_1^\xmom{1}(\vprp{k}^2)$ for up-quarks obtained at a pion mass $m_\pi\approx 500 \units{MeV}$ from two different parametrizations. The solid curve, the statistical error band in blue and the error associated with $\Delta[\delta m]$ shown in gray at the top correspond to $g_1^\xmom{1}[\tAmp_6^\text{Gauss}]/f_1^\xmom{1}[\tAmp_2^\text{Gauss}]$, while the dashed curve, the error band outlined by the dotted curves and the gray  error band at the bottom correspond to $g_1^\xmom{1}[\tAmp_{2\pm6}^\text{Gauss}]/f_1^\xmom{1}[\tAmp_{2\pm6}^\text{Gauss}]$. The gray region at large $|\vprp{k}|$ indicates that we qualitatively expect strong parametrization uncertainties beyond $|\vprp{k}|\gtrsim 1/0.25 \units{fm} \approx 0.8\units{GeV}$.}
	\label{fig-g1overf1}
\end{figure}

Let us now study the ratio
\begin{equation}
	\frac{g_1^\xmom{1}(\vprp{k}^2)}{f_1^\xmom{1}(\vprp{k}^2)} = \frac{\rho^{+\xmom{1}}(\vprp{k}^2) - \rho^{-\xmom{1}}(\vprp{k}^2)}{\rho^{+\xmom{1}}(\vprp{k}^2) + \rho^{-\xmom{1}}(\vprp{k}^2)}
	\label{eq-g1of1}
\end{equation}
as a function of $|\vprp{k}|$. In this quantity, both numerator and denominator become very small at large $|\vprp{k}|$. We plot the result in Figure \ref{fig-g1overf1}, again comparing the two alternative parametrizations. The two results 
are in agreement 
for $|\vprp{k}|\lesssim 0.6\units{GeV}$, at large $|\vprp{k}|$ they deviate strongly. Asymptotically, the curve that corresponds to Gaussian $g_1^\xmom{1}$ and $f_1^\xmom{1}$ tends to zero, because $g_1^\xmom{1}$ has a smaller width. The parametrization does not allow a sign change of $g_1^\xmom{1}/f_1^\xmom{1}$. On the other hand, the result obtained with Gaussian $\rho^{+\xmom{1}}$ and $\rho^{-\xmom{1}}$ exhibits a sign change, and tends to $-1$, because the Gaussian describing $\rho^{+\xmom{1}}$ has a smaller width, so that $\rho^{-\xmom{1}}$ ultimately dominates on the right hand side of Eq. \eqref{eq-g1of1}. It is important to point out that the strong disagreement between the two results at large $|\vprp{k}|$ is an unavoidable consequence of the form of the parametrizations, but \emph{does not} point towards any inconsistencies 
of the lattice data. 
In this respect, we would like to stress that the same type of \prmtr uncertainty will at least in principle 
also affect phenomenological \TMD parametrizations
based on experimental data, which are to this date employing mostly Gaussian ansaetze for the $\vprp{k}$-dependence.
In summary, we see evidence that the relative \prmtr uncertainty of the Gaussian ansatz becomes very large at large $|\vprp{k}|$. 
It appears likely that a better, QCD-motivated parametrization of the amplitudes at small $|\elll|$ can improve the situation. 

In Figs. \ref{fig-flavorratios} and \ref{fig-f1pg1uof1pg1d} of the previous section, 
we have always included the result obtained with the alternative parametrization based on $\tAmp_{2\pm6}^\text{Gauss}$.
For $f_1^\xmom{1} \pm h_1^\xmom{1}$, we can introduce an alternative parametrization in analogy to Eq. \eqref{eq-combinedA2A6} based on Gaussian fits to linear combinations $\tAmp_{2\pm\text{9m}} \equiv \tAmp_{2} \pm \tAmp_{\text{9m}}$.
As before, this ansatz seems to be physically better motivated, since the linear combinations 
correspond to (approximately positive definite) densities as discussed at the end of section \ref{sec-densities}.
The two types of parametrizations $\tAmp_2^\text{Gauss}$, $\tAmp_\text{9m}^\text{Gauss}$ vs. $\tAmp_{2\pm\text{9m}}^\text{Gauss}$ are compared in Fig. \ref{fig-f1ph1uof1ph1d}.
In general, we observe a rather small difference between 
them in the range $0 \leq |\vprp{k}| \lesssim 0.7\units{GeV}$.
We also include results for the alternative parametrizations in Tables \ref{tab-gausstmds} and \ref{tab-xktmoms}. For $g_A/g_V$ and $g_T/g_V$, we find differences between the parametrizations that are in general of the order of the statistical errors. For the fits to $u-d$ data, these differences turn out to be larger than for the fits to $u$ and $d$ data.
 
\section{Testing correlations in \texorpdfstring{$x$ and $\vprp{k}$}{x and kt} }
\label{sec-fac}

What can we learn from the combined ($\elll \tcdot P$,$\elll^2$)-de\-pen\-dence of our amplitudes $\tAmp_i(\elll^2,\elll \tcdot P)$
without taking recourse to parametrizations and models?
A highly interesting question is 
if our lattice results for, e.g., $\tAmp_2(\elll^2,\elll \tcdot P)$ (at least approximately) ``factorize'',
\begin{equation}
	\tAmp_2(\elll^2,\elll \tcdot P)\ \mathop{\approx}^{\displaystyle \text{?}}\ \tAmp_2(\elll^2,0)\ \hat{A}_2(\elll \tcdot P)\,, 
	\label{eq-ampfac}
\end{equation}
or in contrast 
show a distinct correlation in $\elll \tcdot P$ and $\elll^2$.
This is directly related to a corresponding possible factorization of the $x$- and $\vprp{k}$-dependences of the TMDs, e.g.
\begin{equation}
	f_1(x,\vprp{k})  \mathop{\approx}^{\displaystyle \text{?}} f_1(x)\ f_1^\xmom{1}(\vprp{k}^2)\ /\ \mathcal{N} \,,
	\label{eq-tmdfac}
\end{equation}
where $\mathcal{N} = \int d^2 \vprp{k} f_1^\xmom{1}(\vprp{k}^2)$ is a normalization factor. 
Model ansaetze based on this assumption are commonly employed in phenomenological applications, typically in combination with the Gaussian parametrization of the $\vprp{k}^2$-dependent part, $f_1^\xmom{1}(\vprp{k}^2)\ /\ \mathcal{N} = \exp( - \vprp{k}^2 / \mu^2 )/\pi\mu^2$. This approach has been used to parametrize experimental data of semi-inclusive scattering experiments, see, e.g. Refs. \cite{D'Alesio:2004up,Anselmino:2005nn}, and to include effects of intrinsic (``primordial'') parton momentum in Monte Carlo event generators, e.g., in PYTHIA and HERWIG++ \cite{Sjostrand:2006za,Bahr:2008pv,Chekanov:2001aq}. 
Factorization in $x$ and $\vprp{k}^2$ is a simplifying assumption lacking 
fundamental theoretical justification. 
Arguments against the validity of this assumption have been found in model calculations, e.g., in a chiral quark soliton model \cite{Wakamatsu:2009fn} and in a diquark spectator model \cite{Bacchetta:2008af}, see our discussion below.   

If one of the Equations \eqref{eq-ampfac} or \eqref{eq-tmdfac} were to hold exactly, it would imply the other one (assuming well-behaved functions and integrals). 
This can be easily seen from Eq. \eqref{eq-fourint}, which consists of two independent Fourier integrals, establishing correspondences $\elll^2 \leftrightarrow \vprp{k}^2$ and $\elll \tcdot P \leftrightarrow x$. The $(\elll^2,\elll \tcdot P)$-factorization thus translates into $(x,\vprp{k}^2)$-factorization of the Fourier-transformed amplitude, and with the help of equation \eqref{eq-tmdsfromamps}, this directly implies $(x,\vprp{k}^2)$-factorization of $f_1$.

Analogous arguments connect hypothetical $(x,\vprp{k}^2)$-factorization of other \TMDs in Eq. \eqref{eq-tmdsfromamps} with $(\elll^2,\elll \tcdot P)$-factorization of corresponding amplitudes $\tAmp_i$.\footnote{For \TMDs given in terms of several amplitudes, the latter would have to fulfill additional relations among each other.} 

As a first conclusion we note that $(x,\vprp{k}^2)$-factorization is obviously not in conflict with Lorentz-invariance per se, since the parametrization in terms of amplitudes $\tAmp_i$ has been worked out in a manifestly Lorentz-covariant framework. We remark that a factorization assumption of the momentum-space amplitudes (as defined in, e.g., \cite{Mulders:1995dh}) of the type $A_i(k^2,k \tcdot P) = a_i(k^2) \hat{a}_i(k \tcdot P)$ is not equivalent to the above equations. As a specific example, the on-shell approximation $A_i(k^2,k \tcdot P) = \delta(k^2) \hat{a}_i(k \tcdot P)$ discussed in Ref. \cite{Ellis:1982cd} contradicts exact factorization of $f_1(x,\vprp{k}^2)$.

To study the possibility of a factorization as in Eq. \eqref{eq-ampfac} numerically, it is convenient to introduce a normalized amplitude
\begin{equation}
\label{eq-SR}
	\tAmp_i^\text{norm}(\elll^2,\elll \tcdot P) \equiv \frac{\tAmp_i(\elll^2,\elll \tcdot P)}{\tAmp_i(\elll^2,0)} \quad \mathop{\approx}^{\displaystyle ?} \quad \hat{A}_i(\elll \tcdot P)
\end{equation}
and to test whether it is independent of $\elll^2$. We point out that the quantity $\tAmp_i^\text{norm}(\elll^2,\elll \tcdot P)$ is renormalization scheme and scale independent for finite values of $\elll^2$, since both the self energy of the gauge link and the quark field renormalization factors of the respective operators cancel in the ratio. As previously, we discard data for very small quark separations, $|\elll| < 0.25\units{fm}$, to avoid possible lattice cutoff effects.
In the following, we work with the coarse\nbdash06 ensemble at $m_\pi \approx 600\units{MeV}$, where we have better statistics than on the coarse\nbdash04 ensemble due to the heavier quark mass and due to a larger number of gauge configurations. To reduce discretization errors, we use symmetry improved combinations of operators, as explained in appendix \ref{sec-symimpop}. The effect of this improvement turns out to be particularly important for the double ratios discussed below. Moreover, we make sure that the combination of link paths used in the numerator and the denominator of Eq. \eqref{eq-SR} are the same up to transformations under the hypercubic group H(4). This ensures that $\delta m$ is exactly the same for numerator and denominator; differences in $\delta m$ associated with the detailed pattern of the link path at the scale of the lattice spacing cancel in the ratio.

\begin{figure*}[p]
	\centering%
	\subfloat[][]{%
		\label{fig-A2normReUp}%
		\includegraphics[width=0.45\linewidth]{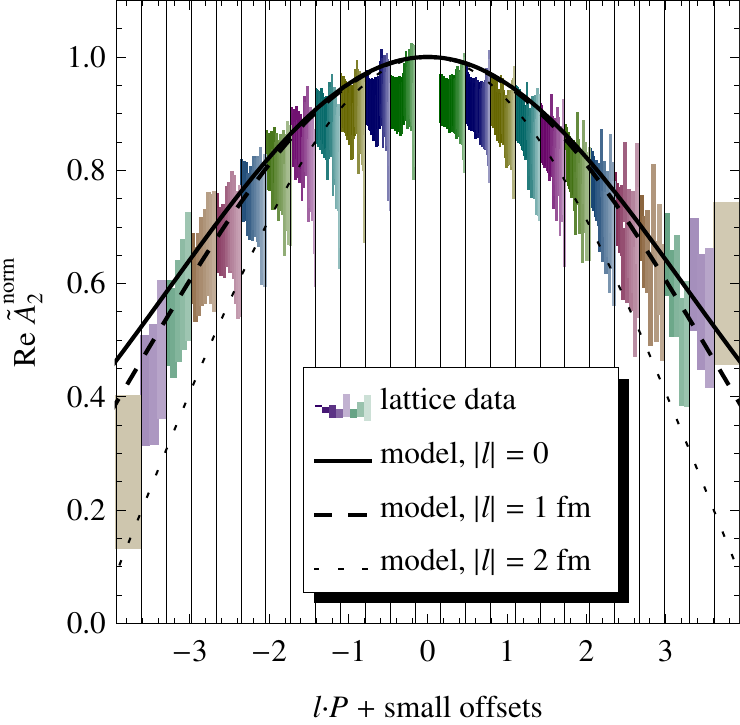}
		}\hfill%
	\subfloat[][]{%
		\label{fig-A2normImUp}%\
		\includegraphics[width=0.45\linewidth]{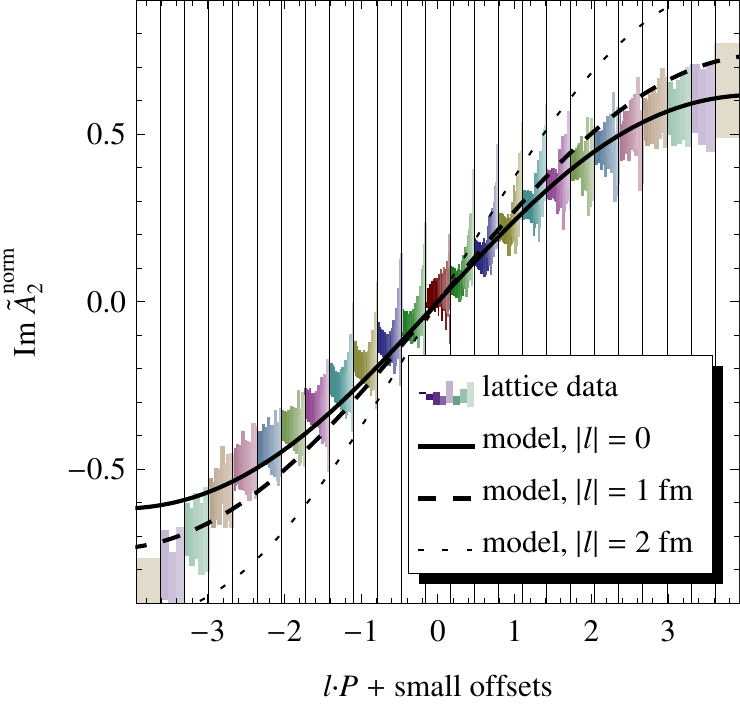}
		}\\[1em]%
	\subfloat[][]{%
		\label{fig-A2normReDown}%
		\includegraphics[width=0.45\linewidth]{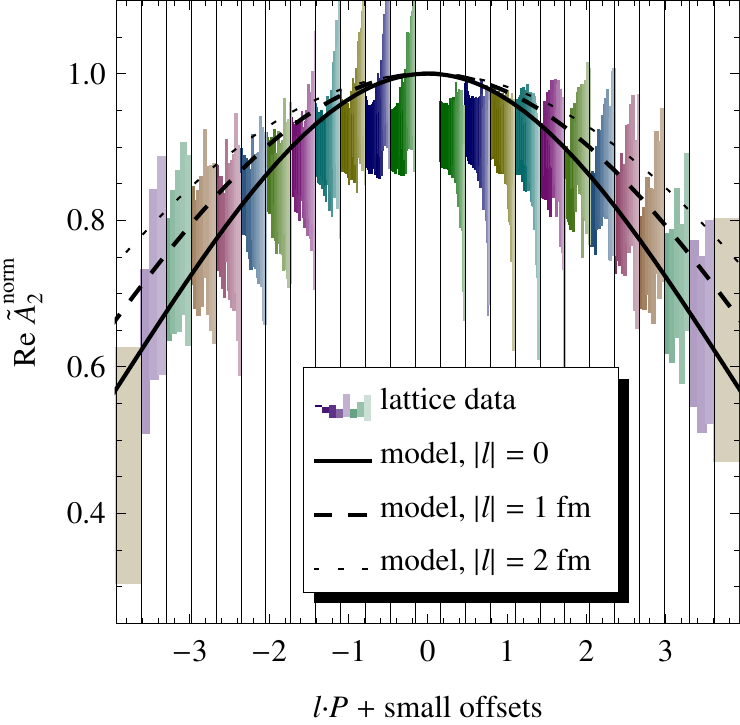}
		}\hfill%
	\subfloat[][]{%
		\label{fig-A2normImDown}%\
		\includegraphics[width=0.45\linewidth]{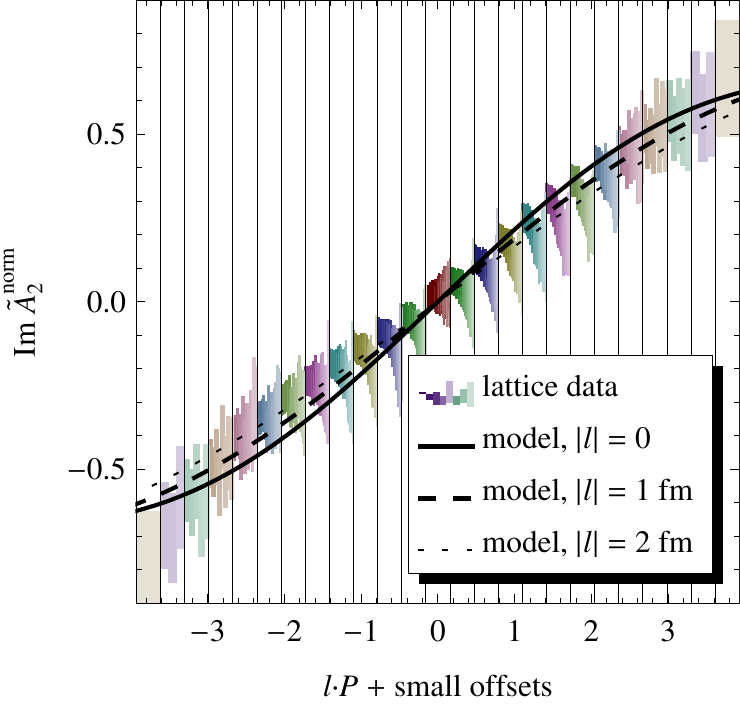}
		}%		
	\caption[fig-A2normUp]{%
		Lattice results for the normalized amplitude $\tAmp_2^\text{norm}(\elll^2,\elll \tcdot P)$, obtained from the coarse\nbdash06 ensemble ($m_\pi \approx  625 \units{MeV}$) with HYP-smeared gauge configurations. Each vertical stripe shows results at constant $\elll \tcdot P$, with values of $|\elll|$ ascending from left to right. The solid and dashed curves show $\tAmp_2^\text{norm}$ as a function of $\elll \tcdot P$ as obtained from a spectator diquark model \cite{Bacchetta:2008af} for several values of $|\elll|$. %
		\subref{fig-A2normReUp}\ up quarks, real part, \ \subref{fig-A2normImUp}\ up quarks, imaginary part, \subref{fig-A2normReDown}\ down quarks, real part, \ \subref{fig-A2normImDown}\ down quarks, imaginary part.  
		\label{fig-A2normUp}
		}
\end{figure*}

Figure \ref{fig-A2normUp} shows our lattice results for $\tAmp_2^\text{norm}(\elll^2,\elll \tcdot P)$. In each vertical stripe of the plots we show the data at constant values of $\elll \tcdot P$, which is dimensionless in natural units and can adopt values that are multiples of $2\pi a /L$ with our lattice method. In each stripe, we display the results for all available values of $|\elll|=\sqrt{-\elll^2}$ in the range $0.25\units{fm} \leq |\elll|  < 1.5\units{fm}$, with increasing $|\elll|$ from left to right. For larger $|\elll \tcdot P|$ only results at larger $|\elll|$ are available, due to the constraint Eq. \eqref{eq-lPdomain}. The data are displayed as filled rectangles representing the statistical error bounds, and are drawn with lighter colors for bigger errors. 

We find that the data is surprisingly constant within the stripes. Taking into account that the errors are correlated, no statistically significant non-trivial dependence on $\elll^2$ can be observed in the these plots. Such a dependence on $\elll^2$ would be in conflict with the factorization
displayed in Eq. \eqref{eq-ampfac} and Eq. \eqref{eq-tmdfac}. 

Together with the lattice data, we also display results from a diquark spectator model \cite{Jakob:1997wg, Bacchetta:2008af}, using formulae and parameters given in Ref. \cite{Bacchetta:2008af}.
To this end, we calculate $\tAmp_2^\text{norm}(\elll^2,\elll \tcdot P)$ by performing the inverse Fourier transform of the analytic model result for $f_1(x,\vprp{k}^2)$ numerically. 
The choice of the straight gauge link on the lattice might be a concern when comparing to models, 
however for time reversal even quantities the model calculations
so far do not explicitly include any gauge links.
Hence, it is difficult to tell at this moment if and how this affects the comparison.
We remark, however, that 
the lattice calculation has been performed at an unphysically large pion mass of about $600\units{MeV}$, and has not been extrapolated to the physical point so far. 
Nevertheless, we observe 
a close similarity of the model curves and the trend of lattice data. 
Interestingly, the model results for $\tAmp_2^\text{norm}$ as a function of $\elll \tcdot P$ lie relatively close together for $|\elll| = 0$ and $|\elll| = 1 \units{fm}$. 
This means that the model, when transformed to $(\elll^2,\elll \tcdot P)$-space, also exhibits an approximate 
compatibility with factorization of $\tAmp_2(\elll^2,\elll \tcdot P)$ as in  Eq. \eqref{eq-ampfac}, 
at least in the parameter range where lattice data is currently available.
For larger values of $|\elll|$, a possible deviation from the factorization may become more visible.

In order to see more concretely what we can learn in principle about the simultaneous dependence
of the lattice amplitudes on $(\elll^2,\elll \tcdot P)$ and possible "violations" of the approximate factorization,
it is advantageous to define a double ratio (of, e.g., the real parts of amplitudes) 
\begin{equation}
R_D(\elll^2,\elll \tcdot P;\elll^2_\text{min})\equiv \frac{\myRe\,\tAmp_i^\text{norm}(\elll^2,\elll \tcdot P)}{\myRe\,\tAmp_i^\text{norm}(\elll^2_\text{min},\elll \tcdot P) } \, ,
\label{eq-DR}
\end{equation}
where $\elll^2_\text{min}$ is the minimal value of $\elll^2$ that is available for a given $\elll \tcdot P$ and $\vect{P}$ in our calculation.
Clearly, the double ratio is strictly equal to unity in the case that the dependences on $\elll^2$ and $\elll \tcdot P$ factorize.
We may therefore use its variation from unity, $1-R_D$, 
as a quantitative measure of a potential "violation" of the naive multiplicative factorization displayed in Eq.~\eqref{eq-ampfac}.
Furthermore, a cancellation of systematic uncertainties and statistical fluctuations is even more likely in $R_D$ than in $\tAmp_i^\text{norm}$.

\begin{figure}
	\centering
	\includegraphics[width=\linewidth]{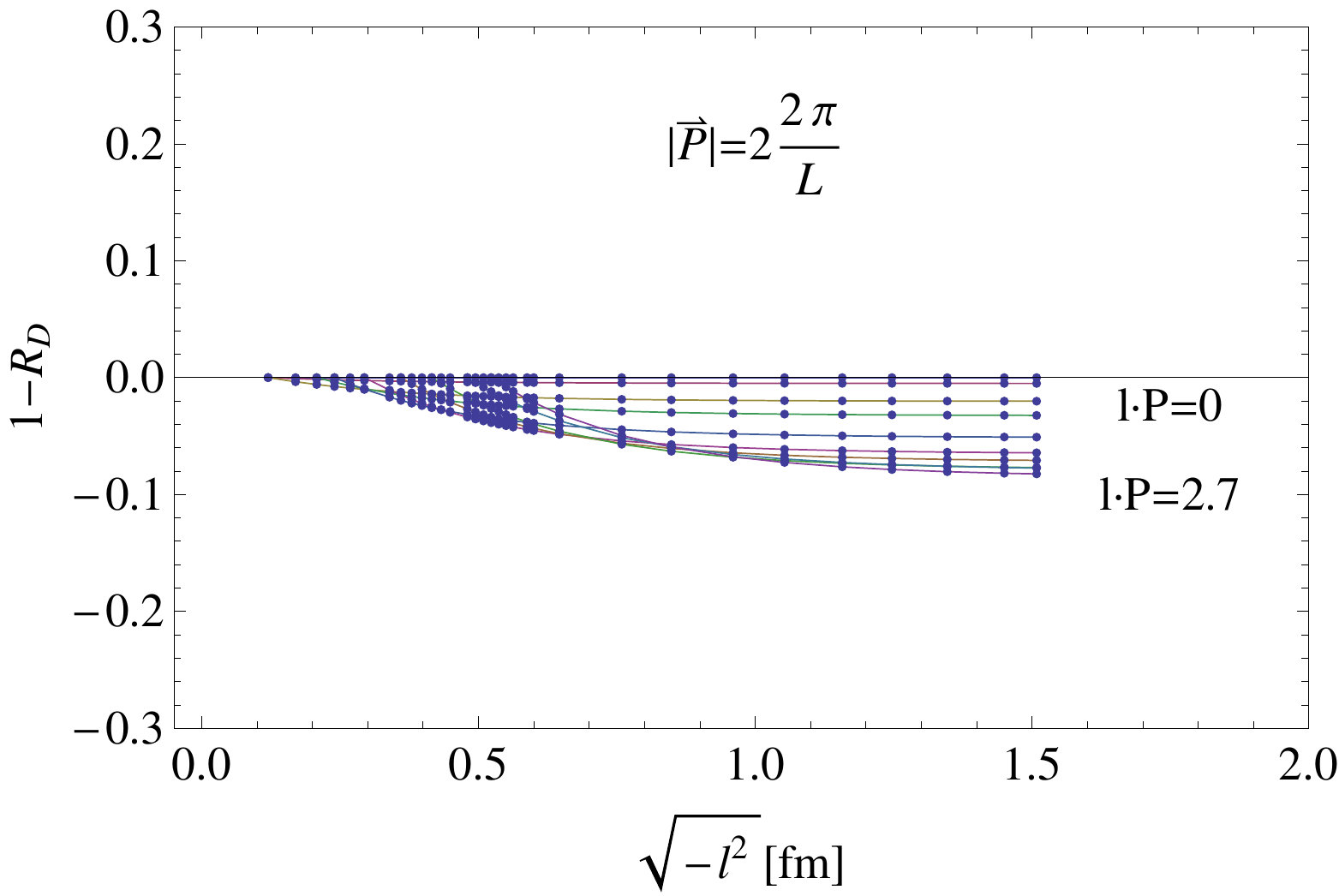}
	\caption{% 
	Unity minus the double ratio for an exponential ansatz for the $(\vprp{k}^2)$-dependence of the unpolarized TMD for up-quarks
	(employing the parametrization of the GPD $H(x,t)$ of Ref.~\cite{Diehl:2004cx}).\label{fig-DRExp}}
\end{figure}
\begin{figure}
	\centering
	\includegraphics[width=\linewidth]{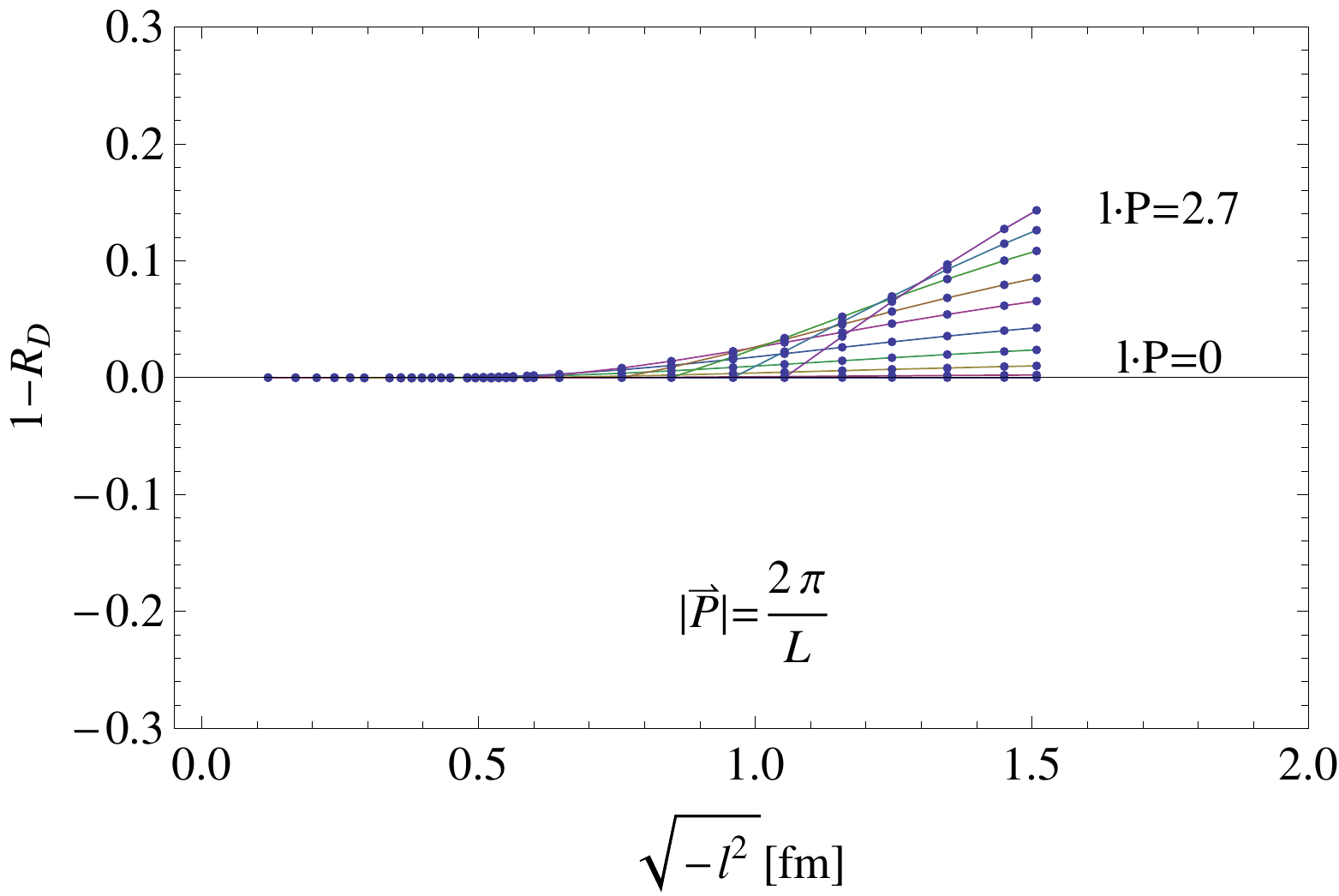}
	\caption{Unity minus the double ratio for the diquark spectator model calculation of $f_1(x,\vprp{k}^2)$
	for up-quarks \cite{Bacchetta:2008af}.\label{fig-DRModel}}
\end{figure}

To get an idea about what we might expect for the deviation of the double ratio from unity, we
show in Figs.~\ref{fig-DRExp} and \ref{fig-DRModel} $1-R_D$ as a function of $|\elll|$ for different
values of $\elll \tcdot P$, as obtained for two different model-ansaetze for the corresponding unpolarized TMD $f_1(x,\vprp{k}^2)$
for up-quarks in the proton.
For a comparison with the lattice results, we have, as before, (numerically) Fourier-transformed the model-ansaetze to 
$(\elll^2,\elll \tcdot P)$-space (neglecting sea quark contributions by setting $f_1(x<0,\vprp{k}^2)=0$),
and then constructed the double ratio mimicking the restrictions in our lattice calculation, i.e. setting $\vect{P}=2\pi/L(n,0,0)$,
employing typical lattice distance vectors $\elll$, and ensuring that $|\elll \tcdot P|\le \sqrt{-\elll^2}|P|$.

The curves in Fig.~\ref{fig-DRExp} are based on an exponential ansatz for the $(\vprp{k}^2)$-dependence and include correlations
of $x$ and $\vprp{k}^2$ in the form $\exp(-f(x)\vprp{k}^2)$. For definiteness, we have chosen the functional form and parameters
obtained in Ref.~\cite{Diehl:2004cx} for the parametrization of the GPD $H^u_v(x,t)$, where we have replaced the squared momentum transfer $t$ 
by $-\vprp{k}^2$. This exponential ansatz has the right properties in the framework of GPDs, but is unphysical in the case of TMDs,
and used here just for illustrational purposes, i.e., as an example for the type of correlations in $x$ and $\vprp{k}^2$ that 
would be surprising to see in our study. 
As can be seen from Fig.~\ref{fig-DRExp}, a non-trivial signature of the exponential (GPD-like) ansatz in $1-R_D$ 
shows up for $\vect{P}=2\pi/L(2,0,0)$ (for $\vect{P}=2\pi/L(1,0,0)$, $1-R_D$ is approximately zero in the accessible range
of variables), 
where one finds increasingly negative values at larger $|\elll \tcdot P|$  and $|\elll|$.

A distinctly different signature in $1-R_D$ is found for the TMD $f_1(x,\vprp{k}^2)$ for up-quarks 
from the diquark-spectator model calculation of Ref.~\cite{Bacchetta:2008af}.
In this case, comparatively strong deviations from the factorized case, i.e., $1-R_D = 0$,
are visible already for $|\vect{P}|=2\pi/L$, 
which are, however, positive and hence opposite in sign compared to the GPD-like ansatz displayed in Fig.~\ref{fig-DRExp}.
Interestingly, no such clear signature is visible for the corresponding down-quark distribution.
We suppose that this is directly related to the fact that the TMD $f_1(x,\vprp{k}^2)$ of Ref.~\cite{Bacchetta:2008af}
for up-quarks has a non-monotonic dependence on $\vprp{k}^2$ at low $x$, which in turn can be traced back to 
contributions of wave functions with non-zero relative orbital angular momentum $\Delta L_z=\pm1$.
Such contributions are absent in this model for $f_1(x,\vprp{k}^2)$ for down-quarks.

Without going into any details, we note that the TMDs obtained in the light-cone quark model calculation of Ref.~\cite{Pasquini:2008ax}
also do not factorize, but that at least $f_1(x,\vprp{k}^2)$ shows a less distinctive signature with respect to $1-R_D$ compared to the 
diquark-spectator model results discussed before. In particular, in the model of 
Ref.~\cite{Pasquini:2008ax}, there is no difference between up- and down-quark distributions regarding
correlations in $x$ and $\vprp{k}^2$.

\begin{figure*}[p]
	\centering
	\includegraphics[width=\linewidth]{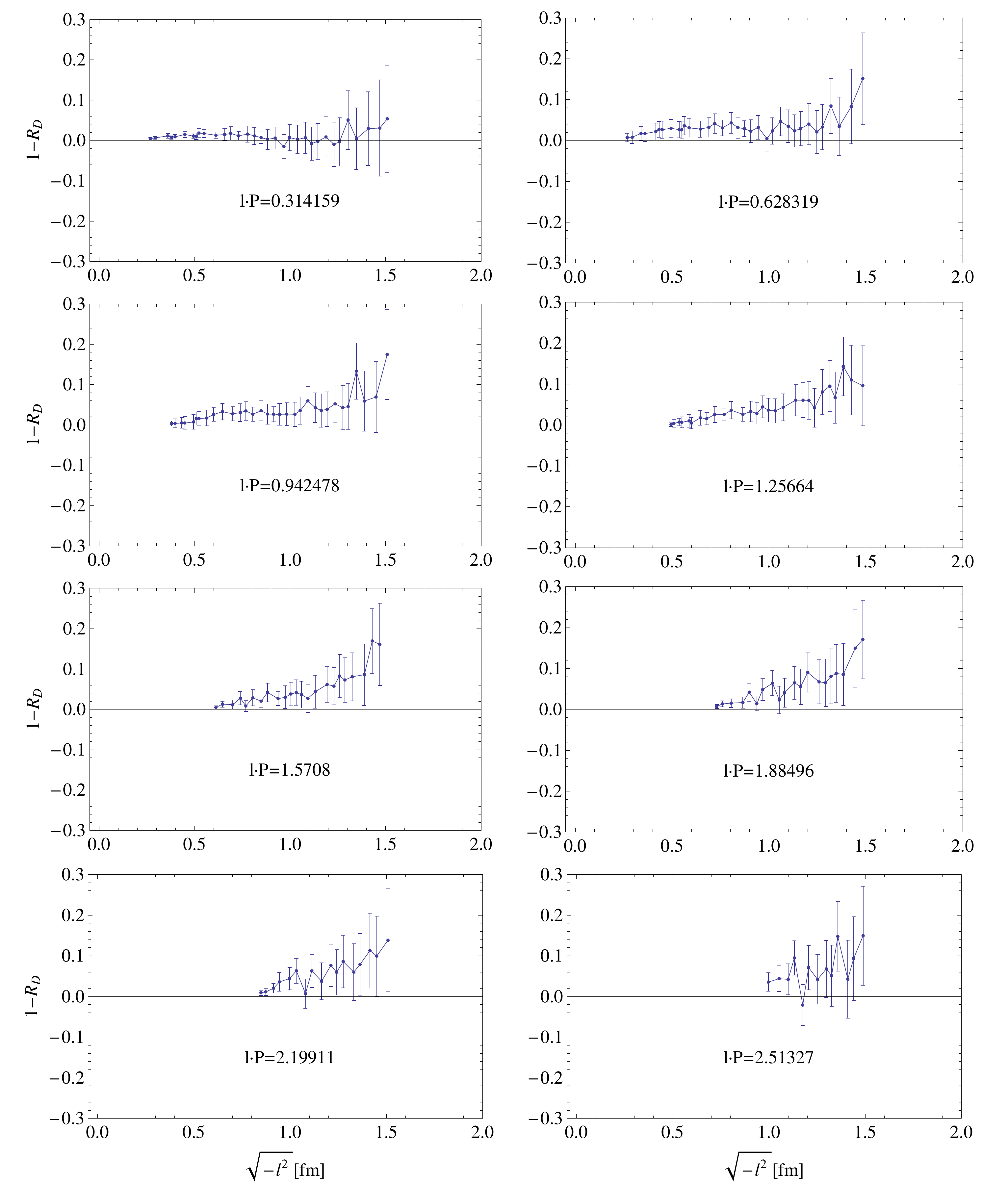}
	\caption{Lattice results for unity minus the double ratio for the real part of the
	amplitude $\tilde A_2$ for up-quarks, for a pion mass of $\approx625\mev$. 
	Note that the non-zero nucleon momentum is $\vect{P}=2\pi/L(-1,0,0)$. 
	\label{fig-DRlattice}}
\end{figure*}

Finally, Fig.~\ref{fig-DRlattice} displays the lattice results for the $|\elll|$-dependence of $1-R_D$ for eight different 
values of $|\elll \tcdot P|$ from $\pi/10$ to $8\pi/10$.
Using lattice data points for $|\elll|>0.2\fm$, 
we have constructed $R_D$ for all accessible values of $\elll^2$, $\elll \tcdot P$
and the corresponding $\elll^2_\text{min}$.
Within statistical uncertainties, we observe numerically the expected symmetry in $\elll \tcdot P\leftrightarrow -\elll \tcdot P$, i.e. 
$R_D(\elll^2,\elll \tcdot P)=R_D(\elll^2,-\elll \tcdot P)$, cf. Eq.~\eqref{eq-conjugated}.
For the final results, we average over positive and negative values to increase the statistics.

Interestingly, the central values of the lattice results for $1-R_D$ for up-quarks in Fig.~\ref{fig-DRlattice} show a
trend towards sizable, positive values for increasing $\elll^2$ at larger $\elll \tcdot P$, which is 
compatible with the results for the diquark spectator TMD model in Figs.~\ref{fig-DRModel}.
However, within the statistical uncertainties, the data points are also still mostly consistent with zero.
Therefore, at present we cannot rule out an at least approximate factorization of the $\elll \tcdot P$-, $\elll^2$-dependences
of the amplitudes, and the $x$-, $\vprp{k}^2$-dependences of the corresponding TMDs, respectively.
As a side remark, we note that corresponding lattice results for the down-quarks do not show any specific trend of the central
values at all.
It will be highly interesting to repeat this study with increased statistics and for larger nucleon momenta with, e.g.,
$|\vect{P}|=\sqrt{2}\times2\pi/L,2\times2\pi/L$, and to see if the 
trend of the central values, pointing towards a significant correlation in $x$ and $\vprp{k}^2$
as expected from certain TMD model calculations, can be firmly established or rejected.

\section{Outlook}

One of the most exciting challenges for lattice calculations of \TMDs is to go beyond the direct, straight gauge link between the quark fields. 
This is clearly necessary for an understanding of the physics of eikonal phases in processes that involve transverse momentum.
The long-term goal is to make contact with experimental measurements of semi-inclusive deep inelastic scattering and Drell-Yan production.
Since these experiments are very challenging, progress on the lattice in this direction would be even more important.
What are the principal limitations of such calculations? We will need to create a staple-like gauge link that resembles the one in Fig. \ref{fig-link-staple}, 
i.e., that generically runs in a direction $v$ along (or close to) the lightcone to infinity and back.
First of all, 
the extent of the staple in $v$-direction, given by the four-vector $\eta v$, 
will always be finite in any practical lattice calculation due to the finite lattice volume.
By increasing $\eta$ step by step, we may hope to find that the data converges to a plateau value, which we might interpret as representing the limit $\eta \rightarrow \infty$. The idea to define the matrix element through the limit $\eta \rightarrow \infty $ has already been mentioned in Ref. \cite{Collins:2008ht}. 
Furthermore, on the lattice, we are restricted to gauge link structures that have no temporal extent, $\elll^0 = v^0 = 0$. 
At a first glance, this might seem to imply that lattice calculations with ``realistic'' gauge links are impossible.
However, as in the case of straight gauge links, we need to establish the connection to \TMDs using a frame independent parametrization.  
As discussed in Ref. \cite{Goeke:2005hb} and appendix \ref{sec-symtraf}, with an additional $v$-dependence,  we now have to deal with 
32 independent invariant amplitudes, which can depend on the invariants $\elll^2$, $\elll \tcdot P$, $\eta v \tcdot \elll$, $(\eta v)^2$, and $\eta v \tcdot P$. 
The amplitudes defined in the limit $\eta \rightarrow \infty$ can only depend on $\eta$-independent combinations of these invariants.
The direction of $v$ relative to the nucleon momentum $P$ is essentially\footnote{The role of the sign of $\eta v \tcdot P$ is discussed in appendix \ref{sec-symtraf}.} given by $\zeta \equiv (2  v \tcdot P)^2 / v^2$, formed from $\eta v \tcdot P$ and $(\eta v)^2$. 
For finite $v \tcdot P$, the limit of a lightlike staple direction $v$ is characterized by $|\zeta| \rightarrow \infty$. 
On the other hand, inserting a spatial lattice vector $v$, we find that $\zeta$ is bounded by 
$0 \leq -\zeta \leq |2\vect{P}|^2$, where $\vect{P}$ is the three-momentum of the nucleon on the lattice. So although lightlike staple links cannot be realized directly on the lattice, the limit $|\zeta| \rightarrow \infty$ can still be approached at least in principle by choosing larger and larger lattice nucleon momenta. Importantly, and as already mentioned in the introduction, one approach to regularize rapidity divergences in the definition of \TMDs
is to introduce gauge links that are slightly off the light cone right from the start, i.e. with $v^2\not=0$, and hence a finite $\zeta$.
\TMDs defined in such a way even follow a known evolution equation in the parameter $\zeta$, see, e.g., Refs. \cite{Collins:1981uk,Collins:1981uw,Ji:2004wu}, which allows to evolve to 
arbitrarily large $|\zeta|$. 
Based on the above observations, we plan to extend our calculations to include staple-shaped Wilson lines with varying staple-extents $\eta$,
for different values of $\zeta$ employing a larger number of non-zero lattice nucleon momenta.
To get into contact with the process-related TMDs, we will then attempt to extrapolate the lattice results 
to large $\eta$ and large $|\zeta|$, the latter possibly with the help of the above mentioned evolution equations.
This approach should lead to results which may be compared in a meaningful manner with corresponding results from
experimental and phenomenological TMD-studies of, e.g., the Sivers effect.
To recapitulate, within such a formalism, the calculation of \TMDs relevant for SIDIS or Drell-Yan processes on the lattice could become feasible, at least in principle. 
In practice, one of the foreseeable technical challenges that one has to face in this case are diminishing signal to noise ratios for increasing nucleon momenta.
Another one is the statistical noise created by the long gauge link. Furthermore, at present, there are also a number of conceptual details concerning renormalization of the matrix elements that need to be worked out. 
As pointed out in Ref. \cite{Collins:2008ht}, embedding certain soft factors in the definition of the correlator could cancel the self-energies of the gauge link in an appropriate way. 
Even without detailed knowledge about soft factors, it might be possible to estimate ratios of certain $\vprp{k}$-moments such as those in Eq. \eqref{eq-gA}-\eqref{eq-shiftTU}, exploiting the cancellation of self-energies on the right hand side of Eq. \eqref{eq-regratio}. Especially the transverse momentum shift $\langle \vect{k}_y \rangle_{TU}$ caused by the Sivers function is a promising and prominent candidate to investigate with extended gauge links on the lattice.

\begin{acknowledgments}

Special thanks are due to the LHP lattice collaboration for providing their lattice quark propagators to us, and for technical advice.
We thank Harut Avakian, Gunnar Bali, Alexei Bazavov, Vladimir Braun, Markus Diehl, Robert Edwards, Meinulf G\"ockeler, Barbara Pasquini, Alexei Prokudin, David Richards and Dru Renner for helpful discussions and suggestions. 
We are grateful to the MILC collaboration, in particular to Carleton DeTar, Doug Toussaint and Robert Suger, for granting us access to their gauge configurations. 
For our calculations, we have been using computing resources at Jefferson Lab. 
A.S. and Ph.H. thank the Yukawa Institute of Kyoto University for hospitality during the HESI10 Workshop.
The authors acknowledge support by the Emmy-Noether program and the cluster of excellence ``Origin and Structure of the Universe'' of the DFG 
(Ph.H. and B.M.), SFB/TRR-55 (A.S.) and the US Department of Energy grant DE-FG02-94ER40818 (J.N.). 
Notice: Authored by Jefferson Science Associates, LLC under U.S. DOE Contract No. DE-AC05-06OR23177. 
The U.S. Government retains a non-exclusive, paid-up, irrevocable, world-wide license to publish or reproduce this manuscript for U.S. Government purposes.
\end{acknowledgments}

%%%%%%%%%%%%%%%%%%%%%%%%%%%%%%%%%%%%%%%%%%%%%%%%%%%%%%%%%%%%%%%%%

\appendix

\section{Conventions and definitions}
\label{sec-conv}

Whenever the four-vector $\elll$ fulfills $\elll^2 \leq 0$, we shall make use of the abbreviation $|\elll| \equiv \sqrt{-\elll^2}$.

In the continuum, a ``gauge link'' or ``Wilson'' line is given by the path-ordered exponential
\begin{align}
	\WlineC{\mathcal{C}_\elll} \ & \equiv\ \mathcal{P}\ \exp\left( -ig \int_{\mathcal{C}_\elll} d \xi^\mu\ A_\mu(\xi) \right) \nonumber \\
	& =  \mathcal{P}\ \exp\left( -ig \int_0^1 d\lambda\ A\!\left(\mathcal{C}_\elll(\lambda)\right) \cdot \dot{\mathcal{C}}_\elll(\lambda) \right)\, .
	\label{eq-wlinecont}
\end{align}
Here the path is specified by a continuous, piecewise differentiable function $\mathcal{C}_\elll$ with derivative $\dot{\mathcal{C}}_\elll$ and with $\mathcal{C}_\elll(0)=\elll$, $\mathcal{C}_\elll(1)=0$. 

For an arbitrary four-vector $w$, we introduce light cone coordinates $w^+ = (w^0 + w^3)/\sqrt{2}$, $w^- = (w^0 - w^3)/\sqrt{2}$ and the transverse projection $w_\prp = (0,w^1, w^2,0)$, which can also be represented as a Euclidean two-component vector $\vprp{w}=(\vect{w}_1,\vect{w}_2)\equiv(w^1,w^2)$, $\vprp{w}\tcdot\vprp{w} \geq 0$. The basis vectors corresponding to the $+$ and $-$ components shall be denoted $\nplus$ and $\nminus$, respectively, and fulfill $\nplus \cdot \nminus = 1$. 
The nucleon moving in $z$-direction has momentum $P =  P^+ \nplus + (m_N^2/2 P^+) \nminus$ and spin $S =  \Lambda (P^+/ m_N) \nplus -  \Lambda (m_N/2 P^+) \nminus + S_\prp$, $S^2 = -1$. We use the convention $\epsilon^{0123}=1$ for the totally antisymmetric Levi-Civita symbol, and introduce $\myeps_{i j} \equiv \epsilon^{-+ij}$ such that $\myeps_{1 2} = 1$.

\section{Naive Continuum Limit of the Lattice Gauge Link}
\label{sec-convproof}

In this section, we show that the discretized Wilson line, given by a
product of link variables as shown in Eq.~\eqref{eq-lat-gaugelink},
approaches
the continuum Wilson line  Eq. \eqref{eq-wlinecont} in the naive continuum
limit.

Consider a lattice path $\mathcal{C}^\lat_\elll = (x^{(n)},\ldots,x^{(0)})$
that ``approximates'' a continuous, piecewise smooth path $\mathcal{C}_\elll$
of fixed length $\len$. By ``approximates'' we refer to the following
criterium: The path $\mathcal{C}_\elll$ can be subdivided into $n$ sections
that connect mutually different, path ordered points $y^{(n)}$, $\ldots$,
$y^{(0)}$  on $\mathcal{C}_\elll$, such that $|y^{(i)}-x^{(i)}| =
\mathcal{O}(a)$ for all $i=0..n$.

Provided the lattice path is not intersecting with itself ( $x^{(i)} \neq x^{(j)}$ for all $i \neq j$ ), $n$ must be of order
$\len/a$ for fixed $\len$, since there are $\mathcal{O}(\len/a)$ lattice sites a distance of
$\mathcal{O}(a)$ away from $\mathcal{C}_\elll$. 
Thus, $n$ grows as $a^{-1}$ in the continuum limit. 
For reasons of definiteness, we now divide the lattice path
$\mathcal{C}^\lat_\elll$ into approximately $\sqrt{n}$ sections,
each section connecting approximately the same number of consecutive points $x^{(i)}$.
%each section connecting $m+1$ consecutive points $x^{(i)}$, so that $m =
%\mathcal{O}(\sqrt{n})$. 
Consider one of these sections, for example, the section running from a point $x^{(m)}$ to $x^{(0)}$.
The number of points in this section is $m+1 = \mathcal{O}(\sqrt{n})$.
For an individual link variable of this section, we write
\begin{align}
	U(x^{(i)},x^{(i-1)}) & =  \Eins + i g\, \Delta x^{(i)} \cdot
A(x^{(i)}) + \mathcal{O}(a^2) \nonumber\\
	& =  \Eins + i g\, \Delta x^{(i)} \cdot A(\bar x) +
\mathcal{O}(a^2\sqrt{n})\, ,
\end{align}
where $\Delta x^{(i)} \equiv x^{(i-1)} - x^{(i)} = \mathcal{O}(a)$ with
$i=1\ldots m$, and where 
we used a Taylor-expansion of the gauge field $A_\mu(x)$ around
$\bar{x}\equiv\frac{1}{m+1}\sum_{i=0}^{m} x^{(i)}$:
\begin{align}
	A_\mu(x) & = A_\mu(\bar x) + (x-\bar x)_\nu \partial_\nu A_\mu(\bar x)
+ \ldots \nonumber \\ & = A_\mu(\bar x) + \mathcal{O}(a\sqrt{n}) ,
\end{align}
which holds since $|x-\bar{x}| =  \mathcal{O}(a\sqrt{n})$. 
For clarity, we have kept $\sqrt{n}$ explicit in our notation, but keep in
mind that we could formally replace $\mathcal{O}(\sqrt{n})$ by
$\mathcal{O}(a^{-1/2})$.
For the product of $m = \mathcal{O}(\sqrt{n})$ link variables we then find 
\begin{align}
	& U(x^{(m)},x^{(m-1)})\cdots U(x^{(1)},x^{(0)}) \nonumber  \\  
	=\ & \Eins + i g\, (x^{(0)}-x^{(m)}) \cdot A(\bar x) +
\mathcal{O}(a^2 n) \nonumber \\
	=\ &\Eins + i g\, (y^{(0)}-y^{(m)}) \cdot A(\bar y) +
\mathcal{O}(a^2 n) \ .
\label{eq-UlatExpansion}
\end{align}

The corresponding section $\mathcal{C}_\elll^{(m,0)}$ of the continuous path
$\mathcal{C}_\elll$, running between $y^{(m)}$ and $y^{(0)}$, reads in
expanded form\footnote{where, as before, $|y-\bar{y}| =
\mathcal{O}(a\sqrt{n})$}
\begin{align}
	\WlineC{\mathcal{C}_\elll^{(m,0)}} & = \mathcal{P} \exp \left(
i\,g\int_{\mathcal{C}_\elll^{(m,0)}}\ d\xi_\mu \left\{\Afield_\mu(\bar y) +
\mathcal{O}(a\sqrt{n})\right\}\right) \nonumber \\
	& = \Eins + i g \, (y^{(0)}-y^{(m)}) \cdot A(\bar y) +
\mathcal{O}(a^2 n)\, .
\label{UcontExpansion}
\end{align}
Comparing this with Eq.~\eqref{eq-UlatExpansion}, we get
\begin{equation}
	U(x^{(m)},x^{(m-1)})\cdots U(x^{(1)},x^{(0)}) =
\WlineC{\mathcal{C}_\elll^{(m,0)}} + \mathcal{O}(a^2 n)\, .
\end{equation}
Analogous relations hold for the other subsections of the lattice path and
their continuous counterparts. Forming the product of these
$\mathcal{O}(\sqrt{n})$ subsections, we finally obtain
\begin{equation}
	\WlineClat{\mathcal{C}^\lat_\elll} =  \WlineC{\mathcal{C}_\elll} +
\mathcal{O}(a^2 n^{3/2}) \xrightarrow{a\rightarrow 0}
\WlineC{\mathcal{C}_\elll}\, ,
\end{equation}
since formally $\mathcal{O}(a^2 n^{3/2})=\mathcal{O}(a^{1/2})$ for fixed
length $\len$. 
%Note that in this basic proof, we have made use of the proximity criterium
%$|y^{(i)}-x^{(i)}| = \mathcal{O}(a)$ only in $\mathcal{O}(\sqrt{n})$ places,
%namely at the ends of each subdivision of the gauge link."

\newcommand{\slfrac}[2]{\left.#1\middle/#2\right.}
\section{Properties under symmetry transformations}
\label{sec-symtraf}

First, consider a general prescription $\mathcal{C}$ for the gauge paths. 
Applying Lorentz transformations $(L[\Lambda])$, parity transformation \conpar\, time reversal \contime, and complex conjugation \conherm, we obtain the following relations:

\begin{align}
	\widetilde \Phi^{[\GammaOp]}(\elll,P,S;\mathcal{C})  
	&= \widetilde \Phi^{[\Lambda_{\slfrac{1}{2}}^{-1}\GammaOp\Lambda_{\slfrac{1}{2}}^{\phantom{-1}}]}(\Lambda \elll,\Lambda P, \Lambda S;\mathcal{C}^{(L[\Lambda])})  \ ,\label{eq-lortrans} \displaybreak[0] \\
	\widetilde \Phi^{[\GammaOp]}(\elll,P,S;\mathcal{C})  
	&= \widetilde \Phi^{[\gamma^0\GammaOp\gamma^0]}(\overline{\elll},\overline{P},-\overline{S};\mathcal{C}^{\conpar})  \ ,\label{eq-conpar} \displaybreak[0] \\
	\left[ \widetilde \Phi^{[\GammaOp]}(\elll,P,S;\mathcal{C})  \right]^* 
	&= \widetilde \Phi^{[\gamma^1 \gamma^3 \GammaOp^* \gamma^3 \gamma^1]}(-\overline{\elll},\overline{P},\overline{S};\mathcal{C}^{\contime})  \ , \label{eq-contime} \displaybreak[0] \\
	\left[  \widetilde \Phi^{[\GammaOp]}(\elll,P,S;\mathcal{C})  \right]^* 
	&= \widetilde \Phi^{[\gamma^0 \GammaOp^\dagger \gamma^0]}(-\elll,P,S;\mathcal{C}^{\conherm})  \ .\label{eq-conherm}
	\end{align}	
Here the matrices $\Lambda$ and $\Lambda^{\phantom{-1}}_{\slfrac{1}{2}}$ describe Lorentz transformations of vectors $x^\mu \rightarrow {\Lambda^\mu}_\nu x^\nu$ and spinors $\psi \rightarrow \Lambda^{\phantom{-1}}_{\slfrac{1}{2}} \psi$. For any Minkowski vector $w=(w^0,\vect{w})$ the space inverted vector is defined as $\overline{w}\equiv(w^0,-\vect{w})$ . The transformed link paths are defined as
\begin{align}
	\mathcal{C}^{(L[\lambda])}_\elll(\lambda) & \equiv \Lambda \mathcal{C}_{\Lambda^{-1} \elll}(\lambda)\ ,\nonumber\\
	\mathcal{C}^{\conpar}_\elll(\lambda) & \equiv \overline{\mathcal{C}_{\overline{\elll}}(\lambda)}\, , \nonumber\\
	\mathcal{C}^{\contime}_\elll(\lambda) & \equiv -\overline{\mathcal{C}_{-\overline{\elll}}(\lambda)}\ ,\nonumber\\ 
	\mathcal{C}^{\conherm}_\elll(\lambda) & \equiv \mathcal{C}_{-\elll}(1-\lambda) + \elll\, . \label{eq-pathtrans}
\end{align} 

For straight gauge links $\WlineC{\mathcal{C}_\elll} = \WlineC{\elll,0}$, we get $\mathcal{C} = \mathcal{C}^{(L[\Lambda])} =\mathcal{C}^{\conpar} = \mathcal{C}^{\contime} = \mathcal{C}^{\conherm}$ , i.e., the link prescription $\mathcal{C}$ is invariant. Equation \eqref{eq-lortrans} then tells us that the correlator can be decomposed into Lorentz-covariant structures weighted by amplitudes $\tAmp_i(\elll^2,\elll \tcdot P)$. 
Equation \eqref{eq-conherm} establishes the relation Eq. \eqref{eq-conjugated} between $\tAmp_i^*(\elll^2,\elll \tcdot P)$ and $\tAmp_i(\elll^2,-\elll \tcdot P)$. Further relations derived from Eqns. \eqref{eq-conpar} and \eqref{eq-contime} reduce the number of possible non-zero amplitudes, eventually leading to the parametrization Eq. \eqref{eq-phitildetraces}.

As a side remark, we briefly discuss the case of staple shaped gauge links in direction $v$. The paths transform according to 
\begin{align}
[\mathcal{C}^{(\eta v)}]^{(L[\Lambda])} & = \mathcal{C}^{(\Lambda \eta v)}, &  [\mathcal{C}^{(\eta v)}]^{\conpar} & = \mathcal{C}^{(\eta \overline{v})}, \nonumber \\
[\mathcal{C}^{(\eta v)}]^{\contime} & = \mathcal{C}^{(-\eta \overline{v})}, &  [\mathcal{C}^{(\eta v)}]^{\conherm} & = \mathcal{C}^{(\eta v)}\ .
\label{eq-stapletrans}
\end{align}
The dependence of the correlator on the direction $v$ leads to the appearance of new amplitudes \cite{Goeke:2005hb}, in total we now have 32.
Moreover, the amplitudes now depend on the Lorentz-invariants $\elll^2$, $\elll \tcdot P$, $ \eta v \tcdot \elll$, $(\eta v)^2$ and $\eta v \tcdot P$. 
The amplitudes $\tAmp_i$ in the limit $\eta \rightarrow \infty$ can only depend on variables that are $|\eta|$-independent combinations of these invariants \cite{MuschThesis2,Accardi:2009au}. To obtain a complete set of such variables, we divide the invariants by appropriate powers of $|\eta v \tcdot P|$, a quantity that remains finite in the limiting case $v = \pm n$, $P^+ \gg m_N$ relevant for the discussion of SIDIS or the Drell-Yan process. We can thus write the amplitudes 
as functions $\widetilde{A}_i(\elll^2, \elll \tcdot P, v \tcdot \elll/|v\tcdot P|, \zeta^{-1}, v \tcdot P/|v \tcdot P|)$, with $\zeta^{-1} \equiv v^2/|2 v\tcdot P|^2$. Inserting Eq. \eqref{eq-stapletrans} into Eqns. \eqref{eq-lortrans}--\eqref{eq-conherm}, we find that the transformations \conherm\ and \conpar\ leave $v^2$ and $v \tcdot P$ invariant, unlike \contime, which changes the sign of $v \tcdot P$. Therefore, time reversal \contime, rather than restricting the number of amplitudes, establishes relations between amplitudes $\widetilde{A}_i(\ldots,+1)$ and $\widetilde{A}_i(\ldots,-1)$. The amplitude with $\sgn(v\tcdot P)=1$ corresponds to SIDIS, the amplitude with $\sgn(v\tcdot P) = -1$ describes DY. Some amplitudes are independent of $\sgn(v \tcdot P)$, others switch sign. Those latter amplitudes lead to ``time-reversal odd'', process-dependent \TMDs like the Sivers function $f_{1T}^\prp$. 

\begin{table}
\begin{centering}
\renewcommand{\arraystretch}{2}
\begin{tabular}{|c|c|l|}
\hline
$\GammaOp$ {\small(Eucl.)} & $\GammaOp$ {\small(Mink.)} & $\frac{1}{2} \bar{R}[O^\ren_\Gamma[\mathcal{C}_\elll]](\vect{P}) $ \\ \hline 
$\Eins$ & $\Eins$ & $\displaystyle \frac{m_N}{E(P)}\,\tilde A_1$ \\
$\gamma_\Eu{1}$ & $- i \gamma^1$ & 
	$\displaystyle - \frac{i}{E(P)}\,\tilde A_2\,\vect{P}_1 + \frac{m_N^2}{E(P)}\, \tilde A_3\, \vect{\elll}_1$ \\
$\gamma_\Eu{2}$ & $- i \gamma^2$ & 
	$\displaystyle \frac{m_N^2}{E(P)}\,\tilde A_3\,\vect{\elll}_2 $ \\
$\frac{1}{2}[\gamma_\Eu{1},\gamma_\Eu{2}]$ & $-i(i \sigma^{03}\gamma^5)$ & 
	$\displaystyle -i\, \tilde A_9 - i m_N^2\, \tilde A_{11} \,(\vect{\elll}_3)^2$ \\
$\gamma_\Eu{3}$ & $- i \gamma^3$ & 
	$\displaystyle \frac{m_N^2}{E(P)}\,\tilde A_3\,\vect{\elll}_3$ \\
$\frac{1}{2}[\gamma_\Eu{1},\gamma_\Eu{3}]$ & $i (i\sigma^{02}\gamma^5)$ & 
	$\displaystyle  i m_N^2\, \tilde A_{11} \,\vect{\elll}_2 \vect{\elll}_3$ \\
$\frac{1}{2}[\gamma_\Eu{2},\gamma_\Eu{3}]$ & $-i(i \sigma^{01}\gamma^5)$ & 
	$\displaystyle - i m_N^2\, \tilde A_{11} \,\vect{\elll}_1 \vect{\elll}_3$ \\
$- \gamma_\Eu{4} \gamma_\Eu{5}$ & $\gamma^0 \gamma^5$ & 
	$\displaystyle  i m_N\, \tilde A_7\, \vect{\elll}_3$ \\
$\gamma_\Eu{4}$ & $\gamma^0$ & 
	$\displaystyle \tilde A_2 $ \\
$\frac{1}{2}[\gamma_\Eu{1},\gamma_\Eu{4}]$ & $i\sigma^{23}\gamma^5$ & 
	$\displaystyle \frac{i m_N^2}{E(P)}\,\tilde A_{10}\,\vect{\elll}_2$ \\
$\frac{1}{2}[\gamma_\Eu{2},\gamma_\Eu{4}]$ & $-i\sigma^{13}\gamma^5$ & 
	$\displaystyle - \frac{1}{E(P)}\,\tilde A_9\,\vect{P}_1 $ \\
& & $\displaystyle  - \frac{i m_N^2}{E(P)}\,\tilde A_{10}\,\vect{\elll}_1 - \frac{m_N^2}{E(P)}\,\tilde A_{11}\,(\vect{\elll}_3)^2 \vect{P}_1$ \\
$\gamma_\Eu{3} \gamma_\Eu{5}$ & $i \gamma^3 \gamma^5$ & 
	$\displaystyle -\frac{i m_N}{E(P)}\,\tilde A_6 - \frac{i m_N^3}{E(P)}\,\tilde A_8\,(\vect{\elll}_3)^2$ \\
$\frac{1}{2}[\gamma_\Eu{3},\gamma_\Eu{4}]$ & $i\sigma^{12}\gamma^5$ & 
	$\displaystyle \frac{m_N^2}{E(P)}\,\tilde A_{11}\,\vect{\elll}_2 \vect{\elll}_3 \vect{P}_1$ \\
$-\gamma_\Eu{2} \gamma_\Eu{5}$ & $-i \gamma^2 \gamma^5$ & 
	$\displaystyle \frac{i m_N^3}{E(P)}\,\tilde A_8\,\vect{\elll}_2 \vect{\elll}_3$ \\
$\gamma_\Eu{1} \gamma_\Eu{5}$ & $i \gamma^1 \gamma^5$ & 
	$\displaystyle - \frac{i m_N^3}{E(P)}\,\tilde A_8\,\vect{\elll}_1 \vect{\elll}_3 - \frac{m_N}{E(P)}\,\tilde A_7\,\vect{\elll}_3 \vect{P}_1$ \\
$\gamma_\Eu{5}$ & $-\gamma^5$ & $\displaystyle 0$ \\
\hline
\end{tabular} \par
\renewcommand{\arraystretch}{1}%
\end{centering}%
\caption{
Plateau values of the ratios $\bar{R}[O^\ren_\Gamma[\mathcal{C}_\elll]](\vect{P})$ for straight gauge links $\mathcal{C}_\elll$ in terms of the amplitudes $\tilde A_i$. Here we employ the LHPC conventions for $\GammaTwo = \GammaThr=(\Eins + \gamma_4)(1+i\gamma_5\gamma_3)/2$, i.e. the nucleons are spin-projected along the $z$-axis. We choose the nucleon momentum $\vect{P} = ( \vect{P}_1, 0, 0 )$, and the quark separation is $\vect{\elll} = (\vect{\elll}_1, \vect{\elll}_2, \vect{\elll}_3)$, $\elll_\Eu{4} = 0$. }
\label{tab-ratios}%
\end{table}

\section{Symmetry improved operators}
\label{sec-symimpop}
Looking at Eqns. \eqref{eq-pathtrans}, we see that the symmetry transformation of the link prescription features a common structure consisting of a backward and a forward transformation that leaves, as a whole, the vector between start and end point of the link invariant. In mathematical terms 
\begin{equation}
	\mathcal{C}_\elll \rightarrow D(g)\, \mathcal{C}_{\tilde D(g^{-1})\, \elll}\ . \label{eq-trafostruc}
\end{equation}
Here $g$ is a group element of one of the respective symmetry groups, i.e., Lorentz-transformations, parity transformation, time reversal or Hermitian conjugation. The representation $D(g)$ of that group element acts on the link path, while the representation $\tilde D(g)$ acts on vectors. The representation $\tilde D$ can be deduced from $D$ by looking at the transformation behavior of the vector between start and end point of a link path, i.e.,
\begin{equation}
	\tilde D(g)\, \elll := \left[ D(g)\, \mathcal{C}_\elll \right](0) -  \left[ D(g)\, \mathcal{C}_\elll \right](1)\, .
\end{equation}
When constructing a discretized version of the gauge link operator, we can reduce discretization artefacts by preserving those symmetry transformation properties of the link path that have a correspondence in discrete Euclidean space. In this context, it is convenient to represent the discrete link path as a sequence of shifts of one lattice unit. Let $\widehat 1$, $\widehat{2}$, $\widehat{3}$, $\widehat 4$ denote vectors of length $a$ along the four lattice axes. A lattice link path may thus be represented as $\mathcal{C}^\lat_\elll = [s^{(n)},\ldots,s^{(1)}]$, with $s^{(i)}\in\{-\hat{4},...,-\hat{1},\hat{1},\ldots,\hat{4}\}$. The sample link path of Fig. \ref{fig-steplike} is given by $\mathcal{C}^\lat_\elll = [\widehat{1},\widehat{2},\widehat{1},\widehat{1},\widehat{2},\widehat{1},\widehat{1},\widehat{2},\widehat{1}]$. On the lattice, the Lorentz group and parity are replaced by the hypercubic group \cite{Baake:1981qe,Gockeler:1996mu}
\begin{equation}
	\mathrm{H}(4) = \left\{(b,\pi)\  \big|\ b_1,b_2,b_3,b_4 \in \{0,1\},\  \pi \in S_4\right\}\, ,
\end{equation}
where $S_4$ is the set of permutations of $\{1,2,3,4\}$. The action of a given group element $h=(b,\pi)$ of $\mathrm{H}(4)$ on a link path is given by
\begin{equation}
	s^{(i)} \rightarrow s'^{(i)} = \left\{ \begin{array}{lcl}
		-(-1)^{b_4}\,\widehat{\pi(4)} & : & s^{(i)} = -\widehat{4} \\
		& \ldots & \\
		-(-1)^{b_1}\,\widehat{\pi(1)} & : & s^{(i)} = -\widehat{1} \\
		\phantom{-}(-1)^{b_1}\,\widehat{\pi(1)} & : & s^{(i)} = \widehat{1} \\
		& \ldots  & \\
		\phantom{-}(-1)^{b_4}\,\widehat{\pi(4)} & : & s^{(i)} = \widehat{4} 
	\end{array} \right.\, ,
\end{equation}  
i.e., $\mathrm{H}(4)$ permutes axis labels and inverts the direction of lattice axes. This defines $D(h)$. Hermitian conjugation $(\dagger)$ of the matrix element reverses the ordering of the shifts and negates them:
\begin{equation}
	D(\dagger)\,  [s^{(n)},\ldots,s^{(1)}] = [-s^{(1)},\ldots,-s^{(n)}]\, .
\end{equation}
The representation $\tilde D$ is deduced from the transformation behavior of $\elll = \sum_{i=1}^n s^{(i)}$:
\begin{align}
	% \tilde D(h)\, \elll & = \left( (-1)^{b_{\pi^{-1}(1)}} \elll_{\pi^{-1}(1)}, \ldots, (-1)^{b_{\pi^{-1}(4)}} \elll_{\pi^{-1}(4)} \right) \ , \\
	\tilde D(h^{-1})\, \elll & = \left( (-1)^{b_1} \elll_{\pi(1)}, \ldots, (-1)^{b_4} \elll_{\pi(4)} \right) \ , \\
	\tilde D(\dagger)\, \elll & = -\elll \ .	
\end{align}
The operation $\dagger$ is its own inverse and commutes with any $h\in\mathrm{H}(4)$. Thus we can define a larger group
\begin{equation}
	G \equiv \bigcup_{h\in\mathrm{H}(4)} \left\{ h, \dagger \circ h \right\}\ .
\end{equation} 
The function $C^\lat$ we use to determine the link path for a given vector $\elll$ is a Bresenham-like algorithm that produces a step-like path close to the straight continuum line. It turns out that, in general, this alorithm is not invariant under transformations of the form Eq. \eqref{eq-trafostruc}. However, it is simple to form a superposition of gauge links that has the desired properties:  
\begin{equation}
	\overline{\mathcal{U}}_\elll \equiv  \frac{1}{\# G} \sum_{\tilde g \in G} \mathcal{U}\left[ D(\tilde g)\, \mathcal{C}^\lat_{\tilde D(\tilde g^{-1})\,\elll} \right]\, . \label{eq-imprsuper}
\end{equation}
All gauge links in the above superposition run from $\elll$ to $0$.
Thanks to the properties of the algorithm $C^\lat$, the above sum does not contain link paths that have an extent in the Euclidean 4-direction.
Performing the substitution Eq. \eqref{eq-trafostruc} on the right hand side of the equation above, we obtain
\begin{align}
	&  \frac{1}{\# G} \sum_{\tilde g \in G} \mathcal{U}\left[ D(\tilde g)\,D(g)\, \mathcal{C}^\lat_{\tilde D(g^{-1})\,\tilde D(\tilde g^{-1})\,\elll} \right] \nonumber \\
	= &  \frac{1}{\# G}  \sum_{\tilde g \in G} \mathcal{U}\left[ D(\tilde g \circ g)\, \mathcal{C}^\lat_{\tilde D((\tilde g \circ g)^{-1})\,\elll} \right] \nonumber \\
	= &  \frac{1}{\# G} \sum_{\hat g \in G} \mathcal{U}\left[ D(\hat g)\, \mathcal{C}^\lat_{\tilde D(\hat g^{-1})\,\elll} \right]  = \overline{\mathcal{U}}_\elll\, ,
\end{align}
because $G \circ g = G$. So $\overline{\mathcal{U}}_\elll$ is indeed invariant under transformations Eq. \eqref{eq-trafostruc} for any $g \in G$.

In practice, the sum of Eq. \eqref{eq-imprsuper} contains typically only a few distinct link paths. We evaluate three-point functions for all these different paths. In the final analysis, we form the superpositions using appropriate weights for the individual paths corresponding to their multiplicities in the sum.

\section{Charge conjugated operator}
\label{sec-chargecon}

In the presence of a general link path $\mathcal{C}$,  a gauge invariant definition of the correlator $\Phi^c$ of Ref. \cite{Mulders:1995dh} is obtained by applying charge conjugation $C$ to the whole operator: 
\begin{align}
  \Phi^{c[\GammaOp]} (k,P,S;\mathcal{C}) & \equiv \int \frac{d^4 \elll}{(2\pi)^4} \ 
  e^{-ik \cdot \elll} \nonumber \\ & \times 
  \frac{1}{2} \bra{\nucl{P,S}}\ C\ \bar \quark(\elll)\, \GammaOp\ \WlineC{\mathcal{C}_\elll}\ \quark(0)\ C\ \ket{\nucl{P,S}} \nonumber\\
  & = \Phi^{[-\gamma^0\gamma^2\GammaOp^\transp \gamma^2\gamma^0]} (-k,P,S;\mathcal{C}^{(\dagger)}) \ .
  \label{eq-Ccorr}
\end{align}
where the conjugated link path $\mathcal{C}^{(\dagger)}$ is defined in Eq. \eqref{eq-pathtrans}. The straight gauge link and the staple-shaped gauge link turn out to be unaffected by the charge conjugation, $\mathcal{C}^{\text{sW}(\dagger)}=\mathcal{C}^{\text{sW}}$, $\mathcal{C}^{(v)(\dagger)}=\mathcal{C}^{(v)}$.

For completeness, we show the proof of the third line of the above equation. Using $C A^\mu(x) C = - A^\mu(x)$, $C q(x) C = i\gamma^0 \gamma^2 \bar{q}^\transp(x)$, $C \bar{q}(x) C = q^\transp(x) i\gamma^0 \gamma^2 $, where $\scriptstyle{\transp}$ is acting on Dirac and color indices only, we get
\begin{align}
	 \MoveEqLeft \phantom{-} C\ \bar \quark(\elll)\, \GammaOp\ \WlineC{\mathcal{C}_\elll}\ \quark(0)\ C \nonumber \\
	 = & \phantom{-} q^\transp(\elll) i\gamma^0 \gamma^2\ \GammaOp\ \WlineC{\mathcal{C}_\elll}^*\ i\gamma^0 \gamma^2 \bar{q}^\transp(0) \nonumber \\
	 = & - \bar{q}(0)\ \left( i\gamma^0 \gamma^2 \GammaOp i\gamma^0 \gamma^2 \right)^\transp\ \WlineC{\mathcal{C}_\elll}^\dagger\ q(\elll)  \, .
\end{align}
In the last line, we have used that fermion fields anti-commute. Denoting reverse path-ordering $\bar{\mathcal{P}}$, we find that the Hermitian conjugate of the gauge link reverses its direction:
\begin{align}
	 \WlineC{\mathcal{C}_\elll}^\dagger 
	 & =  \left[ \mathcal{P}\ \exp\left( -ig \int_0^1 d\lambda\ A\!\left(\mathcal{C}_\elll(\lambda)\right) \cdot \dot{\mathcal{C}}_\elll(\lambda) \right) \right]^\dagger \nonumber \\
	 & = \bar{\mathcal{P}}\ \exp\left( +ig \int_0^1 d\lambda\ A\!\left(\mathcal{C}_\elll(\lambda)\right) \cdot \dot{\mathcal{C}}_\elll(\lambda) \right) \nonumber \\
	 & = \mathcal{P}\ \exp\left( +ig \int_0^1 d\tilde\lambda\ A\!\left(\mathcal{C}_\elll(1-\tilde\lambda)\right) \cdot \dot{\mathcal{C}}_\elll(1-\tilde\lambda) \right) \nonumber \\
	 & = \mathcal{P}\ \exp\left( -ig \int_0^1 d\tilde\lambda\ A\!\left(\tilde{\mathcal{C}}_\elll(\tilde\lambda)\right) \cdot \dot{\tilde{\mathcal{C}}}_\elll(\tilde\lambda) \right)  \nonumber \\
	 & =   \WlineC{\tilde{\mathcal{C}}_\elll}\, ,
\end{align}
where $\tilde{\mathcal{C}}_\elll(\tilde\lambda) \equiv \mathcal{C}_\elll(1-\tilde\lambda)$. Using translation invariance, we obtain
\begin{align}
	\MoveEqLeft \widetilde \Phi^{c[\GammaOp]}(\elll,P,S;\mathcal{C}) \nonumber \\
	= & \frac{1}{2} \bra{\nucl{P,S}} \bar{q}(0)\ \left( - \gamma^0 \gamma^2 \GammaOp^\transp \gamma^0 \gamma^2\right)\ \WlineC{\tilde{\mathcal{C}}_\elll}\ q(\elll) \ket{\nucl{P,S}} \nonumber \\
	= & \frac{1}{2} \bra{\nucl{P,S}} \bar{q}(-\elll)\ \left( - \gamma^0 \gamma^2 \GammaOp^\transp \gamma^0 \gamma^2\right)\ \WlineC{\tilde{\mathcal{C}}_{\elll} -\elll}\ q(0) \ket{\nucl{P,S}} \nonumber \\
	= & \widetilde \Phi^{c[- \gamma^0 \gamma^2 \GammaOp^\transp \gamma^0 \gamma^2]}(-\elll,P,S;\mathcal{C}^{(\dagger)})\, ,
\end{align}
because $\tilde{\mathcal{C}}_{\elll}(\lambda) - \elll = \mathcal{C}_{-(-\elll)}(1-\lambda) + (-\elll) = \mathcal{C}^{(\dagger)}_{-\elll}(\lambda)$. Carrying out the Fourier transform with respect to $\elll$, we arrive at Eq. \eqref{eq-Ccorr}.

\section{Implementation details of link renormalization}
\label{sec-renimpl}

\begin{figure}[tbh]
	\centering%
	\includegraphics[scale=0.45]{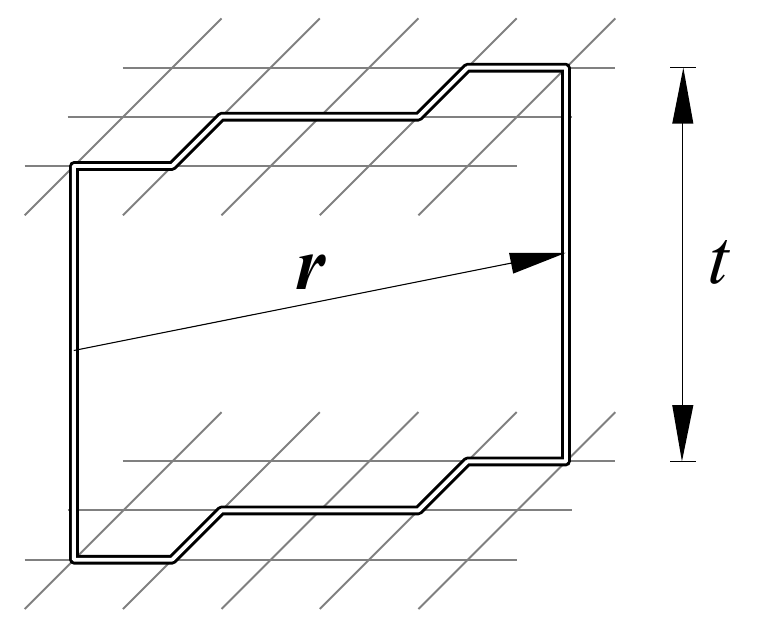}%
	\caption{Rectangular Wilson loop in the calculation of the static quark potential.\label{fig-wloopoblique}}
\end{figure}

We calculate rectangular Wilson loops on the lattice
\begin{align}
	W^\lat(\vect{r},T) \equiv \frac{1}{3} \dlangle \Tr_c\ \WlineClat{\mathcal{C}_{\vect{r},T}} \drangle
\end{align}
for closed paths $\mathcal{C}_{\vect{r},T}$ as depicted in Fig. \ref{fig-wloopoblique}.
Here $\vect{r}$ is a spatial vector between lattice sites. For the corresponding spatial sections of the gauge link, we use step-like paths as in Section \ref{sec-nonloc}.
For large enough $T$, 
\begin{equation}
	W^\lat(\vect{r},T) \approx c(\vect{r}) \exp\left(- V^\lat(\vect{r})\, T\right) \, .
\end{equation}
Taking lattice data at fixed $\vect{r}$ and a range of values $T$ enables us to determine $V^\lat(\vect{r})$ and $c(\vect{r})$ from an exponential fit.
To obtain a smooth interpolating curve of the static quark potential as a function of $R \equiv |\vect{r}|$, and to reduce discretization errors, we follow Refs. \cite{Booth:1992bm,Aubin:2004wf} and fit the functional form 
\begin{equation}
	\hat{V}^\lat(\vect{r}) = \underbrace{\hat{\sigma} \hat{R} - \alpha/\hat{R} + \hat{C}}_{\displaystyle \equiv \hat V(R)} - \lambda \left( \hat{V}^\lat_\text{pert}(\hat{\vect{r}}) - 1/\hat{R} \right)
	\label{eq-potfitfn}
\end{equation}
to the data obtained for $\hat{V}^\lat(\vect{r})$. Here the hat $\hat{\phantom{m}}$ indicates that the respective dimensionful quantity is expressed in lattice units. The potential $\hat{V}^\lat_\text{pert}(\vect{r})$ is  obtained from single gluon exchange between the temporal links in lattice perturbation theory. The corrective term \cite{Michael:1992nj} proportional to $\lambda$, associated with breaking of rotational invariance, becomes negligible for $R \gtrsim 3 a$. For the calculation of $\hat{V}^\lat_\text{pert}(\vect{r})$ we use the inverse gluon propagator of the MILC action \cite{Bistro}, and, if the potential is calculated on smeared gauge configurations, the appropriate HYP smearing coefficients $\tilde h_{\bar \mu, \bar \nu}(k)$ from Ref. \cite{DeGrand:2002va}. Once the fit parameters $\hat{\sigma}$, $\alpha$, $\hat{C}$ and $\lambda$ have been determined, we obtain $\delta \hat m$ from equating the renormalized potential $\hat V^\ren(R) = \hat V(R) + 2\, \delta \hat m$ with the string potential $\hat V_\text{string}(R) = \hat \sigma \hat R - \pi / 12 \hat R$ at a matching point $\hat R = 1.5\, \hat r_0 = 1.5\, \sqrt{(1.65-\alpha)/\sigma}$ :
\begin{equation}
	2 \delta \hat m = - \hat C +  \frac{1}{1.5} \sqrt{\frac{\hat \sigma}{1.65 - \alpha}}\left( \alpha - \frac{\pi}{12}\right) \, .
\end{equation}  

\section{Estimating discretization errors from the gauge link}
\label{sec-discerr}

\begin{figure}[tb]
	\centering%
	\subfloat[][]{%
		\label{fig-Ddeltam-rdepend}%
		\includegraphics[width=\linewidth]{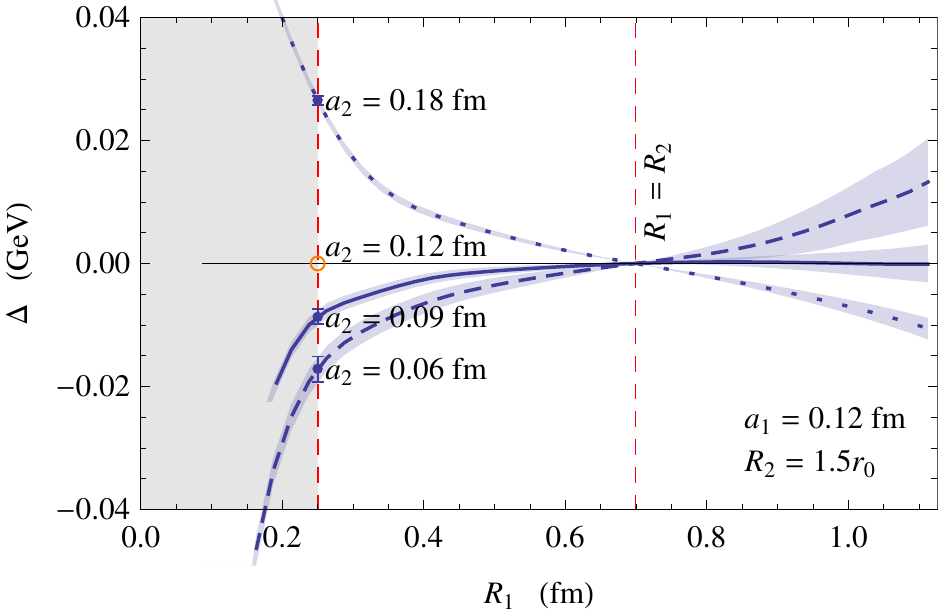}
		}\\%
	\subfloat[][]{%
		\label{fig-Ddeltam-adepend}%\
		\includegraphics[width=\linewidth]{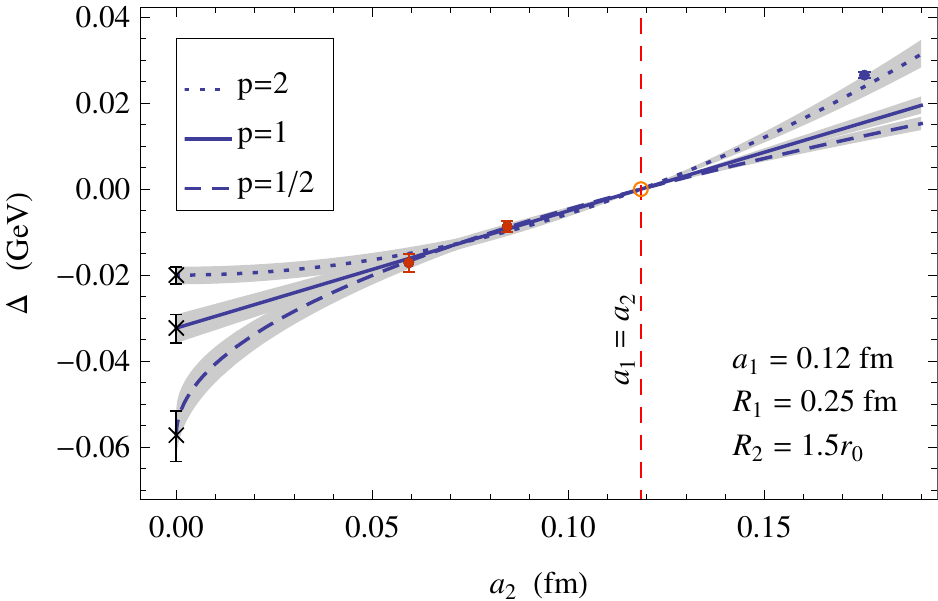}
		}\\[1em]%
	\caption[fig-A2normUp]{%
		\subref{fig-Ddeltam-rdepend}\ $R_1$-dependence of $\Delta(R_1,R_2,a_1,a_2)$ for fixed $a_1$ and $R_2$. The dashed line corresponds to the superfine\nbdash04 ensemble with $a_2\approx0.06\units{fm}$, the solid line to the fine\nbdash04 ensemble with $a_2\approx0.09\units{fm}$ and the dotted line to the extracoarse\nbdash04 ensemble with $a_2\approx0.18\units{fm}$.\ \subref{fig-Ddeltam-adepend}\ $a_2$-dependence of $\Delta(R_1,R_2,a_1,a_2)$ for fixed $a_1$, $R_1$ and $R_2$. The solid data points correspond to the data points extracted at $R_1=0.25\units{fm}$ in the figure above. The curves with statistical error bands are fits to $\Delta$ assuming discretization errors $\sim a^p$. The data point at $a\approx0.18\units{fm}$ has been excluded from the fit. The data points with crosses indicate the extrapolated values $\Delta[\delta m]_\text{dis}$ at $a_2=0$. Note: An error of $\Delta[\delta m]_\text{dis}=0.01 \units{GeV}$ corresponds to an uncertainty of about $2\%$ in the width $2/\sigma_{2,u}$ of the Gaussian we obtain for the $x$-integrated unpolarized distribution $f_{1,u}^\xmom{1}(\vprp{k}^2)$. 
		\label{fig-Ddeltam}
		}
\end{figure}

A comparison of $Y_\text{line}(R)$ defined in Eq. \eqref{eq-Ylinedef} for different lattice spacings allows us to get an idea about the size of discretization errors coming from the gauge link.
While the renormalized quantity $Y_\text{line}^\ren(R)$ must be independent of the lattice action, smearing and the lattice spacing,  $\delta m$ and $Y_\text{line}(R)$ can assume different numerical values for different lattice spacings:
\begin{equation*}
Y_\text{line}^\ren(R) = Y_\text{line}(R;a_1) + \delta m(a_1) = Y_\text{line}(R;a_2) + \delta m(a_2)
\end{equation*}
Thus the right hand side of the difference
\begin{equation}
\delta m(a_2) - \delta m(a_1) = Y_\text{line}(R;a_1) - Y_\text{line}(R;a_2)
\label{eq-DdmDYline}
\end{equation}
should be $R$-independent up to lattice artefacts. We estimate the latter by comparing two different link lengths $R_1$ and $R_2$:
\begin{align}
	\Delta(R_1,R_2,a_1,a_2) \equiv &  \left( Y_\text{line}(R_1;a_1) - Y_\text{line}(R_1;a_2) \right) \nonumber \\
	- & \left( Y_\text{line}(R_2;a_1) - Y_\text{line}(R_2;a_2) \right) \ .
\end{align}
As a technical note, we mention that we employ a spline interpolation in order to be able to evaluate $Y_\text{line}(R)$ at arbitrary values $R$. 
If there were no discretization errors at all, $\Delta$ would be zero for any choice of $a_1$, $a_2$, $R_1$ and $R_2$. 
We remark that $Y_\text{line}(R)$ can naturally provide an alternative way to fix $\delta m$, e.g., with a (gauge dependent) renormalization condition $Y_\text{line}^\ren(R_0) = 0$ for some fixed length $R_0$. This has already been suggested long ago in Ref. \cite{Martinelli:1995vj,Crisafulli:1995pg}.
Comparing with Eq. \eqref{eq-DdmDYline}, we learn that $\Delta$ can be understood as a discrepancy in the values $\delta m$ needed to renormalize $Y_\text{line}$ at two different link lengths $R_1$ and $R_2$.
Our goal here is to estimate discretization errors for the coarse\nobreakdash-\hspace{0pt}04 lattice, so we need to compare $Y_\text{line}$ determined on the coarse\nobreakdash-\hspace{0pt}04 lattice ($a_1 \approx 0.12 \units{fm}$) with the other other \makebox{-04} ensembles ($a_2 \approx 0.06, 0.09$, and $0.18 \units{fm})$. We choose $R_2 = 1.5 r_0 = 0.70 \units{fm}$, the same length scale we use as a matching point in our determination of $\delta m$ from the static quark potential. Figure \ref{fig-Ddeltam-rdepend} shows $\Delta(R_1,R_2{=}1.5r_0,a_1{=}0.12\units{fm},a_2)$ for the different available lattice spacings $a_2$ as a function of $R_1$. 
We find that the magnitude of $\Delta$ and its slope are largest when $R_1$ is small, i.e., when $R_1$ is of the order of a few lattice spacings $a_1$ or $a_2$. 
This finding corresponds to the discrepancies already observed in Fig. \ref{fig-YlineBazRen} in the region $R \lesssim 0.25 \units{fm}$ and leads to the conclusion that very short gauge links suffer from significant discretization errors.
We now choose $R_1 = 0.25 \units{fm}$, i.e., the shortest length of gauge links we accept in our \TMD analysis. The corresponding values $\Delta(R_1{=}0.25\units{fm},R_2{=}1.5r_0,a_1{=}0.12\units{fm},a_2)$ give rise to the data points with statistical error bars at the dashed vertical line on the left in Figure \ref{fig-Ddeltam-rdepend}. The same data points are plotted with respect to $a_2$ in Fig. \ref{fig-Ddeltam-adepend}. Assuming discretization errors of $\mathcal{O}(a^p)$, we have performed one-parameter fits of the form 
\begin{equation}
	\Delta(R_1,R_2,a_1,a_2)  \approx  c ( {a_2}^p - {a_1}^p )\, 
\end{equation}
to the data points in Fig. \ref{fig-Ddeltam-adepend}.
At present, we do not know the order of convergence $p$. 
Appendix \ref{sec-convproof} shows that $p \geq 1/2$ in the naive continuum limit. 
We have tried out fits with $p=1/2$, $p=1$ and $p=2$, always excluding the data point from the extracoarse\nbdash04 lattice from the fit.
In order to estimate discretization errors for the coarse\nbdash04 lattice, we use the above fits to extrapolate $\Delta$ to $a_2 = 0$~:
\begin{equation}
	\Delta[\delta m]_\text{dis} \equiv \left| \lim_{a_2 \rightarrow 0} \Delta(R_1,R_2,a_1,a_2) \right| \, ,
\end{equation}
where $R_1=0.25\units{fm}$, $R_2=1.5r_0$ and $a_1\approx 0.12\units{fm}$ are kept fixed.
We can interpret $\Delta[\delta m]_\text{dis}$ as the size of a spurious R-dependence of $\delta m$ that appears when we match $Y_\text{line}$ at finite lattice spacing $a_1$ to $Y_\text{line}$ in the continuum over a range of link lengths between $R_1$ and $R_2$. Thus $\Delta[\delta m]_\text{dis}$ can be effectively treated as an uncertainty in $\delta m$.
For the three different values of $p$, we obtain from the fits  $\Delta[\delta m]_\text{dis} = 0.0573(59)_\text{stat}\units{GeV}$,  $\Delta[\delta m]_\text{dis} = 0.0323(34)_\text{stat}\units{GeV}$, and  $\Delta[\delta m]_\text{dis} = 0.0200(21)_\text{stat}\units{GeV}$, respectively. 
For our presentation of numerical results in section \ref{sec-lowmom}, we select the value obtained from the assumption of $\mathcal{O}(a)$ convergence: $\Delta[\delta m]_\text{dis} = 0.0323\units{GeV}$, or $\Delta[\delta \hat m]_\text{dis} = 0.0194$ in lattice units. With respect to our analysis based on a Gaussian parametrization, the main effect of $\Delta[\delta m]_\text{dis}$ is an additional uncertainty in the widths $\sigma_{i,q}$ of the amplitudes $\tAmp_{i,q}(\elll^2,0)$.

We remark that our determination of $\Delta[\delta m]_\text{dis}$ is based on open gauge links $\WlineC{\mathcal{C}_\elll}$ evaluated on a gauge fixed ensemble. 
Discretization effects of the complete gauge invariant operator $\bar{q}(\elll) \Gamma \WlineC{\mathcal{C}_\elll} q(0)$ might be different, especially for short gauge links.
Our value $\Delta[\delta m]_\text{dis}$ determined with open Wilson lines can thus only serve as an order of magnitude estimate of potential discretization errors.

%\end{comment}

\section{Expansion in terms of local lattice operators}
\label{sec-localops}

The nonlocal lattice operators studied in this work can be written as weighted sums of local operators involving higher derivatives.
It is well known that due to the loss of translational and rotational symmetries on the lattice in particular, the
operators with two or more derivatives will mix with operators of lower mass dimension under renormalization. This type of
mixing involves inverse powers of the lattice spacing, and hence the respective contributions have to be subtracted explictly
before the continuum limit can be taken, which is in practice a difficult task.
The question then naturaly arises if and how these observations can be reconciled with the known renormalization properties
of a manifestly non-local operator as explained and used in section \ref{sec-contrenorm}.
Although we are not able in the course of this exploratory study to provide a definite answer,
we will briefly explore this question in the following and at least show that our renormalization prescription of the non-local operator on the one hand, and operator-mixing \emph{within} an expansion in terms of local operators on the other, are not in any apparent contradiction to each other.

To keep the discussion simple, we consider here a  non-local operator with a straight-link of length $\len$ in the direction of the unit vector $\hat{e}_\mu$
\begin{equation}
	O_\GammaOp(\len \hat{e}_\mu) \equiv \bar \quark(0)\, \GammaOp\, \Wline{0,\len \hat{e}_\mu}\, q(\len \hat{e}_\mu)\ .
\end{equation}
Our discrete representation of $O_\GammaOp(\len \hat{e}_\mu)$ on the lattice is 
\begin{equation}
	[O_{\GammaOp}(n\hat \mu)]^\lat \equiv \bar \quark(0)\, \GammaOp\, U(0,\hat \mu) \cdots U((n-1)\hat \mu,n\hat \mu)\, \quark(n\hat \mu) \, ,
\end{equation} 
where $n = \len / a$. Together with a discretization prescription for the covariant derivative on the lattice, e.g.
\begin{equation}
	{D}_\mu f(x) \equiv \frac{1}{a}\big\{ U(x,x+\hat \mu) f(x+\hat \mu) - f(x) \big\} \, ,
\end{equation}
we can  write $[O_{\GammaOp}(n\hat \mu)]^\lat$ as a weighted sum of local lattice operators:
\begin{align}
	[O_{\GammaOp}(n\hat \mu)]^\lat & = \bar \quark(0)\, \GammaOp\, \left( a D_\mu + 1 \right)^n\, \quark(0) \nonumber \\
	& = \sum_{k=0}^n  \binom{n}{k} a^k \ \underbrace{\bar \quark(0)\, \GammaOp\, D_\mu^k \, \quark(0)  }_{\displaystyle \equiv [O^{\mu,k}_{\GammaOp}]^\lat }\ .
	\label{eq-latexpan}
\end{align}

To simplify the discussion of operator mixing, we only consider mixing of operators $[O^{\mu,k}_{\GammaOp}]^\lat$ among themselves:
\begin{align}
	[O^{\mu,k}_{\GammaOp}]^\lat & = \sum_{j=0}^\infty Z_{kj}\, a^{j-k}\, [O^{\mu,j}_{\GammaOp}]^\ren \nonumber \\
	& = a^{-k}\, Z_{k0} \, [O^{\mu,0}_{\GammaOp}]^\ren + \ldots +  Z_{kk} \, [O^{\mu,k}_{\GammaOp}]^\ren + \ldots \ .
	\label{eq-mixing}
\end{align}
Here the powers of $a$ required to render the mixing coefficients $Z_{kj}$ dimensionless can become negative, the ``worst case'' being the potential mixing with the derivative-free operator $[O^{\mu,0}_{\GammaOp}]^\lat$.
Inserting the above expression into the second line of Eq. \eqref{eq-latexpan} yields
\begin{align}
	[O_{\GammaOp}(n\hat \mu)]^\lat & = \sum_{j=0}^\infty\ \underbrace{ \left\{ \sum_{k=0}^n \binom{n}{k} Z_{kj} \right\}\, n^{-j} }_{\displaystyle \equiv c_j(n)}\, \len^j [O^{\mu,j}_{\GammaOp}]^\ren
	\label{eq-latmixsubst}
\end{align}
For the discussion of the continuum limit, it is at this point important to distinguish  two cases: 
\begin{enumerate}
\item keeping $n=\len/a$ fixed as $a \rightarrow 0$, 
\item keeping $\len$ fixed as $a \rightarrow 0$, i.e. sending $n\rightarrow\infty$.
\end{enumerate}

In the first case, it is easy to see that \emph{within} the operator expansion in Eq.~\ref{eq-latmixsubst},
inverse powers of $a$ due to mixing are not an issue, since they no longer show up explicitly. 
Clearly, in the continuum limit, the physical extent $\len$ shrinks to zero, and only the operator $[O^{\mu,0}_{\GammaOp}]^\text{ren}$ contributes on the right hand side in Eq.\ref{eq-latmixsubst}, while $[O_{\GammaOp}(n\hat \mu)]^\lat$ for 
fixed $n$ is just the discrete representation of a \emph{local} continuum operator.
This \emph{local} interpretation of $[O_{\GammaOp}(n\hat \mu)]^\lat$ is, however, not the one relevant for this study.

We now turn to the second case, where the length $\len$ is kept fixed.
As $a \rightarrow 0$, $n\rightarrow\infty$, the number of terms in Eq. \eqref{eq-latmixsubst} increases,
and due to the quickly growing binomial coefficients, 
the coefficients $c_j$ eventually receive infinitely large contributions.
Without detailed knowledge about the mixing coefficients $Z_{kj}$, we cannot derive the renormalization properties of the non-local lattice operator from Eq. \eqref{eq-latmixsubst}. 
It is essential to realize, however, that the renormalized form of the non-local operator is known, both in the continuum \cite{Dotsenko:1979wb,Arefeva:1980zd,Craigie:1980qs,Stefanis:1983ke,Dorn:1986dt}, and on the lattice from heavy quark effective theory 
in the static quark limit \cite{Eichten:1989kb,Boucaud:1989ga,Maiani:1991az,Martinelli:1995vj,Martinelli:1998vt}. 
Restating Eq. \eqref{eq-opren}, the non-local operator can be written in terms of the renormalized operators as 
\begin{align}
	 [O_{\GammaOp}(n\hat \mu)]^\lat = Z_{\Psi,z}\, e^{n \, \delta \hat m }\, [O_\GammaOp(\len \hat{e}_\mu)]^\ren\,.
\end{align}
Inserting this into Eq.~\eqref{eq-latmixsubst}, we find that
\begin{align}
	[O_\GammaOp(\len \hat{e}_\mu)]^\ren& = \sum_{j=0}^\infty\ Z^{-1}_{\Psi,z}\, e^{-n \, \delta \hat m }\,c_j(n\!=\!\len/a)\, 
	\len^j [O^{\mu,j}_{\GammaOp}]^\ren\,.
	\label{eq-latmixsubst2}
\end{align}
With linearly independent $[O^{\mu,j}_{\GammaOp}]^\ren$, and assuming a marginal (not power-like) $a$-dependence of the renormalized operators, one finds that for fixed $\len$ the coefficients $c_j$ have to 
scale in unison with the lattice spacing $a$ \emph{independent of $j$}, according to
\begin{align}
	c_j \propto Z_{\Psi,z}\, e^{\delta \hat m\, \len/a}\, .
\end{align}
Such an exponential scaling of the $c_j$ is indeed not an implausible scenario and can be driven by the binomial coefficients, cf. Eq.~\ref{eq-latmixsubst}.
%\begin{align}
%	\sum_{k=0}^n \binom{n}{k} = 2^n = e^{\ln(2)\, \len/a} \ .
%\end{align}
We conclude that a simple dimensional analysis does not reveal any obvious conflict between mixing of local operators and the renormalization properties of our non-local operator.
By evaluating the non-local operator directly, 
%\xout{rather than \uline{separately the individual} \xout{its expansion in terms of} 
%local operators \uline{in the expansion}, 
we apparently bypass the severe $1/a^n$-mixing problem that complicates the computation of individual local operators with higher derivatives on the lattice. 

As a final side remark, we note that the last line of Eq. \eqref{eq-latexpan} can be simply rewritten as
\begin{align}
	[O_{\GammaOp}(n\hat \mu)]^\lat
	& = \sum_{k=0}^{\len/a} \len^k \ \underbrace{ \binom{\len/a}{k} (\len/a)^{-k} }_{\displaystyle \mathclap{\tilde C_k^\lat(\len/a) }} \ \underbrace{ \bar \quark(0)\, \GammaOp\, D_\mu^k \, \quark(0) }_{\displaystyle \equiv [O^{\mu,k}_{\GammaOp}]^\lat }\ .
	\label{eq-latexpantwo}
\end{align}
which has the form of an operator product expansion (OPE) \cite{Wilson64,Wilson:1969zs} in terms of a complete set of local operators $O_i(0)$
and dimensionless coefficients $\tilde C_i(\len \lambda)$ (see, e.g., chapter 18.3 of Ref. \cite{Peskin:1995ev})
\begin{equation}
	[O_\GammaOp(\len \hat{e}_\mu)]^\ren = \sum_{i}  \len^{d_i-3}\, \tilde C_i(\len\lambda)\, [O_i(0)]^\ren \ .
	\label{eq-ope}
\end{equation}
Here, $d_i$ denotes the canonical mass dimension of operator $O_i$,
and all renormalized operators in the above equation depend implicitly on the renormalization scale $\lambda$.
Unlike an OPE in the continuum, 
the expansion on the lattice Eq. \eqref{eq-latexpantwo} terminates after a finite number of operators, but is nevertheless an exact identity among lattice operators.
For $k \ll \len/a$, the binomial coefficient is $\tilde C_k^\lat(\len/a) \approx 1/k!$ such that 
the first terms in the sum remind us of a regular Taylor expansion. 

Interestingly, a strategy proposed to overcome issues of operator mixing in the calculation of higher moments of structure functions \cite{Detmold:2005gg} involves lattice correlators that are quite similar to those employed in the study at hand. This strategy introduces a bi-local operator $\bar q(\elll) \gamma^\mu \Psi(\elll)\,\bar \Psi(0) \gamma^\nu q(0)$ with a fictitious heavy quark field $\Psi$. 
The connection to our approach can be seen in the static quark limit $m_\Psi \rightarrow \infty$, where the field $\Psi$ can be integrated out and $\Psi(\elll)\,\bar \Psi(0)$ essentially becomes a Wilson line in 4-direction. The strategy of Ref. \cite{Detmold:2005gg} requires a continuum extrapolation and interpretation of the bi-local operator \emph{before} local operators are determined from the matching to an OPE, thus avoiding complications related to the reduced symmetries of the lattice. 
%Similarly, as corroborated by our numerical studies, our approach is only meaningful in the near-continuum situation, $\len \gg a$.

\FloatBarrier

\bibliography{TMDs}

\end{document}